\documentclass[10pt]{article}

\usepackage{amsthm,amssymb}
\usepackage{amsmath}
\usepackage{bm}   
\usepackage[top=2.5cm, bottom=2.5cm, left=2cm,right=2cm]{geometry}
\usepackage{indentfirst}

\numberwithin{equation}{section}

\newtheorem*{remark*}{Remark}
\newtheorem{theorem}{Theorem}

\usepackage[colorlinks=true,linkcolor=blue,citecolor=blue]{hyperref}
\usepackage{cite}
\usepackage{graphicx}
\usepackage{subfigure}
\usepackage{caption}
\usepackage{float}
\usepackage{authblk}
\usepackage{abstract}
\usepackage{soul}
\usepackage{xcolor}

\usepackage{dutchcal}
\usepackage{varwidth}

\title{Cauchy matrix structures and solutions to the nonisospectral three-component mKdV equations}
\author[a]{Mengli Tian}
\author[b]{Chunxia Li}
\author[a]{Yue Li}
\author[a]{Fei Li}
\author[a]{Yuqin Yao*}
\affil[a]{College of Science, China Agricultural University, Beijing 100083, China}
\affil[b]{School of Mathematical Sciences, Capital Normal University, Beijing 100048, China  \authorcr Email: yaoyq@cau.edu.cn (Yuqin Yao, *Corresponding author)}

\date{}

\begin{document}
\maketitle
\vspace{-6em}

\begin{onecolabstract}
\noindent{}\normalsize{\textbf{Abstract:} Nonisospectral integrable systems can describe solitary waves in nonuniform media. In this paper, we apply the Cauchy matrix approach to construct three types of nonisospectral matrix modified Korteweg-de Vries (mKdV) eqautions and present their Cauchy matrix structures and solutions. Further, through complex reduction, we further obtain three nonisospectral three-component mKdV (NTCmKdV) equations, which can be regarded as novel members of the nonisospectral Ablowitz-Kaup-Newell-Segur (AKNS) hierarchy. In particular, the explicit solutions are given for the soliton solutions, and the double-pole solutions, respectively. The dynamical behaviors of these solutions are analyzed to reveal the influence of nonisospectral terms on the solution structure.}\par \vspace{1em}
\noindent{}\normalsize{\textbf{Keywords:} Nonisospectral three-component mKdV equations; Cauchy matrix approach; reduction; soliton solution; double-pole solution}

\noindent{}\normalsize{\textbf{MSC:} 35A25; 37K10 }

\end{onecolabstract}

\section{Introduction}\label{sec1}
Nonisospectral integrable equations \cite{non1,non2,non3} are characterized by the involvement of a time-dependent spectral parameter $z$ in the associated Lax pair formulation. More specifically, this feature is reflected in the appearance of $x$-coefficient in the equation, along with the spectral parameter $z_{t}\ne 0$, which leads to the time-dependent soliton characteristics such as amplitude, velocity, or width. These equations are capable of describing solitary waves in nonuniform media \cite{non4,non5} and these kind of solutions are of significant importance in both physics and mathematics. To date, extensive research has been conducted on nonisospectral integrable systems. For example, the extension of the traditional inverse scattering transform (IST) method to nonisospectral integrable equations \cite{nony1,nony2}; the effects of local and nonlocal reductions of negative-order nonisospectral equations on the dynamics of explicit solutions \cite{nony3}; the generation of nonisospectral integrable systems via vector product loop algebra and the derivation of their Hamiltonian structures \cite{nony4}; the construction of a method for generating 
(2+1)-dimensional nonisospectral super-integrable hierarchies based on extended Lie superalgebras\cite{nony5}; and the derivation of Wronskian form solution for nonisospectral Kadomtsev–Petviashvili (KP) equations through the Hirota method and Wronskian technique \cite{nony6}.

Multi-component equations are have richer internal degrees of freedom and can be reduced to  a variety of novel
integrable equations. Motivated by this, we focus on the multi-component modified Korteweg–de Vries (mKdV) equation. Starting from the following higher-order Ablowitz-Kaup-Newell-Segur (AKNS) spectral problem
\begin{equation}\label{spectral}
 \Phi _{x}=\left[\begin{array}{cc}
		-i z\bm{I}_{m} & \bm{u} \\
		\bm{v}^{\mathrm{T}} & i z\bm{I}_{n}
	\end{array}\right] \Phi , 
\end{equation}
where $\bm{u}$ and $\bm{v}$ are $m \times n$ matrix functions and $\bm{I}_{m},~\bm{I}_{n}$ are $m$th- and $n$th-order identity matrices, the isospectral matrix mKdV equation
\begin{subequations}\label{mkdv}
	\begin{equation}
		\bm{u}_{t}+\bm{u}_{x x x}-3 \bm{u}_{x} \bm{v}^{\mathrm{T}} \bm{u}-3 \bm{u} \bm{v}^{\mathrm{T}} \bm{u}_{x}=\bm{0},
	\end{equation}
	\begin{equation}
		-\bm{v}_{t}-\bm{v}_{x x x}+3 \bm{v}_{x} \bm{u}^{\mathrm{T}} \bm{v}+3 \bm{v} \bm{u}^{\mathrm{T}} \bm{v}_{x}=\bm{0},
	\end{equation}
\end{subequations}
is presented in refs. \cite{yang2010,tsuchida1998}. Within the framework of the Cauchy matrix structure, we constructed the isospectral matrix mKdV equations \eqref{mkdv} and studied its local reductions and nonlocal reductions \cite{3mkdv1,3mkdv2}. Nevertheless, current literature reveals no existing research regarding nonisospectral multi-component mKdV equation.

In this paper, we would like to derive three kinds of nonisospectral three-component mKdV (NTCmKdV) equations using the Cauchy matrix approach, which was proposed by Fokas and Ablowitz \cite{fokas19811} and later further developed by Nijhoff and his collaborators \cite{N1}. The three nonisospectral equations are respectively the NTCmKdV-$\mathrm{\uppercase\expandafter{\romannumeral1}}$
\begin{equation}\label{NTCmKdV1}
\begin{aligned}
	& q_{1,t}-q_{1,xxx}-24q_{1,x}(|q_{1}|^{2}+|q_{0}|^{2})-24q_{0,x}(q_{0}^{*}q_{1}+q_{2}^{*}q_{0})-2\mathrm{i}\alpha xq_{1}=0, \\
	& q_{2,t}-q_{2,xxx}-24q_{2,x}(|q_{0}|^{2}+|q_{2}|^{2})-24q_{0,x}(q_{0}^{*}q_{2}+q_{1}^{*}q_{0})-2\mathrm{i}\alpha xq_{2}=0,\\
	&\begin{aligned} q_{0,t}-q_{0,xxx}&-12q_{0,x}(2|q_{0}|^{2}+|q_{2}|^{2}+|q_{1}|^{2})-12q_{1,x}(q_{0}^{*}q_{2}+q_{1}^{*}q_{0}) \\
	&-12q_{2,x}(q_{0}^{*}q_{1}+q_{2}^{*}q_{0})-2\mathrm{i}\alpha xq_{0}=0,
\end{aligned}
\end{aligned}
\end{equation}
the NTCmKdV-$\mathrm{\uppercase\expandafter{\romannumeral2}}$
\begin{equation}\label{NTCmKdV2}
	\begin{aligned}
		& q_{1,t}-q_{1,xxx}-24q_{1,x}(|q_{1}|^{2}+|q_{0}|^{2})-24q_{0,x}(q_{0}^{*}q_{1}+q_{2}^{*}q_{0})+\beta\left(x q_{1}\right)_{x}=0, \\
		& q_{2,t}-q_{2,xxx}-24q_{2,x}(|q_{0}|^{2}+|q_{2}|^{2})-24q_{0,x}(q_{0}^{*}q_{2}+q_{1}^{*}q_{0})+\beta\left(x q_{2}\right)_{x}=0,\\
		&\begin{aligned} q_{0,t}-q_{0,xxx}&-12q_{0,x}(2|q_{0}|^{2}+|q_{2}|^{2}+|q_{1}|^{2})-12q_{1,x}(q_{0}^{*}q_{2}+q_{1}^{*}q_{0}) \\
			&-12q_{2,x}(q_{0}^{*}q_{1}+q_{2}^{*}q_{0})+\beta\left(x q_{0}\right)_{x}=0,
		\end{aligned}
	\end{aligned}
\end{equation}
and the NTCmKdV-$\mathrm{\uppercase\expandafter{\romannumeral3}}$
\begin{equation}\label{NTCmKdV3}
\begin{aligned}
&\begin{aligned} q_{1,t}-q_{1,xxx}&-24q_{1,x}(|q_{1}|^{2}+|q_{0}|^{2})-24q_{0,x}(q_{0}^{*}q_{1}+q_{2}^{*}q_{0})\\&-\frac{1}{2} \mathrm{i} \gamma x\left(q_{1,xx}+8q_{1}\left(|q_{1}|^{2}+2|q_{0}|^{2}\right)+8q_{0}^{2}q_{2}^{*}\right)- \mathrm{i} \gamma\left(q_{1,x}+4q_{1}\partial^{-1}_{x}\left(|q_{1}|^{2}+|q_{0}|^{2}\right)+4q_{0}\partial^{-1}_{x}\left(q_{0}^{*}q_{1}+q_{2}^{*}q_{0}\right)\right)=0,
\end{aligned}\\
&\begin{aligned} q_{2,t}-q_{2,xxx}&-24q_{2,x}(|q_{0}|^{2}+|q_{2}|^{2})-24q_{0,x}(q_{0}^{*}q_{2}+q_{1}^{*}q_{0})\\&-\frac{1}{2} \mathrm{i} \gamma x\left(q_{2,xx}+8q_{2}\left(|q_{2}|^{2}+2|q_{0}|^{2}\right)+8q_{0}^{2}q_{1}^{*}\right)- \mathrm{i} \gamma\left(q_{2,x}+4q_{2}\partial^{-1}_{x}\left(|q_{2}|^{2}+|q_{0}|^{2}\right)+4q_{0}\partial^{-1}_{x}\left(q_{0}^{*}q_{2}+q_{1}^{*}q_{0}\right)\right)=0,
\end{aligned}\\
&\begin{aligned} q_{0,t}-q_{0,xxx}&-12q_{0,x}(2|q_{0}|^{2}+|q_{2}|^{2}+|q_{1}|^{2})-12q_{1,x}(q_{0}^{*}q_{2}+q_{1}^{*}q_{0})-12q_{2,x}(q_{0}^{*}q_{1}+q_{2}^{*}q_{0})\\&-\frac{1}{2} \mathrm{i} \gamma x\left(q_{0,xx}+8q_{0}\left(|q_{0}|^{2}+|q_{1}|^{2}+|q_{2}|^{2}\right)+8q_{0}^{*}q_{1}q_{2}\right)\\&- \mathrm{i} \gamma\left(q_{0,x}+2q_{1}\partial^{-1}_{x}\left(q_{0}^{*}q_{2}+q_{1}^{*}q_{0}\right)+2q_{2}\partial^{-1}_{x}\left(q_{0}^{*}q_{1}+q_{2}^{*}q_{0}\right)+2q_{0}\partial^{-1}_{x}\left(|q_{0}|^{2}+|q_{1}|^{2}\right)+2q_{0}\partial^{-1}_{x}\left(|q_{0}|^{2}+|q_{2}|^{2}\right)\right)=0,
\end{aligned}
	\end{aligned}
\end{equation}
where $*$ stands for conjugation, $q_{i}(i=1,2,3)$ represents the functions of $x$ and $t$. These equations can be regarded as new members of the nonisospectral AKNS hierarchy. As will be demonstrated in the following analysis, nonisospectral matrix integrable equations exhibit significant differences from the isospectral equations. Further, we aim to study the Cauchy matrix structure and classification of solutions of the NTCmKdV equations \eqref{NTCmKdV1}-\eqref{NTCmKdV3}.

This paper is organized as follows: In Section \ref{sec2}, we propose four kinds of time-dependent evolution relations and derive three unreduced nonisospectral matrix mKdV systems. In Section \ref{sec3}, we construct explicit $N$-soliton and multi-pole solutions for the unreduced nonisospectral mKdV equations $\mathrm{\uppercase\expandafter{\romannumeral1}}$, $\mathrm{\uppercase\expandafter{\romannumeral2}}$, and $\mathrm{\uppercase\expandafter{\romannumeral3}}$ by considering the matrix $\bm{K}$ in Sylvester equation as the diagonal matrix and the lower triangular Toeplitz matrix. In Section \ref{sec4}, we propose a reduction condition to obtain the NTCmKdV equations, and also described the internal parameter requirements for the solutions. In Section \ref{sec5}, we present the explicit one-soliton solutions,  two-soliton solutions, and double-pole solutions of the NTCmKdV equations, and analyze the dynamical behavior of the solutions.  A conclusion is provided in Section \ref{sec6}.

\section{Cauchy matrix structure of the unreduced nonisospectral mKdV equations}\label{sec2}
In this section, we propose four kinds of nonisospectral equations and establish their Cauchy matrix structures.
\subsection{Sylvester equation and master function}\label{sec21}
The form of the Sylvester equation that we need is
\begin{equation}\label{S}
\bm{KM}-\bm{MK}=\bm{rs^{\mathrm{T}}},
\end{equation}
where
\begin{equation}\label{jz}
 \bm{K}=\left(\begin{array}{cc}
	 \bm{K}_{1} &  \bm{0} \\
	\bm{0} & \bm{K}_{2}
\end{array}\right), \quad \bm{M}=\left(\begin{array}{cc}
	\bm{0} & \bm{M}_{1} \\
	\bm{M}_{2} & \bm{0}
\end{array}\right),\quad \bm{r}=\left(\begin{array}{cc}
	\bm{r}_{1} & \bm{0} \\
	\bm{0} & \bm{r}_{2}
\end{array}\right),\quad \bm{s}=\left(\begin{array}{cc}
	\bm{0} & \bm{s}_{1} \\
	\bm{s}_{2} & \bm{0}
\end{array}\right),
\end{equation}
 $\bm{K}_{i} \in \mathbb{C}_{N \times N}[t], \bm{M}_{i} \in \mathbb{C}_{N \times N}[x, t],\text{ and } \bm{r}_{i}, \bm{s}_{i} \in \mathbb{C}_{N \times 2}[x, t],i=1,2.$ In this paper, we denote $\mathcal{E}\left(\bm{A}\right)$ as the set of eigenvalues of matrix $\bm{A}$. Assuming $\mathcal{E}\left(\bm{K}_{1}\right) \cap \mathcal{E}\left(\bm{K}_{2}\right)=\varnothing$, the Sylvester equation \eqref{S} for given $\bm{K}, \bm{r}, \bm{s}$ has a unique solution $\bm{M}$\cite{Sylvester1}. Furthermore, expanding $\bm{r}_{i}$ and $\bm{s}_{i}$ in columns, we have
\begin{equation}\label{rsfl}
\bm{r}_{i}=(\bm{r}_{i}^{(1)},\bm{r}_{i}^{(2)}), \quad \bm{s}_{i} =(\bm{s}_{i}^{(1)},\bm{s}_{i}^{(2)}), \quad i=1,2.
\end{equation}
Therefore, the Sylvester equation \eqref{S} can be regarded as
\begin{subequations}\label{Kmfk}
\begin{align}
		\label{Kmfk1}
	&\bm{K}_{1} \bm{M}_{1}-\bm{M}_{1} \bm{K}_{2}=\bm{r}_{1} \bm{s}_{2}^{\mathrm{T}}=\bm{r}_{1}^{(1)}(\bm{s}_{2}^{(1)})^{\mathrm{T}}+\bm{r}_{1}^{(2)}(\bm{s}_{2}^{(2)})^{\mathrm{T}},
\\
\label{Kmfk2}
	&\bm{K}_{2} \bm{M}_{2}-\bm{M}_{2} \bm{K}_{1}=\bm{r}_{2} \bm{s}_{1}^{\mathrm{T}}=\bm{r}_{2}^{(1)}(\bm{s}_{1}^{(1)})^{\mathrm{T}}+\bm{r}_{2}^{(2)}(\bm{s}_{1}^{(2)})^{\mathrm{T}}.
\end{align}
\end{subequations}

We define the matrix master function $\bm{S}^{(i, j)}$ as 
\begin{equation}\label{sij}
	\bm{S}^{(i, j)}=\bm{s}^{\mathrm{T}} \bm{K}^{j}\left(\bm{I}_{2 N}+\bm{M}\right)^{-1} \bm{K}^{i} \bm{r}=\left(\begin{array}{cc}
		\bm{S}_{1}^{(i, j)} & \bm{S}_{2}^{(i, j)} \\
		\bm{S}_{3}^{(i, j)} & \bm{S}_{4}^{(i, j)}
	\end{array}\right), \quad\quad i, j \in \mathbb{Z},
\end{equation}
where $\bm{I}_{2N}$ represents a $2N$ order identity matrix, and $\bm{S}^{(i, j)}$ is a $4\times4 $ matrix master function. Using Eqs. \eqref{S} and \eqref{sij}, a direct calculation leads to the relation
\begin{equation}\label{sijgx}
	\bm{S}^{(i+1, j)}-\bm{S}^{(i, j+1)}
	 =\bm{s}^{\mathrm{T}}\bm{K}^{j}(\bm{I}_{2 N}+\bm{M})^{-1}(\bm{KM}-\bm{M K})(\bm{I}_{2 N}+\bm{M})^{-1} \bm{K}^{i} \bm{r} =\bm{S}^{(0, j)}\bm{S}^{(i, 0)}.
\end{equation}

Based on \eqref{jz} and \eqref{sij}, we can obtain the explicit expression for $\bm{S}_{m}^{(i, j)} (m=1,2,3,4)$, and each is a $2\times2 $ matrix function:
\begin{subequations}\label{sd}
	\begin{align}
\label{sd1}
	&\bm{S}_{1}^{(i, j)}=-\bm{s}_{2}^{\mathrm{T}} \bm{K}_{2}^{j}\bm{M}_{2}\left(\bm{I}_{N}-\bm{M}_{1} \bm{M}_{2}\right)^{-1} \bm{K}_{1}^{i} \bm{r}_{1},\\
\label{sd2}
	&\bm{S}_{2}^{(i, j)}=\bm{s}_{2}^{\mathrm{T}} \bm{K}_{2}^{j}\left(\bm{I}_{N}-\bm{M}_{2} \bm{M}_{1}\right)^{-1} \bm{K}_{2}^{i} \bm{r}_{2},\\
	\label{sd3}
	&\bm{S}_{3}^{(i, j)}=\bm{s}_{1}^{\mathrm{T}} \bm{K}_{1}^{j}\left(\bm{I}_{N}-\bm{M}_{1}\bm{M}_{2}\right)^{-1}\bm{K}_{1}^{i}\bm{r}_{1},\\
\label{sd4}
	&\bm{S}_{4}^{(i, j)}=-\bm{s}_{1}^{\mathrm{T}} \bm{K}_{1}^{j}\bm{M}_{1}\left(\bm{I}_{N}-\bm{M}_{2} \bm{M}_{1}\right)^{-1} \bm{K}_{2}^{i} \bm{r}_{2}.
\end{align}
\end{subequations}
\subsection{Unreduced nonisospectral mKdV equation}\label{sec22}

Generally, when $\bm{K}_{t}=0$, the isospectral integrable system can be derived. $\bm{K}_{t}\ne 0$ resulting  in nonisospectral  integrable systems. As for mKdV system, we can take 
\begin{equation}\label{zKt}
\bm{K}_{t}=C\bm{K}^{n}(t)\quad (n=0,1,2,3),
\end{equation}
where $C$ is a determined constant.

\textbf{Case A.} $n=0$, i.e. $\bm{K}_{t}=C$. In this case  the dispersion relations for $\bm{r}$ and $\bm{s}$ are taken as 
\begin{equation}\label{dis1}
	\begin{array}{ll}
		\bm{r}_{x}=\bm{Kr\sigma}, & \bm{s}_{x}=-\bm{K}^{\mathrm{T}}\bm{s\sigma}, \\
		\bm{r}_{t}=\left(4\bm{K}^{3}-\mathrm{i}\alpha x\right)\bm{r\sigma}, & \bm{s}_{t}=\left(-4\left(\bm{K}^{\mathrm{T}}\right)^{3}+\mathrm{i}\alpha x\right)\bm{s\sigma},
	\end{array}
\end{equation}
where $\bm{\sigma}=\operatorname{diag}\left(\bm{I}_{2},-\bm{I}_{2}\right)$, $\alpha$ is a real constant and ``$\mathrm{i}$'' is the imaginary unit. $\bm{K}$ is a matrix function related to time $t$. According to the compatibilities $\left(\bm{r}_{x}\right)_{t}=\left(\bm{r}_{t}\right)_{x}$ and  $\left(\bm{s}_{x}\right)_{t}=\left(\bm{s}_{t}\right)_{x}$, it is easy to find $\bm{K}=\bm{K}(t)$ satisfies 
\begin{equation}\label{Kt1}
\bm{K}_{t}=\frac{\mathrm{d}\bm{K}(t)}{\mathrm{d} t}=-\mathrm{i}\alpha\bm{I}_{2N\times2N}, \quad \alpha \in \mathbb{R}.
\end{equation}

Next, we respectively take the derivatives of Sylvester  equation \eqref{S} with respect to $x$ and $t$ to obtain the evolution formula of $\bm{M}$. The $x$-derivative and $t$-derivative of Eq. \eqref{S} give
\begin{subequations}
\begin{align}
	\label{kx}
&\bm{K}\bm{M}_{x}-\bm{M}_{x}\bm{K}=\bm{r}_{x}\bm{s}^{\mathrm{T}}+\bm{r}\bm{s}_{x}^{\mathrm{T}},\\
	\label{kt1}
&\bm{K}_{t}\bm{M}+\bm{K}\bm{M}_{t}-\bm{M}_{t}\bm{K}-\bm{M}\bm{K}_{t}=\bm{r}_{t}\bm{s}^{\mathrm{T}}+\bm{r}\bm{s}_{t}^{\mathrm{T}}.
\end{align}
\end{subequations}
Substituting Eq. \eqref{dis1} into Eq. \eqref{kx}, we have
\begin{equation*}
\bm{K}\left(\bm{M}_{x}-\bm{r\sigma s}^{\mathrm{T}}\right)-\left(\bm{M}_{x}-\bm{r\sigma s}^{\mathrm{T}}\right)\bm{K}=\bm{0}.
\end{equation*}
Thus, we have
\begin{equation}\label{Mx}
\bm{M}_{x}=\bm{r\sigma s}^{\mathrm{T}}.
\end{equation}
Substituting Eqs. \eqref{dis1} and \eqref{Kt1} into Eq. \eqref{kt1}, we have
\begin{equation*}
	\bm{K}\left[\bm{M}_{t}-4\left(\bm{K}^{2} \bm{r\sigma s}^{\mathrm{T}}+\bm{r\sigma s}^{\mathrm{T}}\bm{K}^{2}+\bm{K}\bm{r\sigma s}^{\mathrm{T}}\bm{K}\right)\right]-\left[\bm{M}_{t}-4\left(\bm{K}^{2} \bm{r\sigma s}^{\mathrm{T}}+\bm{r\sigma s}^{\mathrm{T}}\bm{K}^{2}+\bm{K}\bm{r\sigma s}^{\mathrm{T}}\bm{K}\right)\right]\bm{K}=\bm{0}.
\end{equation*}
Thus, we have
\begin{equation}\label{Mt1}
	\bm{M}_{t}=4\left(\bm{K}^{2} \bm{r\sigma s}^{\mathrm{T}}+\bm{r\sigma s}^{\mathrm{T}}\bm{K}^{2}+\bm{K}\bm{r\sigma s}^{\mathrm{T}}\bm{K}\right).
\end{equation}

Following this, we evolve the matrix master function $\bm{S}^{(i, j)}$, starting with the definition of the auxiliary vector function
\begin{equation}\label{fzu}
\bm{u}^{(i)}=\left(\bm{I}_{2 N}+\bm{M}\right)^{-1} \bm{K}^{i} \bm{r},
\end{equation}
that is,
\begin{equation}\label{Sugx}
\bm{S}^{(i, j)}= \bm{s}^{\mathrm{T}} \bm{K}^{j}\left(\bm{I}_{2 N}+\bm{M}\right)^{-1} \bm{K}^{i} \bm{r} =\bm{s}^{\mathrm{T}} \bm{K}^{j} \bm{u}^{(i)} .
\end{equation}
For auxiliary function \eqref{fzu}, it can be naturally obtained that
\begin{equation}\label{uiqd}
	\left(\bm{I}_{2 N}+\bm{M}\right)\bm{u}^{(i)}=\bm{K}^{i} \bm{r}.
\end{equation}
By taking the  $x$-derivative of Eq. \eqref{uiqd} and using Eqs. \eqref{dis1} and \eqref{Mx}, we obtain $\bm{u}^{(i)}_{x}$
\begin{equation}\label{ux}
\bm{u}_{x}^{(i)}=\bm{u}^{(i+1)}\bm{\sigma} -\bm{u}^{(0)} \bm{\sigma} \bm{S}^{(i, 0)}.
\end{equation}
Similarly, using Eq. \eqref{Kt1}, we can obtain the  $t$-derivative of $\bm{u}^{(i)}$
\begin{equation}\label{ut}
\bm{u}_{t}^{(i)}=4\left(\bm{u}^{(i+3)}\bm{\sigma} -\bm{u}^{(2)} \bm{\sigma} \bm{S}^{(i,0)} -\bm{u}^{(0)} \bm{\sigma} \bm{S}^{(i, 2)} -\bm{u}^{(1)} \bm{\sigma} \bm{S}^{(i, 1)}\right)-i\mathrm{i}\alpha\bm{u}^{(i-1)}-i\alpha x \bm{u}^{(i)}\bm{\sigma}.
\end{equation}
It should be noted that $\bm{K}$ is a matrix function related to $t$. It is straightforward to derive the derivative and then utilize Eqs. \eqref{dis1}, \eqref{ux} and \eqref{ut} to obtain 
\begin{subequations}
\begin{equation}\label{Sx}
\bm{S}^{(i, j)}_{x} =\left(\bm{s}^{\mathrm{T}} \bm{K}^{j} \bm{u}^{(i)}\right)_{x}=-\bm{\sigma}\bm{S}^{(i,j+1)}+ \bm{S}^{(i+1,j)}\bm{\sigma}-\bm{S}^{(0,j)}\bm{\sigma}\bm{S}^{(i,0)},
\end{equation}
\begin{equation}\label{St}
	\begin{aligned}
		\bm{S}^{(i, j)}_{t} =\left(\bm{s}^{\mathrm{T}} \bm{K}^{j} \bm{u}^{(i)}\right)_{t}=\bm{s}^{\mathrm{T}}_{t} \bm{K}^{j} \bm{u}^{(i)}+\bm{s}^{\mathrm{T}} \left(\bm{K}^{j}\right)_{t} \bm{u}^{(i)}+\bm{s}^{\mathrm{T}} \bm{K}^{j} \bm{u}^{(i)}_{t}.
	\end{aligned}
\end{equation}
\end{subequations}
Substituting Eq. \eqref{Kt1} into Eq. \eqref{St} gives
\begin{equation}
		\begin{aligned}
		\bm{S}^{(i, j)}_{t} =&-4\left(\bm{\sigma}\bm{S}^{(i,j+3)}-\bm{S}^{(i+3,j)}\bm{\sigma}+\bm{S}^{(2,j)}\bm{\sigma}\bm{S}^{(i,0)}+\bm{S}^{(0,j)}\bm{\sigma}\bm{S}^{(i,2)}+\bm{S}^{(1,j)}\bm{\sigma}\bm{S}^{(i,1)}\right)\\
		&+\mathrm{i}\alpha \left[x \left(\bm{\sigma}\bm{S}^{(i,j)}-\bm{S}^{(i,j)}\bm{\sigma}\right)-j\bm{S}^{(i,j-1)}-i\bm{S}^{(i-1,j)}\right].
			\end{aligned}
\end{equation}
By reusing Eq. \eqref{Sx}, we obtain
\begin{subequations}
	\begin{align}
		&\begin{aligned}
		\label{Sxx}
		\bm{S}^{(i, j)}_{xx} &=\bm{S}^{(i,j+2)}+\bm{S}^{(i+2,j)}+\bm{S}^{(0,j)}\bm{S}^{(i,1)}-\bm{S}^{(1,j)}\bm{S}^{(i,0)}-2\bm{\sigma}\bm{S}^{(i+1,j+1)}\bm{\sigma}+2\bm{\sigma}\bm{S}^{(0,j+1)}\bm{\sigma}\bm{S}^{(i,0)}\\&-2\bm{S}^{(0,j)}\bm{\sigma}\bm{S}^{(i+1,0)}\bm{\sigma}+2\bm{S}^{(0,j)}\bm{\sigma}\bm{S}^{(0,0)}\bm{\sigma}\bm{S}^{(i,0)},
	\end{aligned}\\
		&\begin{aligned}
				\label{Sxxx}
			\bm{S}^{(i, j)}_{xxx} &= -\bm{\sigma}\bm{S}^{(i,j+3)}+\bm{S}^{(i+3,j)}\bm{\sigma}+3\bm{S}^{(i+1,j+2)}\bm{\sigma}-3\bm{\sigma}\bm{S}^{(i+2,j+1)}-3\bm{S}^{(0,j+2)}\bm{\sigma}\bm{S}^{(i,0)}-3\bm{S}^{(0,j)}\bm{\sigma}\bm{S}^{(i+2,0)}\\&-3\bm{\sigma}\bm{S}^{(0,j+1)}\bm{S}^{(i,1)}+3\bm{\sigma}\bm{S}^{(1,j+1)}\bm{S}^{(i,0)}+3\bm{S}^{(0,j)}\bm{S}^{(i+1,1)}\bm{\sigma}-3\bm{S}^{(1,j)}\bm{S}^{(i+1,0)}\bm{\sigma}+2\bm{S}^{(1,j)}\bm{\sigma}\bm{S}^{(i,1)}\\&-\bm{S}^{(0,j)}\bm{\sigma}\bm{S}^{(i,2)}-\bm{S}^{(2,j)}\bm{\sigma}\bm{S}^{(i,0)}+3\bm{S}^{(0,j)}\bm{\sigma}\bm{S}^{(1,0)}\bm{S}^{(i,0)}-3\bm{S}^{(0,j)}\bm{\sigma}\bm{S}^{(0,0)}\bm{S}^{(i,1)}\\&-3\bm{S}^{(0,j)}\bm{S}^{(0,1)}\bm{\sigma}\bm{S}^{(i,0)}+3\bm{S}^{(1,j)}\bm{S}^{(0,0)}\bm{\sigma}\bm{S}^{(i,0)}+6\bm{\sigma}\bm{S}^{(0,j+1)}\bm{\sigma}\bm{S}^{(i+1,0)}\bm{\sigma}-6\bm{\sigma}\bm{S}^{(0,j+1)}\bm{\sigma}\bm{S}^{(0,0)}\bm{\sigma}\bm{S}^{(i,0)}\\&+6\bm{S}^{(0,j)}\bm{\sigma}\bm{S}^{(0,0)}\bm{\sigma}\bm{S}^{(i+1,0)}\bm{\sigma}-6\bm{S}^{(0,j)}\bm{\sigma}\bm{S}^{(0,0)}\bm{\sigma}\bm{S}^{(0,0)}\bm{\sigma}\bm{S}^{(i,0)}.
	 \end{aligned}
	\end{align}
\end{subequations}

Here, we make $i=j=0$ in the master function $\bm{S}^{(i, j)}$ and define
\begin{equation}\label{uj}
	\bm{U}\doteq \bm{S}^{(0,0)}=\left(\begin{array}{ll}\bm{S}_{1}^{(0,0)} & \bm{S}_{2}^{(0,0)} \\ \bm{S}_{3}^{(0,0)} & \bm{S}_{4}^{(0,0)}\end{array}\right)=\left(\begin{array}{ll}\bm{U}_{1} & \bm{U}_{2} \\ \bm{U}_{3} & \bm{U}_{4}\end{array}\right).
\end{equation}
Thus, the evolution of $\bm{S}^{(0, 0)}$ can be written as
\begin{subequations}\label{yxdU}
	\begin{align}
		&\begin{aligned}
		\label{Ux}
	\bm{U}_{x} =-\bm{\sigma}\bm{S}^{(0,1)}+ \bm{S}^{(1,0)}\bm{\sigma}-\bm{U}\bm{\sigma}\bm{U},\\
\end{aligned}\\
	&\begin{aligned}
	\label{Ut}
	\bm{U}_{t} &=-4\left(\bm{\sigma}\bm{S}^{(0,3)}-\bm{S}^{(3,0)}\bm{\sigma}+\bm{S}^{(2,0)}\bm{\sigma}\bm{U}+\bm{U}\bm{\sigma}\bm{S}^{(0,2)}+\bm{S}^{(1,0)}\bm{\sigma}\bm{S}^{(0,1)}\right)+\mathrm{i}\alpha x \left(\bm{\sigma}\bm{U}-\bm{U}\bm{\sigma}\right),
\end{aligned}\\
&\begin{aligned}
\label{Uxx}
	\bm{U}_{xx} &=\bm{S}^{(0,2)}+\bm{S}^{(2,0)}+\bm{U}\bm{S}^{(0,1)}-\bm{S}^{(1,0)}\bm{U}-2\bm{\sigma}\bm{S}^{(1,1)}\bm{\sigma}+2\bm{\sigma}\bm{S}^{(0,1)}\bm{\sigma}\bm{U}-2\bm{U}\bm{\sigma}\bm{S}^{(1,0)}\bm{\sigma}+2\bm{U}\bm{\sigma}\bm{U}\bm{\sigma}\bm{U},
\end{aligned}\\
&\begin{aligned}
\label{Uxxx}
	\bm{U}_{xxx} &= -\bm{\sigma}\bm{S}^{(0,3)}+\bm{S}^{(3,0)}\bm{\sigma}+3\bm{S}^{(1,2)}\bm{\sigma}-3\bm{\sigma}\bm{S}^{(2,1)}-3\bm{S}^{(0,2)}\bm{\sigma}\bm{U}-3\bm{U}\bm{\sigma}\bm{S}^{(2,0)}-3\bm{\sigma}\left(\bm{S}^{(0,1)}\right)^{2}\\&+3\bm{\sigma}\bm{S}^{(1,1)}\bm{U}+3\bm{U}\bm{S}^{(1,1)}\bm{\sigma}-3\left(\bm{S}^{(1,0)}\right)^{2}\bm{\sigma}+2\bm{S}^{(1,0)}\bm{\sigma}\bm{S}^{(0,1)}-\bm{U}\bm{\sigma}\bm{S}^{(0,2)}-\bm{S}^{(2,0)}\bm{\sigma}\bm{U}\\&+3\bm{U}\bm{\sigma}\bm{S}^{(1,0)}\bm{U}-3\bm{U}\bm{\sigma}\bm{U}\bm{S}^{(0,1)}-3\bm{U}\bm{S}^{(0,1)}\bm{\sigma}\bm{U}+3\bm{S}^{(1,0)}\bm{U}\bm{\sigma}\bm{U}\\&+6\bm{\sigma}\bm{S}^{(0,1)}\bm{\sigma}\bm{S}^{(1,0)}\bm{\sigma}-6\bm{\sigma}\bm{S}^{(0,1)}\bm{\sigma}\bm{U}\bm{\sigma}\bm{U}+6\bm{U}\bm{\sigma}\bm{U}\bm{\sigma}\bm{S}^{(1,0)}\bm{\sigma}-6\bm{U}\bm{\sigma}\bm{U}\bm{\sigma}\bm{U}\bm{\sigma}\bm{U}.
\end{aligned}
	\end{align}
\end{subequations}

From Eqs. \eqref{sijgx} and \eqref{uj}, we have
\begin{equation}\label{Sijgx}
	\begin{aligned}
		&\bm{S}^{(0,1)}= \bm{S}^{(1,0)}-\bm{U}^{2}, \quad  \bm{S}^{(2,0)} = \bm{S}^{(1,1)}+\bm{U}\bm{S}^{(1,0)}, \quad \bm{S}^{(0,2)}= \bm{S}^{(1,1)}-\bm{S}^{(1,0)}\bm{U}+\bm{U}^{3},\\&\bm{S}^{(2,1)}=\bm{S}^{(1,2)}+\left(\bm{S}^{(1,0)}\right)^{2}-\bm{U}^{2}\bm{S}^{(1,0)},\quad  \bm{S}^{(0,3)}=\bm{S}^{(1,2)}-\bm{S}^{(1,1)}\bm{U}+\bm{S}^{(1,0)}\bm{U}^{2}-U^{4},\\& \bm{S}^{(3,0)}= \bm{S}^{(1,2)}+\left(\bm{S}^{(1,0)}\right)^{2}+\bm{U}\bm{S}^{(1,1)}.
	\end{aligned}
\end{equation}
Substituting Eq. \eqref{Sijgx} into Eq. \eqref{yxdU} gives
\begin{subequations}
	\begin{align}
		&\begin{aligned}
			\label{xUx}
			\bm{U}_{x} =-\bm{\sigma}\bm{S}^{(1,0)}+ \bm{S}^{(1,0)}\bm{\sigma}+\bm{\sigma}\bm{U}^{2}-\bm{U}\bm{\sigma}\bm{U},
		\end{aligned}\\
		&\begin{aligned}
			\label{xUt}
			\bm{U}_{t} &=-4\left(\bm{\sigma}\bm{S}^{(1,2)}-\bm{\sigma}\bm{S}^{(1,1)}\bm{U}+\bm{\sigma}\bm{S}^{(1,0)}\bm{U}^{2}-\bm{\sigma}\bm{U}^{4}-\bm{S}^{(1,2)}\bm{\sigma}-\left(\bm{S}^{(1,0)}\right)^{2}\bm{\sigma}-\bm{U}\bm{S}^{(1,1)}\bm{\sigma}\right.\\&+\left.\bm{S}^{(1,1)}\bm{\sigma}\bm{U}+\bm{U}\bm{S}^{(1,0)}\bm{\sigma}\bm{U}+\bm{U}\bm{\sigma}\bm{S}^{(1,1)}-\bm{U}\bm{\sigma}\bm{S}^{(1,0)}\bm{U}+\bm{U}\bm{\sigma}\bm{U}^{3}+\bm{S}^{(1,0)}\bm{\sigma}\bm{S}^{(1,0)}-\bm{S}^{(1,0)}\bm{\sigma}\bm{U}^{2}\right)\\&+\mathrm{i}\alpha x \left(\bm{\sigma}\bm{U}-\bm{U}\bm{\sigma}\right),
		\end{aligned}\\
		&\begin{aligned}
			\label{xUxx}
			\bm{U}_{xx} &=2\bm{S}^{(1,1)}-2\bm{S}^{(1,0)}\bm{U}+2\bm{U}\bm{S}^{(1,0)}-2\bm{\sigma}\bm{S}^{(1,1)}\bm{\sigma}+2\bm{\sigma}\bm{S}^{(1,0)}\bm{\sigma}\bm{U}-2\bm{U}\bm{\sigma}\bm{S}^{(1,0)}\bm{\sigma}-2\bm{\sigma}\bm{U}^{2}\bm{\sigma}\bm{U}\\&+2\bm{U}\bm{\sigma}\bm{U}\bm{\sigma}\bm{U},
		\end{aligned}\\
		&\begin{aligned}
			\label{xUxxx}
			\bm{U}_{xxx}&= -2\bm{\sigma}\bm{U}^{4}+4\bm{S}^{(1,2)}\bm{\sigma}-4\bm{\sigma}\bm{S}^{(1,2)}+4\bm{\sigma}\bm{S}^{(1,1)}\bm{U}+4\bm{U}\bm{S}^{(1,1)}\bm{\sigma}-4\bm{U}\bm{\sigma}\bm{S}^{(1,1)}-4\bm{S}^{(1,1)}\bm{\sigma}\bm{U}\\&-2\left(\bm{S}^{(1,0)}\right)^{2}\bm{\sigma}-6\bm{\sigma}\left(\bm{S}^{(1,0)}\right)^{2}-2\bm{S}^{(1,0)}\bm{\sigma}\bm{U}^{2}+2\bm{U}\bm{\sigma}\bm{U}^{3}-6\bm{U}\bm{\sigma}\bm{U}\bm{\sigma}\bm{U}\bm{\sigma}\bm{U}+6\bm{U}\bm{\sigma}\bm{U}\bm{\sigma}\bm{S}^{(1,0)}\bm{\sigma}\\&+6\bm{\sigma}\bm{U}^{2}\bm{\sigma}\bm{U}\bm{\sigma}\bm{U}-6\bm{\sigma}\bm{S}^{(1,0)}\bm{\sigma}\bm{U}\bm{\sigma}\bm{U}-6\bm{\sigma}\bm{U}^{2}\bm{\sigma}\bm{S}^{(1,0)}\bm{\sigma}+6\bm{S}^{(1,0)}\bm{U}\bm{\sigma}\bm{U}+4\bm{U}\bm{\sigma}\bm{S}^{(1,0)}\bm{U}\\&-6\bm{U}\bm{\sigma}\bm{U}\bm{S}^{(1,0)}+6\bm{\sigma}\bm{S}^{(1,0)}\bm{\sigma}\bm{S}^{(1,0)}\bm{\sigma}-4\bm{U}\bm{S}^{(1,0)}\bm{\sigma}\bm{U}+2\bm{S}^{(1,0)}\bm{\sigma}\bm{S}^{(1,0)}+6\bm{\sigma}\bm{U}^{2}\bm{S}^{(1,0)}+2\bm{\sigma}\bm{S}^{(1,0)}\bm{U}^{2}.
		\end{aligned}
	\end{align}
\end{subequations}
A direct calculation gives
\begin{equation}\label{eq1}
	\begin{aligned}
\bm{U}_{t}-\bm{U}_{xxx}&=6\left(-\left[\bm{\sigma},\bm{U}\right]\bm{U}\left(\left[\bm{S}^{(1,0)}\bm{\sigma},\bm{\sigma}\right]+\left[\bm{\sigma},\bm{U}\right]\bm{\sigma}\bm{U}\right)+\left[\bm{\sigma},\bm{S}^{(1,0)}\right]\left(\bm{S}^{(1,0)}-\bm{U}^{2}\right)+\left[\bm{S}^{(1,0)},\bm{\sigma}\right]\left(\bm{S}^{(1,0)}\bm{\sigma}-\bm{U}\bm{\sigma}\bm{U}\right)\right)\\&+\mathrm{i}\alpha x \left[\bm{\sigma},\bm{U}\right],
	\end{aligned}
\end{equation}
where $\left[\bm{A},\bm{B}\right]=\bm{AB}-\bm{BA}$. By expanding the matrix form of Eq. \eqref{eq1} using Eq. \eqref{uj}, a set of equations related only to $\bm{U}_{2}$ and $\bm{U}_{3}$ can be obtained, which is called the unreduced nonisospectral mKdV equation-$\mathrm{\uppercase\expandafter{\romannumeral1}}$
\begin{equation}\label{wyh1}
	\begin{aligned}
		&\bm{U}_{2, t}- \bm{U}_{2, x x x}- 12 \bm{U}_{2, x}\bm{U}_{3}\bm{U}_{2}- 12 \bm{U}_{2}\bm{U}_{3}\bm{U}_{2, x}-2\mathrm{i}\alpha x\bm{U}_{2}=\bm{0},\\
		&\bm{U}_{3, t}- \bm{U}_{3, x x x}- 12 \bm{U}_{3, x}\bm{U}_{2}\bm{U}_{3}- 12 \bm{U}_{3}\bm{U}_{2}\bm{U}_{3, x}+2\mathrm{i}\alpha x\bm{U}_{3}=\bm{0}.
	\end{aligned}
\end{equation}

\textbf{Case B.} $n=1$, i.e. $\bm{K}_{t}=C\bm{K}(t)$.  In this case, the dispersion relations for $\bm{r}$ and $\bm{s}$ are
\begin{equation}\label{dis2}
	\begin{array}{ll}
		\bm{r}_{x}=\bm{Kr\sigma}, & \bm{s}_{x}=-\bm{K}^{\mathrm{T}}\bm{s\sigma}, \\
		\bm{r}_{t}=\left(4\bm{K}^{3}-\beta x\bm{K}\right)\bm{r\sigma}-\frac{1}{2} \beta \bm{r}, & \bm{s}_{t}=\left(-4\left(\bm{K}^{\mathrm{T}}\right)^{3}+\beta x\bm{K}^{\mathrm{T}}\right)\bm{s\sigma}-\frac{1}{2} \beta \bm{s},
	\end{array}
\end{equation}
where $\beta$ is a real constant. According to the compatibilities, we can find $C=-\beta $. 

Since $\bm{r}_{x}$ and $\bm{s}_{x}$ have no change compared with Eq. \eqref{dis1}, the expressions for $\bm{M}_{x}$, $\bm{S}^{(i, j)}_{x}$, $\bm{S}^{(i, j)}_{xx}$ and $\bm{S}^{(i, j)}_{xxx}$  are the same as in Case A. For the $t$-derivative of $\bm{M}$, it can be obtained by the same method as in Eq. \eqref{Mt1}
\begin{equation}\label{Mt2}
	\bm{M}_{t}=4\left(\bm{K}^{2} \bm{r\sigma s}^{\mathrm{T}}+\bm{r\sigma s}^{\mathrm{T}}\bm{K}^{2}+\bm{K}\bm{r\sigma s}^{\mathrm{T}}\bm{K}\right)-\beta x\bm{r\sigma s}^{\mathrm{T}} .
\end{equation}
Differentiating both sides of Eq. \eqref{uiqd} with respect to $t$, we have
\begin{equation}\label{uiqd2}
	\bm{M}_{t}\bm{u}^{(i)}+\left(\bm{I}_{2 N}+\bm{M}\right)\bm{u}^{(i)}_{t}=i\bm{K}^{i-1}\bm{K}_{t} \bm{r}+\bm{K}^{i}\bm{r}_{t},
\end{equation}
where $\bm{u}^{(i)}$ is defined in Eq. \eqref{fzu}. By substituting Eqs. \eqref{dis2}, \eqref{Mt2} and $\bm{K}_{t}=-\beta \bm{K}(t)$ into Eq. \eqref{uiqd2}, we obtain
\begin{equation}\label{ut2}
	\bm{u}_{t}^{(i)}=4\left(\bm{u}^{(i+3)}\bm{\sigma} -\bm{u}^{(2)} \bm{\sigma}\bm{S}^{(i,0)}-\bm{u}^{(0)} \bm{\sigma} \bm{S}^{(i, 2)}-\bm{u}^{(1)} \bm{\sigma} \bm{S}^{(i, 1)}\right)-i\beta\bm{u}^{(i)}-\frac{1}{2} \beta\bm{u}^{(i)}+\beta x\left(\bm{u}^{(0)}\bm{\sigma}\bm{S}^{(i,0)}-\bm{u}^{(i+1)}\bm{\sigma} \right).
\end{equation}
Similar to Case A, we can obtain the $t$-derivative of $\bm{S}^{(i, j)}$
\begin{equation}\label{St2}
	\begin{aligned}
		\bm{S}^{(i, j)}_{t} &=-4\left(\bm{\sigma}\bm{S}^{(i,j+3)}-\bm{S}^{(i+3,j)}\bm{\sigma}+\bm{S}^{(2,j)}\bm{\sigma}\bm{S}^{(i,0)}+\bm{S}^{(0,j)}\bm{\sigma}\bm{S}^{(i,2)}+\bm{S}^{(1,j)}\bm{\sigma}\bm{S}^{(i,1)}+\frac{1}{4}\left(1+i+j\right)\beta\bm{S}^{(i, j)}\right)\\
		&+\beta x  \left(\bm{\sigma}\bm{S}^{(i,j+1)}+\bm{S}^{(0,j)}\bm{\sigma}\bm{S}^{(i,0)}-\bm{S}^{(i+1,j)}\bm{\sigma} \right).
	\end{aligned}
\end{equation}
Taking $i=j=0$ in Eq. \eqref{St2}, we obtain 
\begin{equation*}
\begin{aligned}
	\bm{U}_{t} &=-4\left(\bm{\sigma}\bm{S}^{(0,3)}-\bm{S}^{(3,0)}\bm{\sigma}
	+\bm{S}^{(2,0)}\bm{\sigma}\bm{U}+\bm{S}^{(1,0)}\bm{\sigma}\bm{S}^{(0,1)}+\bm{U}\bm{\sigma}\bm{S}^{(0,2)}+\frac{1}{4}\beta\bm{U}\right)+\beta x \left(\bm{\sigma}\bm{S}^{(0,1)}+\bm{U}\bm{\sigma}\bm{U}-\bm{S}^{(1,0)}\bm{\sigma}\right).
\end{aligned}
\end{equation*}
By using Eq. \eqref{Sijgx}, we find 
\begin{equation}
\begin{aligned}
	\label{xUt2}
	\bm{U}_{t} &=-4\left(\bm{\sigma}\bm{S}^{(1,2)}-\bm{\sigma}\bm{S}^{(1,1)}\bm{U}+\bm{\sigma}\bm{S}^{(1,0)}\bm{U}^{2}-\bm{\sigma}\bm{U}^{4}-\bm{S}^{(1,2)}\bm{\sigma}-\left(\bm{S}^{(1,0)}\right)^{2}\bm{\sigma}-\bm{U}\bm{S}^{(1,1)}\bm{\sigma}\right.\\&+\left.\bm{S}^{(1,1)}\bm{\sigma}\bm{U}+\bm{U}\bm{S}^{(1,0)}\bm{\sigma}\bm{U}+\bm{U}\bm{\sigma}\bm{S}^{(1,1)}-\bm{U}\bm{\sigma}\bm{S}^{(1,0)}\bm{U}+\bm{U}\bm{\sigma}\bm{U}^{3}+\bm{S}^{(1,0)}\bm{\sigma}\bm{S}^{(1,0)}-\bm{S}^{(1,0)}\bm{\sigma}\bm{U}^{2}+\frac{1}{4}\beta\bm{U}\right)\\&+\beta x \left(\bm{\sigma}\bm{S}^{(1,0)}-\bm{S}^{(1,0)}\bm{\sigma}+\bm{U}\bm{\sigma}\bm{U}-\bm{\sigma}\bm{U}^{2}\right).
\end{aligned}
\end{equation}
Based on Eqs. \eqref{xUxxx} and \eqref{xUt2}, and using the definition of Eq. \eqref{uj}, a set of equations only related to $\bm{U}_{2}$ and $\bm{U}_{3}$  can be presented
\begin{equation}\label{wyh2}
	\begin{aligned}
		&\bm{U}_{2, t}- \bm{U}_{2, x x x}- 12 \bm{U}_{2, x}\bm{U}_{3}\bm{U}_{2}- 12 \bm{U}_{2}\bm{U}_{3}\bm{U}_{2, x}+\beta \left(x\bm{U}_{2}\right)_{x}=\bm{0},\\
		&\bm{U}_{3, t}- \bm{U}_{3, x x x}- 12 \bm{U}_{3, x}\bm{U}_{2}\bm{U}_{3}- 12 \bm{U}_{3}\bm{U}_{2}\bm{U}_{3, x}+\beta \left(x\bm{U}_{3}\right)_{x}=\bm{0},
	\end{aligned}
\end{equation}
which is called the unreduced nonisospectral mKdV equation-$\mathrm{\uppercase\expandafter{\romannumeral2}}$.

\textbf{Case C.} $n=2$, i.e. $\bm{K}_{t}=C\bm{K}^{2}(t)$. In this case, the dispersion relations for $\bm{r}$ and $\bm{s}$ are
\begin{equation}\label{dis3}
	\begin{array}{ll}
		\bm{r}_{x}=\bm{Kr\sigma}, & \bm{s}_{x}=-\bm{K}^{\mathrm{T}}\bm{s\sigma}, \\
		\bm{r}_{t}=\left(4\bm{K}^{3}-\mathrm{i}\gamma x\bm{K}^{2}\right)\bm{r\sigma}-\mathrm{i}\gamma\bm{K}\bm{r}, & \bm{s}_{t}=\left(-4\left(\bm{K}^{\mathrm{T}}\right)^{3}+\mathrm{i}\gamma x\left(\bm{K}^{\mathrm{T}}\right)^{2}\right)\bm{s\sigma}-\mathrm{i}\gamma\bm{K}^{\mathrm{T}}\bm{s},
	\end{array}
\end{equation}
where $\gamma$ is a real constant and $\mathrm{i}$ is an  imaginary unit. Through the same calculation as the previous two cases, we find $C=-\mathrm{i}\gamma$. 

For the $t$-derivative of $\bm{M}$, we have
\begin{equation}\label{Mt3}
	\bm{M}_{t}=4\left(\bm{K}^{2} \bm{r\sigma s}^{\mathrm{T}}+\bm{r\sigma s}^{\mathrm{T}}\bm{K}^{2}+\bm{K}\bm{r\sigma s}^{\mathrm{T}}\bm{K}\right)-\mathrm{i}\gamma x \left(\bm{K}\bm{r\sigma s}^{\mathrm{T}}+\bm{r\sigma s}^{\mathrm{T}}\bm{K}\right) .
\end{equation}
Thus, for the evolution of $\bm{u}^{(i)}$ and $\bm{S}^{(i, j)}$ with respect to $t$, we have
\begin{subequations}
	\begin{align}
		&\begin{aligned}
			\label{ut3}
			\bm{u}_{t}^{(i)}&=4\left(\bm{u}^{(i+3)}\bm{\sigma} -\bm{u}^{(2)} \bm{\sigma}\bm{S}^{(i,0)}-\bm{u}^{(0)} \bm{\sigma} \bm{S}^{(i, 2)}-\bm{u}^{(1)} \bm{\sigma} \bm{S}^{(i, 1)}\right)-i\mathrm{i}\gamma\bm{u}^{(i+1)}-\mathrm{i}\gamma\bm{u}^{(i+1)}\\
			&+\mathrm{i}\gamma x\left(\bm{u}^{(1)}\bm{\sigma}\bm{S}^{(i,0)}+\bm{u}^{(0)}\bm{\sigma}\bm{S}^{(i,1)}-\bm{u}^{(i+2)}\bm{\sigma} \right),
		\end{aligned}\\
		&\begin{aligned}
			\label{St3}
			\bm{S}^{(i, j)}_{t} &=-4\left(\bm{\sigma}\bm{S}^{(i,j+3)}-\bm{S}^{(i+3,j)}\bm{\sigma}+\bm{S}^{(2,j)}\bm{\sigma}\bm{S}^{(i,0)}+\bm{S}^{(0,j)}\bm{\sigma}\bm{S}^{(i,2)}+\bm{S}^{(1,j)}\bm{\sigma}\bm{S}^{(i,1)}\right.\\
			&\left.+\frac{1}{4}\mathrm{i}\gamma \left(1+j\right)\bm{S}^{(i, j+1)}+\frac{1}{4}\mathrm{i}\gamma \left(1+i\right)\bm{S}^{(i+1, j)}\right)\\
			&+\mathrm{i}\gamma x  \left(\bm{\sigma}\bm{S}^{(i,j+2)}-\bm{S}^{(i+2,j)}\bm{\sigma}+\bm{S}^{(1,j)}\bm{\sigma}\bm{S}^{(i,0)}+\bm{S}^{(0,j)}\bm{\sigma}\bm{S}^{(i,1)} \right).
		\end{aligned}
	\end{align}
\end{subequations}
Taking $i=j=0$ in Eq. \eqref{St3}, we obtain 
\begin{equation*}
	\begin{aligned}
		\bm{U}_{t} &=-4\left(\bm{\sigma}\bm{S}^{(0,3)}-\bm{S}^{(3,0)}\bm{\sigma}
		+\bm{S}^{(2,0)}\bm{\sigma}\bm{U}+\bm{S}^{(1,0)}\bm{\sigma}\bm{S}^{(0,1)}+\bm{U}\bm{\sigma}\bm{S}^{(0,2)}+\frac{1}{4}\mathrm{i}\gamma\bm{S}^{(0,1)}+\frac{1}{4}\mathrm{i}\gamma\bm{S}^{(1,0)}\right)\\
		&+\mathrm{i}\gamma x \left(\bm{\sigma}\bm{S}^{(0,2)}-\bm{S}^{(2,0)}\bm{\sigma}+\bm{S}^{(1,0)}\bm{\sigma}\bm{U}+\bm{U}\bm{\sigma}\bm{S}^{(0,1)}\right),
	\end{aligned}
\end{equation*}
which means 
\begin{equation}
	\begin{aligned}
		\label{xUt3}
		\bm{U}_{t} &=-4\left(\bm{\sigma}\bm{S}^{(1,2)}-\bm{\sigma}\bm{S}^{(1,1)}\bm{U}+\bm{\sigma}\bm{S}^{(1,0)}\bm{U}^{2}-\bm{\sigma}\bm{U}^{4}-\bm{S}^{(1,2)}\bm{\sigma}-\left(\bm{S}^{(1,0)}\right)^{2}\bm{\sigma}-\bm{U}\bm{S}^{(1,1)}\bm{\sigma}\right.\\&+\left.\bm{S}^{(1,1)}\bm{\sigma}\bm{U}+\bm{U}\bm{S}^{(1,0)}\bm{\sigma}\bm{U}+\bm{U}\bm{\sigma}\bm{S}^{(1,1)}-\bm{U}\bm{\sigma}\bm{S}^{(1,0)}\bm{U}+\bm{U}\bm{\sigma}\bm{U}^{3}+\bm{S}^{(1,0)}\bm{\sigma}\bm{S}^{(1,0)}-\bm{S}^{(1,0)}\bm{\sigma}\bm{U}^{2}+\frac{1}{2}\mathrm{i}\gamma\bm{S}^{(1,0)}-\frac{1}{4}\mathrm{i}\gamma\bm{U}^{2}\right)\\&+\mathrm{i}\gamma x \left(\bm{\sigma}\bm{S}^{(1,1)}-\bm{S}^{(1,1)}\bm{\sigma}+\bm{\sigma}\bm{U}^{3}-\bm{U}\bm{\sigma}\bm{U}^{2}+\bm{S}^{(1,0)}\bm{\sigma}\bm{U}-\bm{\sigma}\bm{S}^{(1,0)}\bm{U}+\bm{U}\bm{\sigma}\bm{S}^{(1,0)}-\bm{U}\bm{S}^{(1,0)}\bm{\sigma}\right).
	\end{aligned}
\end{equation}

By expanding the matrix \eqref{xUx}, there is the relation
\begin{equation}
	\bm{U}_{1}=2\partial^{-1}_{x}\left(\bm{U}_{2}\bm{U}_{3}\right), \quad
	\bm{U}_{4}=-2\partial^{-1}_{x}\left(\bm{U}_{3}\bm{U}_{2}\right),
\end{equation}
where
$\partial_{x}^{-1}=\frac{1}{2}\left(\int_{-\infty}^{x} \cdot \mathrm{~d} x-\int_{x}^{+\infty} \cdot \mathrm{~d} x\right)$.

Thus, after a series of lengthy but direct operations, we can get a set of equations only related to $\bm{U}_{2}$ and $\bm{U}_{3}$. It is called the unreduced nonisospectral mKdV equation-$\mathrm{\uppercase\expandafter{\romannumeral3}}$
\begin{subequations}\label{wyh3}
	\begin{align}
	&\begin{aligned}
		\bm{U}_{2, t}- \bm{U}_{2, x x x}- 12 \bm{U}_{2, x}\bm{U}_{3}\bm{U}_{2}- 12 \bm{U}_{2}\bm{U}_{3}\bm{U}_{2, x}&-\mathrm{i}\gamma x \left(\frac{1}{2}\bm{U}_{2, x x}+4\bm{U}_{2}\bm{U}_{3}\bm{U}_{2}\right)\\&-\mathrm{i}\gamma\left(\bm{U}_{2, x}+2\partial^{-1}_{x}\left(\bm{U}_{2}\bm{U}_{3}\right)\bm{U}_{2}+2\bm{U}_{2}\partial^{-1}_{x}\left(\bm{U}_{3}\bm{U}_{2}\right)\right)=\bm{0},
	\end{aligned}\\
	&\begin{aligned}
		\bm{U}_{3, t}- \bm{U}_{3, x x x}- 12 \bm{U}_{3, x}\bm{U}_{2}\bm{U}_{3}- 12 \bm{U}_{3}\bm{U}_{2}\bm{U}_{3, x}&+\mathrm{i}\gamma x \left(\frac{1}{2}\bm{U}_{3, x x}+4\bm{U}_{3}\bm{U}_{2}\bm{U}_{3}\right)\\&+\mathrm{i}\gamma\left(\bm{U}_{3, x}+2\partial^{-1}_{x}\left(\bm{U}_{3}\bm{U}_{2}\right)\bm{U}_{3}+2\bm{U}_{3}\partial^{-1}_{x}\left(\bm{U}_{2}\bm{U}_{3}\right)\right)=\bm{0}.
		\end{aligned}
\end{align}
\end{subequations}

\textbf{Case D.} $n=3$, i.e. $\bm{K}_{t}=C\bm{K}^{3}(t)$.  In this case, the dispersion relations for $\bm{r}$ and $\bm{s}$ are
\begin{equation}\label{dis4}
	\begin{array}{ll}
		\bm{r}_{x}=\bm{Kr\sigma}, & \bm{s}_{x}=-\bm{K}^{\mathrm{T}}\bm{s\sigma}, \\
		\bm{r}_{t}=4x\bm{K}^{3}\bm{r\sigma}+6\bm{K}^{2}\bm{r}, & \bm{s}_{t}=-4x\left(\bm{K}^{\mathrm{T}}\right)^{3}\bm{s\sigma}+6\left(\bm{K}^{\mathrm{T}}\right)^{2}\bm{s}.
	\end{array}
\end{equation}
The compatibilities imply that $C=4$. After a series of operations, the evolutions of $\bm{M}$, $\bm{u}^{(i)}$ and $\bm{S}^{(i, j)}$ with respect to $t$ are
\begin{subequations}
	\begin{align}
		&\begin{aligned}
\label{Mt4}
	\bm{M}_{t}=4x\left(\bm{K}^{2} \bm{r\sigma s}^{\mathrm{T}}+\bm{r\sigma s}^{\mathrm{T}}\bm{K}^{2}+\bm{K}\bm{r\sigma s}^{\mathrm{T}}\bm{K}\right)+2\bm{K}\bm{rs}^{\mathrm{T}}-2\bm{rs}^{\mathrm{T}}\bm{K},
		\end{aligned}\\
		&\begin{aligned}
			\label{ut4}
			\bm{u}_{t}^{(i)}&=4x\left(\bm{u}^{(i+3)}\bm{\sigma} -\bm{u}^{(2)}\bm{\sigma}\bm{S}^{(i,0)}-\bm{u}^{(0)} \bm{\sigma} \bm{S}^{(i, 2)}-\bm{u}^{(1)} \bm{\sigma} \bm{S}^{(i,1)}\right)\\
			&+2\left(\left(2i+3\right)\bm{u}^{(i+2)}-\bm{u}^{(1)}\bm{S}^{(i,0)}+\bm{u}^{(0)}\bm{S}^{(i,1)}\right),
		\end{aligned}\\
		&\begin{aligned}
			\label{St4}
			\bm{S}^{(i, j)}_{t} &=-4x\left(\bm{\sigma}\bm{S}^{(i,j+3)}-\bm{S}^{(i+3,j)}\bm{\sigma}+\bm{S}^{(2,j)}\bm{\sigma}\bm{S}^{(i,0)}+\bm{S}^{(0,j)}\bm{\sigma}\bm{S}^{(i,2)}+\bm{S}^{(1,j)}\bm{\sigma}\bm{S}^{(i,1)}\right)\\
			&+2\left(\left(2i+3\right)\bm{S}^{(i+2,j)}+\left(2j+3\right)\bm{S}^{(i,j+2)}-\bm{S}^{(1,j)}\bm{S}^{(i,0)}+\bm{S}^{(1,j)}\bm{S}^{(i,1)}\right).
		\end{aligned}
\end{align}
\end{subequations}
Taking $i=j=0$ in Eq. \eqref{St4}, we obtain 
\begin{equation}
\begin{aligned}
	\label{xUt4}
	\bm{U}_{t} &=-4x\left(\bm{\sigma}\bm{S}^{(1,2)}-\bm{\sigma}\bm{S}^{(1,1)}\bm{U}+\bm{\sigma}\bm{S}^{(1,0)}\bm{U}^{2}-\bm{\sigma}\bm{U}^{4}-\bm{S}^{(1,2)}\bm{\sigma}-\left(\bm{S}^{(1,0)}\right)^{2}\bm{\sigma}-\bm{U}\bm{S}^{(1,1)}\bm{\sigma}\right.\\&+\left.\bm{S}^{(1,1)}\bm{\sigma}\bm{U}+\bm{U}\bm{S}^{(1,0)}\bm{\sigma}\bm{U}+\bm{U}\bm{\sigma}\bm{S}^{(1,1)}-\bm{U}\bm{\sigma}\bm{S}^{(1,0)}\bm{U}+\bm{U}\bm{\sigma}\bm{U}^{3}+\bm{S}^{(1,0)}\bm{\sigma}\bm{S}^{(1,0)}-\bm{S}^{(1,0)}\bm{\sigma}\bm{U}^{2}\right)\\&+4 \left(3\bm{S}^{(1,1)}+2\bm{U}\bm{S}^{(1,0)}-2\bm{S}^{(1,0)}\bm{U}+\bm{U}^{3}\right).
\end{aligned}
\end{equation}
\begin{remark*}
Since matrix multiplication does not necessarily satisfy the commutative law, that is $\bm{U}\bm{S}^{(1,0)}\ne \bm{S}^{(1,0)}\bm{U}$ in Eq. \eqref{xUt4}. Consequently, it is impossible to formulate nonisospectral equations in this case. However, if the main function $\bm{S}^{(i, j)}$ is a scalar function, the valid equation under this dispersion relation can be found in \cite{nonA2} as
\begin{equation*}
	u_{t}-x\left(u_{x x x}+6 u u_{x}\right)-4 u_{x x}-8 u^{2}-2 u_{x} \partial_{x}^{-1} u=0.
\end{equation*}
This fundamentally distinguishes multi-component systems from single-component systems.
\end{remark*}

\section{Solutions to the unreduced nonisospectral mKdV equations}\label{sec3}
In this section, we will construct the explicit solutions of the unreduced nonisospectral mKdV equations $\mathrm{\uppercase\expandafter{\romannumeral1}}$ \eqref{wyh1}, $\mathrm{\uppercase\expandafter{\romannumeral2}}$ \eqref{wyh2}, and $\mathrm{\uppercase\expandafter{\romannumeral3}}$ \eqref{wyh3}. Their solutions are determined by a unique main function $\bm{S}^{(0, 0)}$. It is worth noting that in the nonisospectral case, $\bm{K}$ no longer behaves as a constant matrix like in the isospectral case \cite{iso1, iso2,iso3}, but becomes a matrix function with respect to $t$. So here, we consider $\bm{K}_{1}$ and $\bm{K}_{2}$ as diagonal matrices to obtain $N$-soliton solutions, and $\bm{K}_{1}$ and $\bm{K}_{2}$ as matrices composed of the lower triangular Toeplitz matrices to obtain multi-pole solutions.

\subsection{\texorpdfstring{$\bm{K}$}{K} is the diagonal matrix}\label{sec31}
When $\bm{K}_{1}$ and $\bm{K}_{2}$ are diagonal matrices,
\begin{equation}\label{k1k2}
		\bm{K}_{1}=\operatorname{diag}\left(k_{1}, k_{2}, \cdots, k_{N}\right), \quad \bm{K}_{2}=\operatorname{diag}\left(g_{1}, g_{2}, \cdots, g_{N}\right),
\end{equation}
where $\bm{K}_{i} \in \mathbb{C}_{N \times N}[t], i= 1, 2$, the $N$-soliton solution of the unreduced nonisospectral matrix mKdV equations can be obtained.

For the unreduced nonisospectral mKdV equation-$\mathrm{\uppercase\expandafter{\romannumeral1}}$
$\eqref{wyh1}$, due to the evolution of $\bm{K}$ in Eq. \eqref{Kt1}, we have
\begin{equation}\label{kjgj1}
k_{j}(t)=-\mathrm{i}\alpha t+a_{j}, \quad g_{j}(t)=-\mathrm{i}\alpha t-b_{j}, \quad a_{j}, b_{j} \in \mathbb{C}, \quad j=1,...,N.
\end{equation}
By using Eq. \eqref{rsfl} to express the dispersion relation \eqref{dis1} in the form of $N$th-order column vector, we get
\begin{equation*}
\begin{array}{llll}
	\bm{r}_{1, x}=\bm{K}_{1} \bm{r}_{1}, & \bm{s}_{1, x}= \bm{K}_{1}^{\mathrm{T}}\bm{s}_{1}, & \bm{r}_{1, t}=4\bm{K}_{1}^{3}\bm{r}_{1}-\mathrm{i}\alpha x \bm{r}_{1}, & \bm{s}_{1, t}=4\left(\bm{K}_{1}^{\mathrm{T}}\right)^{3}\bm{s}_{1}-\mathrm{i}\alpha x \bm{s}_{1}, \\
\bm{r}_{2, x}=-\bm{K}_{2} \bm{r}_{2}, & \bm{s}_{2, x}=- \bm{K}_{2}^{\mathrm{T}}\bm{s}_{2}, & \bm{r}_{2, t}=-4\bm{K}_{2}^{3}\bm{r}_{2}+\mathrm{i}\alpha x \bm{r}_{2}, & \bm{s}_{2, t}=-4\left(\bm{K}_{2}^{\mathrm{T}}\right)^{3}\bm{s}_{2}+\mathrm{i}\alpha x \bm{s}_{2}.
\end{array}
\end{equation*}
Thus, $\bm{r}_{i}$ and $\bm{s}_{i}$ can be written as 
\begin{equation}\label{rs1}
	\begin{array}{ll}
		{\bm{r}_{1}^{(j)}=\left(\delta _{1}^{(j)}\left(k_{1}\right), \delta _{1}^{(j)}\left(k_{2}\right), \cdots, \delta _{1}^{(j)}\left(k_{N}\right)\right)}^{\mathrm{T}},
		&{\bm{r}_{2}^{(j)}=\left(\delta _{2}^{(j)}\left(g_{1}\right), \delta _{2}^{(j)}\left(g_{2}\right), \cdots, \delta _{2}^{(j)}\left(g_{N}\right)\right)}^{\mathrm{T}},\\
		\\
		{\bm{s}_{1}^{(j)}=\left(\omega _{1}^{(j)}\left(k_{1}\right), \omega_{1}^{(j)}\left(k_{2}\right), \cdots, \omega_{1}^{(j)}\left(k_{N}\right)\right)}^{\mathrm{T}}, 
		&{\bm{s}_{2}^{(j)}=\left(\omega_{2}^{(j)}\left(g_{1}\right), \omega_{2}^{(j)}\left(g_{2}\right), \cdots, \omega_{2}^{(j)}\left(g_{N}\right)\right)} ^{\mathrm{T}},
	\end{array}
\end{equation}
where
\begin{subequations}\label{1xjdy}
	\begin{align}\label{xjdy1}
		&\delta_{1}^{(j)}(k)=\exp \left(k\left(t\right)x-\frac{k^{4}\left(t\right)}{\mathrm{i}\alpha }+\varphi_{1}^{(j)}(k)\right), 
		&\delta_{2}^{(j)}(g)=\exp \left(-g\left(t\right) x+\frac{g^{4}\left(t\right)}{\mathrm{i}\alpha }+\varphi_{2}^{(j)}(g)\right), \\
		\label{xjdy3}
		&\omega _{1}^{(j)}(k)=\exp \left(k\left(t\right)x-\frac{k^{4}\left(t\right)}{\mathrm{i}\alpha }+\phi _{1}^{(j)}(k)\right),
		&\omega _{2}^{(j)}(g)=\exp \left(-g\left(t\right) x+\frac{g^{4}\left(t\right)}{\mathrm{i}\alpha }+\phi _{2}^{(j)}(g)\right),
	\end{align}
\end{subequations}
$\varphi_{1}^{(j)}(k)$, $\varphi_{2}^{(j)}(g)$, $\phi_{1}^{(j)}(k)$ and  $\phi_{2}^{(j)}(g)$ are constants. At this point, we have the explicit expressions for $\bm{r}_{i}$ and $\bm{s}_{i}$, which can be substituted into Eq. \eqref{Kmfk} to obtain the expression of the matrix $\bm{M}_{i}(i=1,2)$. We let $(i,j)$ represent the elements in row $i$ and column $j$, we have
\begin{equation}\label{m}
	\begin{array}{l}\left(\bm{M}_{1}\right)_{i, j}=\frac{\delta _{1}^{(1)}\left(k_{i}\right) \omega _{2}^{(1)}\left(g_{j}\right)+\delta _{1}^{(2)}\left(k_{i}\right) \omega _{2}^{(2)}\left(g_{j}\right)}{k_{i}-g_{j}}, \\[1ex]
		\left(\bm{M}_{2}\right)_{i, j}=\frac{\delta _{2}^{(1)}\left(g_{i}\right) \omega _{1}^{(1)}\left(k_{j}\right)+\delta _{2}^{(2)}\left(g_{i}\right) \omega _{1}^{(2)}\left(k_{j}\right)}{g_{i}-k_{j}},\end{array}
\end{equation}
where $ k_{i}-g_{j} \neq 0 (i, j=1,..., N)$. After determining $\bm{K}$, $\bm{M}$, $\bm{r}$, $\bm{s}$, the solutions $\bm{U}_{2}$ and $\bm{U}_{3}$ of the unreduced nonisospectral mKdV equations can be uniquely determined. Therefore, according to Eq. \eqref{sd}, the solutions $\bm{U}_{2}$ and $\bm{U}_{3}$  are 
\begin{subequations}\label{U2U3}
	\begin{align}\label{U22}
		\bm{U}_{2} & =\left(\begin{array}{ll}
			\left(\bm{s}_{2}^{(1)}\right)^{\mathrm{T}}\left(\bm{I}_{N}-\bm{M}_{2} \bm{M}_{1}\right)^{-1} \bm{r}_{2}^{(1)} & \left(\bm{s}_{2}^{(1)}\right)^{\mathrm{T}}\left(\bm{I}_{N}-\bm{M}_{2} \bm{M}_{1}\right)^{-1} \bm{r}_{2}^{(2)} \\
			\left(\bm{s}_{2}^{(2)}\right)^{\mathrm{T}}\left(\bm{I}_{N}-\bm{M}_{2} \bm{M}_{1}\right)^{-1} \bm{r}_{2}^{(1)} & \left(\bm{s}_{2}^{(2)}\right)^{\mathrm{T}}\left(\bm{I}_{N}-\bm{M}_{2} \bm{M}_{1}\right)^{-1} \bm{r}_{2}^{(2)}
		\end{array}\right), \\
		\label{U33}
		\bm{U}_{3} & =\left(\begin{array}{ll}
			\left(\bm{s}_{1}^{(1)}\right)^{\mathrm{T}}\left(\bm{I}_{N}-\bm{M}_{1} \bm{M}_{2}\right)^{-1} \bm{r}_{1}^{(1)} & \left(\bm{s}_{1}^{(1)}\right)^{\mathrm{T}}\left(\bm{I}_{N}-\bm{M}_{1} \bm{M}_{2}\right)^{-1} \bm{r}_{1}^{(2)} \\
			\left(\bm{s}_{1}^{(2)}\right)^{\mathrm{T}}\left(\bm{I}_{N}-\bm{M}_{1} \bm{M}_{2}\right)^{-1} \bm{r}_{1}^{(1)} & \left(\bm{s}_{1}^{(2)}\right)^{\mathrm{T}}\left(\bm{I}_{N}-\bm{M}_{1} \bm{M}_{2}\right)^{-1} \bm{r}_{1}^{(2)}
		\end{array}\right).
	\end{align}
\end{subequations}

For the unreduced nonisospectral mKdV equation-$\mathrm{\uppercase\expandafter{\romannumeral2}}$
$\eqref{wyh2}$, we have
\begin{equation}\label{kjgj2}
	k_{j}(t)=a_{j}\mathrm{e}^{\beta t}, \quad g_{j}(t)=-b_{j}\mathrm{e}^{-\beta t}, \quad a_{j}, b_{j} \in \mathbb{C}, \quad j=1,...,N.
\end{equation}
The dispersion relation \eqref{dis2} with respect to $t$ is
\begin{equation*}
	\begin{array}{ll}
	\bm{r}_{1, t}=\left(4\bm{K}_{1}^{3}-\beta x \bm{K}_{1}\right)\bm{r}_{1}-\frac{1}{2} \beta \bm{r}_{1}, & \bm{s}_{1, t}=\left(4\left(\bm{K}_{1}^{\mathrm{T}}\right)^{3}-\beta x\bm{K}_{1}^{\mathrm{T}}\right)\bm{s}_{1}-\frac{1}{2} \beta\bm{s}_{1}, \\
	\bm{r}_{2, t}=\left(-4\bm{K}_{2}^{3}+\beta x \bm{K}_{2}\right)\bm{r}_{2}-\frac{1}{2} \beta \bm{r}_{2}, & \bm{s}_{2, t}=\left(-4\left(\bm{K}_{2}^{\mathrm{T}}\right)^{3}+\beta x\bm{K}_{2}^{\mathrm{T}}\right)\bm{s}_{2}-\frac{1}{2} \beta\bm{s}_{2}.
	\end{array}
\end{equation*}
Hence, $\bm{r}_{i}$ and $\bm{s}_{i}$ are defined as \eqref{rs1}, and we have 
\begin{subequations}\label{2xjdy}
	\begin{align}\label{2xjdy1}
		&\delta_{1}^{(j)}(k)=\exp \left(k\left(t\right)x-\frac{4k^{3}\left(t\right)}{3\beta}-\frac{1}{2} \beta t+\varphi_{1}^{(j)}(k)\right), 
		&\delta_{2}^{(j)}(g)=\exp \left(-g\left(t\right) x+\frac{4g^{3}\left(t\right)}{3\beta}-\frac{1}{2} \beta t+\varphi_{2}^{(j)}(g)\right), \\
		\label{2xjdy3}
		&\omega _{1}^{(j)}(k)=\exp \left(k\left(t\right)x-\frac{4k^{3}\left(t\right)}{3\beta}-\frac{1}{2} \beta t+\phi _{1}^{(j)}(k)\right),
		&\omega _{2}^{(j)}(g)=\exp \left(-g\left(t\right) x+\frac{4g^{3}\left(t\right)}{3\beta}-\frac{1}{2} \beta t+\phi _{2}^{(j)}(g)\right).
	\end{align}
\end{subequations}
Furthermore, by substituting Eq. \eqref{2xjdy} into Eq. \eqref{U2U3}, the solutions $\bm{U}_{2}$ and $\bm{U}_{3}$ to the unreduced nonisospectral mKdV equation-$\mathrm{\uppercase\expandafter{\romannumeral2}}$ \eqref{wyh2} can be obtained.

For the unreduced nonisospectral mKdV equation-$\mathrm{\uppercase\expandafter{\romannumeral3}}$
$\eqref{wyh3}$, we have
\begin{equation}\label{kjgj3}
	k_{j}(t)=\frac{1}{\mathrm{i}\gamma\left(t+a_{j}\right)}, \quad g_{j}(t)=\frac{1}{\mathrm{i}\gamma\left(t+ b_{j}\right)}, \quad a_{j}, b_{j} \in \mathbb{C}, \quad j=1,...,N.
\end{equation}
The dispersion relation \eqref{dis2} with respect to $t$ is
\begin{equation*}
	\begin{array}{ll}
		\bm{r}_{1, t}=4\bm{K}_{1}^{3}\bm{r}_{1}-\mathrm{i}\gamma x \bm{K}_{1}^{2}\bm{r}_{1}-\mathrm{i}\gamma\bm{K}_{1}\bm{r}_{1}, & \bm{s}_{1, t}=4\left(\bm{K}_{1}^{\mathrm{T}}\right)^{3}\bm{s}_{1}-\mathrm{i}\gamma x\left(\bm{K}_{1}^{\mathrm{T}}\right)^{2}\bm{s}_{1}-\mathrm{i}\gamma\bm{K}_{1}^{\mathrm{T}}\bm{s}_{1}, \\
		\bm{r}_{2, t}=-4\bm{K}_{2}^{3}\bm{r}_{2}+\mathrm{i}\gamma x \bm{K}_{2}^{2}\bm{r}_{2}-\mathrm{i}\gamma\bm{K}_{2}\bm{r}_{2}, & \bm{s}_{2, t}=-4\left(\bm{K}_{2}^{\mathrm{T}}\right)^{3}\bm{s}_{2}+\mathrm{i}\gamma x\left(\bm{K}_{2}^{\mathrm{T}}\right)^{2}\bm{s}_{2}-\mathrm{i}\gamma\bm{K}_{2}^{\mathrm{T}}\bm{s}_{2}.
	\end{array}
\end{equation*}
Hence
\begin{subequations}\label{3xjdy}
	\begin{align}\label{3xjdy1}
		&\delta_{1}^{(j)}(k)=\exp \left(k\left(t\right)x+\ln \left(k(t)\right)-\frac{2k^{2}\left(t\right)}{\mathrm{i}\gamma}+\varphi_{1}^{(j)}(k)\right), 
		&\delta_{2}^{(j)}(g)=\exp \left(-g\left(t\right)x+\ln \left(g(t)\right)+\frac{2g^{2}\left(t\right)}{\mathrm{i}\gamma}+\varphi_{2}^{(j)}(g)\right), \\
		\label{3xjdy3}
		&\omega _{1}^{(j)}(k)=\exp \left(k\left(t\right)x+\ln \left(k(t)\right)-\frac{2k^{2}\left(t\right)}{\mathrm{i}\gamma}+\phi _{1}^{(j)}(k)\right),
		&\omega _{2}^{(j)}(g)=\exp \left(-g\left(t\right)x+\ln \left(g(t)\right)+\frac{2g^{2}\left(t\right)}{\mathrm{i}\gamma}+\phi _{2}^{(j)}(g)\right).
	\end{align}
\end{subequations}
Furthermore, by substituting Eq. \eqref{3xjdy} into Eq. \eqref{U2U3}, the solutions $\bm{U}_{2}$ and $\bm{U}_{3}$ to the unreduced nonisospectral mKdV equation-$\mathrm{\uppercase\expandafter{\romannumeral3}}$ \eqref{wyh3} can be obtained.

\subsection{\texorpdfstring{$\bm{K}$}{K} is the lower triangular Toeplitz matrix}\label{sec32}
In the nonisospectral case, since $\bm{K}$ is a matrix function, it means that solutions cannot be classified by considering the $\bm{K}$ standard form, i.e. diagonal or Jordan forms or their combinations. Below, we introduce the $N \times N $ lower triangular Toeplitz matrix $\bm{F}_{a}^{N}\left[f\left(a\right)\right]$ and the $N \times N $ symmetric matrix $\bm{H}_{a}^{N}\left[f\left(a\right)\right]$ to provide more comprehensive solutions to the unreduced nonisospectral mKdV equations. 

The lower triangular Toeplitz matrix $\bm{F}_{a}^{N}\left[f\left(a\right)\right]$ generated by $f\left(a\right)$
\begin{equation}\label{toep}
	\bm{F}_{a}^{N}\left[f\left(a\right)\right]=\left(\begin{array}{ccccc}
		f & 0 & 0 & \cdots & 0 \\
		\frac{\partial_{a} f}{1!} & f & 0 & \cdots & 0 \\
		\frac{\partial_{a}^{2} f}{2!} & \frac{\partial_{a} f}{1!} & f & \cdots & 0 \\
		\vdots & \vdots & \vdots & \ddots & \vdots \\
		\frac{\partial_{a}^{N-1} f}{(N-1)!} & \frac{\partial_{a}^{N-2} f}{(N-2)!} & \frac{\partial_{a}^{N-3} f}{(N-3)!} & \cdots & f
	\end{array}\right),
\end{equation}
where $f=f\left(a\right)$ is $C^{\infty}$ function of $a$, and the symmetric matrix $\bm{H}_{a}^{N}\left[f\left(a\right)\right]$ generated by $f\left(a\right)$
\begin{equation}\label{sym}
	\bm{H}_{a}^{N}\left[f\left(a\right)\right]=\left(\begin{array}{ccccc}
		f & \frac{\partial_{a} f}{1!} & \frac{\partial_{a}^{2} f}{2!} & \cdots & \frac{\partial_{a}^{N-1} f}{(N-1)!} \\
		\frac{\partial_{a} f}{1!} & \frac{\partial_{a}^{2} f}{2!} & \frac{\partial_{a}^{3} f}{3!} & \cdots & 0 \\
		\frac{\partial_{a}^{2} f}{2!} & \frac{\partial_{a}^{3} f}{3!} & \frac{\partial_{a}^{4} f}{4!} & \cdots & 0 \\
		\vdots & \vdots & \vdots & \ddots & \vdots \\
		\frac{\partial_{a}^{N-1} f}{(N-1)!} & 0 & 0 & \cdots & 0
	\end{array}\right).
\end{equation}

Next, we will continue to use the symbols in Section \ref{sec31} and provide a general solution method for the three unreduced nonisospectral mKdV equations. When $\bm{K}_{1}$ and $\bm{K}_{2}$ are lower triangular Toeplitz matrices
\begin{equation}\label{2k1k2}
\bm{K}_{1}=\bm{F}_{a_{1}}^{[N]}\left[k_{1}\left(a_{1}\right)\right], \quad \bm{K}_{2}=\bm{F}_{b_{1}}^{[N]}\left[g_{1}\left(b_{1}\right)\right] .
\end{equation}
Namely
\begin{equation*}
	\bm{K}_{1}=\left(\begin{array}{ccccc}
		k_{1} & 0 & 0 & \cdots & 0 \\
		\frac{\partial_{a_{1}} k_{1}}{1!} & k_{1} & 0 & \cdots & 0 \\
		\frac{\partial_{a_{1}}^{2} k_{1}}{2!} & \frac{\partial_{a_{1}} k_{1}}{1!} & k_{1} & \cdots & 0 \\
		\vdots & \vdots & \vdots & \ddots & \vdots \\
		\frac{\partial_{a_{1}}^{N-1} k_{1}}{(N-1)!} & \frac{\partial_{a_{1}}^{N-2} k_{1}}{(N-2)!} & \frac{\partial_{a_{1}}^{N-3} k_{1}}{(N-3)!} & \cdots & k_{1}
	\end{array}\right), \quad \bm{K}_{2}=\left(\begin{array}{ccccc}
	g_{1} & 0 & 0 & \cdots & 0 \\
	\frac{\partial_{b_{1}} g_{1}}{1!} & g_{1} & 0 & \cdots & 0 \\
	\frac{\partial_{b_{1}}^{2} g_{1}}{2!} & \frac{\partial_{b_{1}} g_{1}}{1!} & g_{1} & \cdots & 0 \\
	\vdots & \vdots & \vdots & \ddots & \vdots \\
	\frac{\partial_{b_{1}}^{N-1} g_{1}}{(N-1)!} & \frac{\partial_{b_{1}}^{N-2} g_{1}}{(N-2)!} & \frac{\partial_{b_{1}}^{N-3} g_{1}}{(N-3)!} & \cdots & g_{1}
	\end{array}\right).
\end{equation*}
It can be verified that the $\bm{K}=\operatorname{diag}\left(\bm{K}_{1}, \bm{K}_{2}\right)$ formed by Eq. \eqref{2k1k2} satisfies the evolution relation of $\bm{K}$ with respect to $t$, as expressed in Eq. \eqref{zKt}, when $k_{1}$ and $g_{1}$ are defined in Eqs. \eqref{kjgj1}, \eqref{kjgj2}, and \eqref{kjgj3}, respectively.

Here, $\bm{r}_{i}$ and $\bm{s}_{i}$ can be defined as
\begin{equation}\label{rs2}
	\begin{aligned}
\bm{r}_{1}^{(j)}&=\bm{F}_{a_{1}}^{[N]}\left[\delta_{1}^{(j)}\left(k\right)\right] \cdot \bm{e}_{N}=\left(\delta_{1}^{(j)}\left(k\right), \frac{\partial_{a_{1}} \delta_{1}^{(j)}\left(k\right)}{1!}, \ldots, \frac{\partial_{a_{1}} \delta_{1}^{(j)}\left(k\right)}{(N-1)!}\right)^{\mathrm{T}},\\ \bm{s}_{1}^{(j)}&=\bm{H}_{a_{1}}^{[N]}\left[\omega_{1}^{(j)}(k)\right] \cdot \bm{e}_{N}=\left(\omega_{1}^{(j)}(k), \frac{\partial_{a_{1}} \omega_{1}^{(j)}(k)}{1!}, \ldots, \frac{\partial_{a_{1}} \omega_{1}^{(j)}(k)}{(N-1)!}\right)^{\mathrm{T}},\\ \bm{r}_{2}^{(j)}&=\bm{F}_{b_{1}}^{[N]}\left[\delta_{2}^{(j)}(g)\right] \cdot \bm{e}_{N}=\left(\delta_{2}^{(j)}(g), \frac{\partial_{b_{1}} \delta_{2}^{(j)}(g)}{1!}, \ldots, \frac{\partial_{b_{1}} \delta_{2}^{(j)}(g)}{(N-1)!}\right)^{\mathrm{T}}, \\ \bm{s}_{2}^{(j)}&=\bm{H}_{b_{1}}^{[N]}\left[\omega_{2}^{(j)}(g)\right] \cdot \bm{e}_{N}=\left(\omega_{2}^{(j)}(g), \frac{\partial_{b_{1}} \omega_{2}^{(i)}(g)}{1!}, \ldots, \frac{\partial_{b_{1}} \omega_{2}^{(j)}(g)}{(N-1)!}\right)^{\mathrm{T}},
	\end{aligned}
\end{equation}
where $\bm{e}_{N}=(1, 0 \ldots 0)^{\mathrm{T}}$ represents an N-dimensional column vector. According to \cite{fhgx1,fhgx}, we can know that there is a special property between the lower triangular Toeplitz matrix $\bm{F}_{a}^{N}\left[f\left(a\right)\right]$ and the symmetric matrix $\bm{H}_{a}^{N}\left[f\left(a\right)\right]$
\begin{equation}\label{Fff}
\bm{F}_{c}^{[N]}[f(c) g(c)]=\bm{F}_{c}^{[N]}[f(c)] \bm{F}_{c}^{[N]}[g(c)].
\end{equation}
From this, we can verify that Eq. \eqref{rs2} satisfies dispersion relations \eqref{dis1}, \eqref{dis2}, and \eqref{dis3}.

We now determine the explicit forms of $\bm{M}_1$ and $\bm{M}_2$
\begin{equation}\label{m12j}
	\bm{M}_{1}= \bm{M}_{1}^{(1)}+ \bm{M}_{1}^{(2)}, \quad \bm{M}_{2}= \bm{M}_{2}^{(1)}+ \bm{M}_{2}^{(2)},
\end{equation}
where
\begin{equation}\label{m12jn}
	\bm{M}_{1}^{(i)}=\bm{F}_{a_{1}}^{[N]}\left[\delta_{1}^{(i)}\left(k\right)\right] \cdot \bm{G}_{1} \cdot \bm{H}_{b_{1}}^{[N]}\left[\omega_{2}^{(i)}(g)\right] , \quad \bm{M}_{2}^{(i)}=\bm{F}_{b_{1}}^{[N]}\left[\delta_{2}^{(i)}\left(g\right)\right]\cdot \bm{G}_{2} \cdot  \bm{H}_{a_{1}}^{[N]}\left[\omega_{1}^{(i)}(k)\right],\quad i=1,2,
\end{equation}
and $\bm{G}_{1}$ and $\bm{G}_{2}$ are unknown functions. In the following, we take $\bm{G}_{1}$ as an example, and $\bm{G}_{2}$ can be solved in the same way.

For notational simplicity, we denote $\bm{F}_{a_{1}}^{[N]}\left[\delta_{1}^{(i)}\left(k\right)\right]=\bm{F}_{1}^{(i)}$ and $\bm{H}_{b_{1}}^{[N]}\left[\omega_{2}^{(i)}(g)\right]=\bm{H}_{2}^{(i)}$. Substituting Eqs. \eqref{m12j} and \eqref{m12jn} into Eq. \eqref{Kmfk1} gives
\begin{equation*}
\begin{aligned}
	& \bm{K}_{1}\left(\bm{F}_{1}^{(1)}\cdot \bm{G}_{1}\cdot \bm{H}_{2}^{(1)}+\bm{F}_{1}^{(2)}\cdot \bm{G}_{1}\cdot \bm{H}_{2}^{(2)}\right)-\left(\bm{F}_{1}^{(1)}\cdot \bm{G}_{1}\cdot \bm{H}_{2}^{(1)}+\bm{F}_{1}^{(2)}\cdot \bm{G}_{1}\cdot \bm{H}_{2}^{(2)}\right)\bm{K}_{2} \\
	& =\bm{F}_{1}^{(1)}\bm{e}_{N}\cdot(\bm{H}_{2}^{(1)}\cdot \bm{e}_{N})^{\mathrm{T}}+\bm{F}_{1}^{(2)}\bm{e}_{N}\cdot(\bm{H}_{2}^{(2)}\cdot \bm{e}_{N})^{\mathrm{T}}.
\end{aligned}
\end{equation*}
Also, because $\bm{H}$ is a symmetric matrix, there is
\begin{equation}\label{KFH}
	\begin{aligned}
		& \bm{K}_{1}\bm{F}_{1}^{(1)}\bm{G}_{1} \bm{H}_{2}^{(1)}+\bm{K}_{1}\bm{F}_{1}^{(2)} \bm{G}_{1} \bm{H}_{2}^{(2)}-\bm{F}_{1}^{(1)}\bm{G}_{1} \bm{H}_{2}^{(1)}\bm{K}_{2}-\bm{F}_{1}^{(2)} \bm{G}_{1} \bm{H}_{2}^{(2)}\bm{K}_{2} \\
		& =\bm{F}_{1}^{(1)}\bm{e}_{N}\bm{e}_{N}^{\mathrm{T}}\bm{H}_{2}^{(1)}+\bm{F}_{1}^{(2)}\bm{e}_{N}\bm{e}_{N}^{\mathrm{T}}\bm{H}_{2}^{(2)}.
	\end{aligned}
\end{equation}
By using Eqs. \eqref{2k1k2} and \eqref{Fff}, we have
\begin{equation}\label{KFgx}
	\begin{aligned}
&\bm{K}_{1}\bm{F}_{1}^{(i)}=\bm{F}_{a_1}^{\left[N\right]}\left[k_{1}\right] \cdot \bm{F}_{a_{1}}^{\left[N\right]}\left[\delta_{1}^{(i)}\left(k\right)\right]=\bm{F}_{a_1}^{\left[N\right]}\left[k_{1} \delta_{1}^{(i)}\left(k\right)\right]=\bm{F}_{a_{1}}^{\left[N\right]}\left[\delta_{1}^{(i)}\left(k\right)\right] \cdot \bm{F}_{a_1}^{\left[N\right]}\left[k_{1}\right] =\bm{F}_{1}^{(i)} \bm{K}_{1},\\
&\bm{H}_{2}^{(i)}\bm{K}_{2}=\bm{H}_{b_1}^{\left[N\right]}\left[\omega_{2}^{(i)}(g)\right] \cdot \bm{F}_{b_1}^{\left[N\right]}\left[g_{1}\right]=\left(\bm{F}_{b_1}^{\left[N\right]}\left[g_{1}\right]\right)^{\mathrm{T}} \cdot \bm{H}_{b_1}^{\left[N\right]}\left[\omega_{2}^{(i)}(g)\right]=\bm{K}_{2}^{\mathrm{T}}\bm{H}_{2}^{(i)}.
	\end{aligned}
\end{equation}
Therefore, substituting Eq. \eqref{KFgx} into Eq. \eqref{KFH} yields 
\begin{equation*}
	\begin{aligned}
		&\bm{F}_{1}^{(1)}\bm{K}_{1}\bm{G}_{1} \bm{H}_{2}^{(1)}+\bm{F}_{1}^{(2)} \bm{K}_{1}\bm{G}_{1} \bm{H}_{2}^{(2)}-\bm{F}_{1}^{(1)}\bm{G}_{1}\bm{K}_{2}^{\mathrm{T}} \bm{H}_{2}^{(1)}-\bm{F}_{1}^{(2)} \bm{G}_{1}\bm{K}_{2}^{\mathrm{T}} \bm{H}_{2}^{(2)} \\
		& =\bm{F}_{1}^{(1)}\bm{e}_{N}\bm{e}_{N}^{\mathrm{T}}\bm{H}_{2}^{(1)}+\bm{F}_{1}^{(2)}\bm{e}_{N}\bm{e}_{N}^{\mathrm{T}}\bm{H}_{2}^{(2)}.
	\end{aligned}
\end{equation*}
So, Eq. \eqref{KFH} can be reduced to 
\begin{equation}\label{ent}
\bm{K}_{1}\bm{G}_{1}-\bm{G}_{1}\bm{K}_{2}^{\mathrm{T}} =\bm{e}_{N}\bm{e}_{N}^{\mathrm{T}}.
\end{equation}

For clarity of expression, let $\bm{G}_1=\left(\bm{\mathcal{g}}_1,\bm{\mathcal{g}}_2,\cdots,\bm{\mathcal{g}}_N\right)$, $\bm{\mathcal{g}}_j=\left(\mathcal{g}_{1,j},\mathcal{g}_{2,j},\cdots,\mathcal{g}_{N,j}\right)^\mathrm{T}$. Expand Eq. \eqref{ent} in matrix form, where the first column is 
\begin{equation}
\begin{pmatrix}
	k_1\mathcal{g}_{1,1} \\
	\frac{\partial_{a_{1}} k_{1}}{1!} \mathcal{g}_{1,1}+k_1 \mathcal{g}_{2,1} \\
	\frac{\partial_{a_1}^2 k_1}{2!}\mathcal{g}_{1,1}+\frac{\partial_{a_{1}} k_{1}}{1!}\mathcal{g}_{2,1}+ k_1\mathcal{g}_{3,1}\\
	\vdots \\
	\frac{\partial_{a_1}^{N-1}k_1}{(N-1)!}\mathcal{g}_{1,1}+	\frac{\partial_{a_1}^{N-2}k_1}{(N-2)!}\mathcal{g}_{2,1}+\cdots+k_1\mathcal{g}_{N,1}
\end{pmatrix}-g_{1}
\begin{pmatrix}
	\mathcal{g}_{1,1} \\
	\mathcal{g}_{2,1} \\
	\mathcal{g}_{3,1} \\
	\vdots \\
	\mathcal{g}_{N,1}
\end{pmatrix}=
\begin{pmatrix}
	1 \\
	0 \\
	0 \\
	\vdots \\
	0
\end{pmatrix}.
\end{equation}
We can find
\begin{equation*}
		\mathcal{g}_{1,1}=\frac{1}{k_{1}-g_{1}}, \quad \mathcal{g}_{2,1}=-\frac{\partial_{a_{1}} k_{1}}{\left(k_{1}-g_{1}\right)^{2}}, \quad 	\mathcal{g}_{3,1}=\frac{\left(\partial_{a_{1}} k_{1}\right)^{2}}{\left(k_{1}-g_{1}\right)^{3}}-\frac{\partial_{a_{1}}^{2} k_{1}}{2!\left(k_{1}-g_{1}\right)^{2}},  \cdots.
\end{equation*}
Thus, the first column element of $\bm{G}_{1}$ can be shown as
\begin{equation}\label{gm1}
\mathcal{g}_{m, 1}=-\frac{1}{k_{1}-g_{1}}\left(\sum_{j=1}^{m-1} \frac{\partial_{a_{1}}^{j} k_{1}}{j!} g_{m-j, 1}\right), \quad m=2,3, \cdots, N.
\end{equation}
Once we have the first column $\bm{\mathcal{g}}_1$, we can consider the second column $\bm{\mathcal{g}}_2$ of \eqref{ent}
\begin{equation}
	\begin{pmatrix}
		k_1\mathcal{g}_{1,2} \\
		\frac{\partial_{a_{1}} k_{1}}{1!} \mathcal{g}_{1,2}+k_1 \mathcal{g}_{2,2} \\
		\frac{\partial_{a_1}^2 k_1}{2!}\mathcal{g}_{1,2}+\frac{\partial_{a_{1}} k_{1}}{1!}\mathcal{g}_{2,2}+ k_1\mathcal{g}_{3,2}\\
		\vdots \\
		\frac{\partial_{a_1}^{N-1}k_1}{(N-1)!}\mathcal{g}_{1,2}+	\frac{\partial_{a_1}^{N-2}k_1}{(N-2)!}\mathcal{g}_{2,2}+\cdots+k_1\mathcal{g}_{N,2}
	\end{pmatrix}-
	\begin{pmatrix}
		\mathcal{g}_{1,1} \frac{\partial_{b_{1}} g_{1}}{1!}+ \mathcal{g}_{1,2} g_{1} \\
		\mathcal{g}_{2,1} \frac{\partial_{b_{1}} g_{1}}{1!}+ \mathcal{g}_{2,2} g_{1}\\
		\mathcal{g}_{3,1} \frac{\partial_{b_{1}} g_{1}}{1!}+ \mathcal{g}_{3,2} g_{1}\\
		\vdots \\
		\mathcal{g}_{N,1}\frac{\partial_{b_{1}} g_{1}}{1!}+ \mathcal{g}_{N,2} g_{1}
	\end{pmatrix}=
	\begin{pmatrix}
		0 \\
		0 \\
		0 \\
		\vdots \\
		0
	\end{pmatrix}.
\end{equation}
We can find
\begin{equation*}
	\begin{aligned}
&\mathcal{g}_{1,2}=\frac{1}{k_{1}-g_{1}}\left(\partial_{b_{1}} g_{1}\right) \mathcal{g}_{1,1}, \quad \mathcal{g}_{2,2}=\frac{1}{k_{1}-g_{1}}\left(\mathcal{g}_{2,1}\partial_{b_{1}} g_{1}-\left(\partial_{a_{1}} k_{1}\right)\mathcal{g}_{1,2}\right),\\ &\mathcal{g}_{3,2}=\frac{1}{k_{1}-g_{1}}\left(\mathcal{g}_{3,1}\partial_{b_{1}} g_{1}-\left(\frac{\partial_{a_1}^2 k_1}{2!}\right)\mathcal{g}_{1,2}-\left(\partial_{a_{1}}k_{1}\right)\mathcal{g}_{2,2}\right), \cdots.
\end{aligned}
\end{equation*}
Thus, the second column element of $\bm{G}_{1}$ can be shown
\begin{equation}\label{gm2}
	\mathcal{g}_{m, 2}=\frac{1}{k_{1}-g_{1}}\left(\left(\partial_{b_{1}} g_{1}\right) \mathcal{g}_{m, 1}-\sum_{j=1}^{m-1} \frac{\partial_{a_{1}}^{j} k_{1}}{j!} \mathcal{g}_{m-j, 2}\right), \quad m=2,3, \ldots, N.
\end{equation}
Combining Eqs. \eqref{gm1} and \eqref{gm2}, we have
\begin{equation*}
\mathcal{g}_{1,2}=\frac{\partial_{b_{1}} g_{1}}{\left(k_{1}-g_{1}\right)^{2}}, \quad \mathcal{g}_{2,2}=-\frac{2\left(\partial_{a_{1}} k_{1}\right)\left(\partial_{b_{1}} g_{1}\right)}{\left(k_{1}-g_{1}\right)^{3}}, \quad \mathcal{g}_{3,2}=\frac{3\left(\partial_{a_{1}} k_{1}\right)^{2}\left(\partial_{b_{1}} g_{1}\right)}{\left(k_{1}-g_{1}\right)^{4}}-\frac{\left(\partial_{a_{1}}^{2} k_{1}\right)\left(\partial_{b_{1}} g_{1}\right)}{\left(k_{1}-g_{1}\right)^{3}}.
\end{equation*}
Overall, the element in the $n$th column $\bm{\mathcal{g}}_n\left(n> 1\right)$ of $\bm{G}_{1}$ can be shown
\begin{equation}
	\bm{K}_{1} \bm{\mathcal{g}}_{n}-\sum_{j=1}^{n-1} \frac{\partial_{b_{1}}^{j} g_{1}}{j!} \bm{\mathcal{g}}_{n-j}-g_{1} \bm{\mathcal{g}}_{n}=\mathbf{0},
\end{equation}
namely
\begin{equation}
\bm{\mathcal{g}}_{n}=\left(\bm{K}_{\mathbf{1}}-g_{1} \bm{I}_{N}\right)^{-1} \cdot \sum_{j=1}^{n-1} \frac{\partial_{b_{1}}^{j} g_{1}}{j!}\bm{\mathcal{g}}_{n-j}.
\end{equation}
So, with the  explicit expression of $\bm{\mathcal{g}}_1,\bm{\mathcal{g}}_2,\cdots,\bm{\mathcal{g}}_N$, we have provided an explicit representation of $\bm{G}_{1}$. Thus, $\bm{G}_{2}$ can also be obtained. Furthermore, we obtain $\bm{M}$. Considering the $\bm{r}_{i}$ and $\bm{s}_{i}$ given by Eq. \eqref{rs2}, we can use Eq. \eqref{sij} to derive the solutions to the unreduced nonisospectral mKdV equations.

\section{Reduction}\label{sec4}
In order to obtain the NTCmKdV equations from the unreduced system, we have the reduction  
\begin{equation}\label{k1g}
	\bm{K}_{2}=-\bm{K}_{1}^{*}.
\end{equation}
Under this reduction, the dispersion relations $\bm{r_{1}}$ and $\bm{r_{2}}$, as well as $\bm{s_{1}}$ and $\bm{s_{2}}$ become
\begin{equation}\label{r2s2}
	\bm{r}_{2}= \bm{r}_{1}^{*}, \quad \bm{s}_{2}= \bm{s}_{1}^{*}.
\end{equation}
Taking the conjugate for both sides of Eq.\eqref {Kmfk1} and combining it with Eq.\eqref {r2s2}, we have
\begin{equation*}
\left(\bm{K}_{1} \bm{M}_{1}-\bm{M}_{1} \bm{K}_{2}\right)^{*}=\bm{r}_{1}^{*}\bm{s}_{2}^{\mathrm{\dagger}}=\bm{r}_{2}\bm{s}_{1}^{\mathrm{T}}=\bm{K}_{2} \bm{M}_{2}-\bm{M}_{2} \bm{K}_{1},
\end{equation*}
where $\mathrm{\dagger}$  stands for the conjugate transposition. By substituting the reduction \eqref{k1g} into the above equation, it becomes
\begin{equation*}
\bm{K}_{2}\left( \bm{M}_{2}+\bm{M}_{1}^{*}\right)-\left( \bm{M}_{2}+\bm{M}_{1}^{*}\right) \bm{K}_{1}=\bm{0}.
\end{equation*}
So, we have
\begin{equation}\label{m1m2g}
	\bm{M}_{2}= -\bm{M}_{1}^{*}.
\end{equation}
Next, taking $i=j=0$ in Eqs. \eqref{sd2} and \eqref{sd3}, we have
\begin{equation}\label{u21}
	\begin{aligned}
		\bm{U}_{2}^{*}=\left(\bm{S}_{2}^{(0,0)}\right)^{*}=\left(\bm{s}_{2}^{\mathrm{T}} \left(\bm{I}_{N}-\bm{M}_{2}\bm{M}_{1}\right)^{-1}  \bm{r}_{2}\right)^{*}=\bm{s}_{1}^{\mathrm{T}} \left(\bm{I}_{N}-\bm{M}_{1} \bm{M}_{2}\right)^{-1}\bm{r}_{1}=\bm{S}_{3}^{(0,0)}=\bm{U}_{3}.
	\end{aligned}
\end{equation}
It should be noted that when the condition \eqref{r2s2} holds, for the unreduced nonisospectral mKdV equation-$\mathrm{\uppercase\expandafter{\romannumeral1}}$
$\eqref{wyh1}$ and equation-$\mathrm{\uppercase\expandafter{\romannumeral2}}$
$\eqref{wyh2}$, the constants in \eqref{1xjdy} and  \eqref{2xjdy} should satisfy
\begin{equation}\label{lm}
	\varphi _{2}^{(j)}\left(g_{i}\right)=\left(\varphi _{1}^{(j)}\left(k_{i}\right)\right)^{*}, \quad \phi _{2}^{(j)}\left(g_{i}\right)=\left(\phi _{1}^{(j)}\left(k_{i}\right)\right)^{*},\quad i=1, \ldots, N, \quad j=1,2.
\end{equation}
However, for the unreduced nonisospectral mKdV equation-$\mathrm{\uppercase\expandafter{\romannumeral3}}$
$\eqref{wyh3}$, the constants in \eqref{3xjdy} satisfy
\begin{equation}\label{lmb}
	\varphi _{2}^{(j)}\left(g_{i}\right)=\left(\varphi _{1}^{(j)}\left(k_{i}\right)\right)^{*}-\mathrm{i}\pi , \quad \phi _{2}^{(j)}\left(g_{i}\right)=\left(\phi _{1}^{(j)}\left(k_{i}\right)\right)^{*}-\mathrm{i}\pi,\quad i=1, \ldots, N, \quad j=1,2.
\end{equation}
Here, we only consider $\bm{K_{1}}$ and $\bm{K_{2}}$ as diagonal matrices or lower triangular Toeplitz matrices. To make $\bm{U_{2}}$ as a symmetric matrix, we have
\begin{subequations}
	\begin{align}
	\label{lmnb1}
	&\varphi_{2}^{(1)}\left(g_{i}\right)+\phi_{2}^{(2)}\left(g_{j}\right)  =\varphi_{2}^{(2)}\left(g_{i}\right)+\phi_{2}^{(1)}\left(g_{j}\right), \\
	\label{lmnb2}
	&\varphi _{1}^{(1)}\left(k_{i}\right)+\phi _{1}^{(2)}\left(k_{j}\right)  =\varphi_{1}^{(2)}\left(k_{i}\right)+\phi_{1}^{(1)}\left(k_{j}\right), \quad i, j=1, \ldots, N.
	\end{align}
\end{subequations}
Therefore, we can obtain
\begin{equation}\label{u212}
\left(\bm{U}_{2}^{*}\right)^{\mathrm{T}}=\bm{U}_{3}.
\end{equation}
Expanding the $2\times 2$ matrix $\bm{U}_{2}$, we define
\begin{equation}\label{qu2gx}
	\bm{U}_{2}\doteq \bm{Q}=\left(\begin{array}{ll}
		Q_{11} & Q_{12} \\
		Q_{21} & Q_{22}
	\end{array}\right)=\left(\begin{array}{cc}
		q_{1} & q_{0} \\
		q_{0} & q_{2}
	\end{array}\right).
\end{equation}
Using Eqs. \eqref{u21} and \eqref{qu2gx}, the unreduced nonisospectral mKdV equations \eqref{wyh1}, \eqref{wyh2}, and \eqref{wyh3} can be respectively reduced to the NTCmKdV equations \eqref{NTCmKdV1}, \eqref{NTCmKdV2}, and \eqref{NTCmKdV3}.

\section{Solutions to the NTCmKdV equations}\label{sec5}
Based on the Sections \ref{sec3} and \ref{sec4}, we can summarize the following general theorem for the NTCmKdV equations.
\begin{theorem}\label{th1}
When $\bm{K}_{2}=-\bm{K}_{1}^{*}$, $ \alpha, \beta, \gamma \in \mathbb{R}$ and $\mathcal{E}\left(\bm{K}_{1}\right) \cap \mathcal{E}\left(\bm{-{K}_{1}^{*}}\right)=\varnothing$, $\bm{r}_{1}, \bm{s}_{1} \in \mathbb{C}_{N \times 2}[x, t]$,  $\bm{r}_{1}$ and $\bm{s}_{1}$  respectively satisfy the dispersion relations \eqref{dis1}, \eqref{dis2} and \eqref{dis3}, $\bm{K}_{1}$ satisfies the evolution Eq. \eqref{zKt} with respect to  $t$. And, the solution to the Sylvester equation 
\begin{equation}\label{S1}
	\bm{K}_{1}\bm{M}_{1}+\bm{M}_{1}\bm{K}_{1}^{*}=\bm{r}_{1}\bm{s}_{1}^{\mathrm{\dagger}}
\end{equation}
can be uniquely determined. So, the solutions to the NTCmKdV equations can be written as
\begin{equation}\label{1zj1}
\bm{Q}=\bm{s}_{1}^{\dagger}\left(\bm{I}_{N}+ \bm{M}_{1}^{*} \bm{M}_{1}\right)^{-1} \bm{r}_{1}^{*}=\left(\begin{array}{ll}
	Q_{11} & Q_{12} \\
	Q_{21} & Q_{22}
\end{array}\right)=\left(\begin{array}{cc}
q_{1} & q_{0} \\
q_{0} & q_{2}
\end{array}\right),
\end{equation}
with
\begin{equation}\label{zj11}
Q_{i j}=\left(s_{1}^{(i)}\right)^{\dagger}\left(\bm{I}+ \bm{M}_{1}^{*} \bm{M}_{1}\right)^{-1}\left(\bm{r}_{1}^{(j)}\right)^{*}
,\quad i, j=1,2.
\end{equation}
\end{theorem}

\subsection{Solutions to the NTCmKdV-$\mathrm{\uppercase\expandafter{\romannumeral1}}$}\label{sec51}
\subsubsection{One-soliton solution}\label{sec511}
For $N=1$, we have
\begin{equation}\label{cj1}
	\bm{K}_{1}=k_{1}, \quad \bm{r}_{1}=\left(\delta _{1}^{(1)}\left(k_{1}\right), \delta_{1}^{(2)}\left(k_{1}\right)\right), \quad \bm{s}_{1}=\left(\omega _{1}^{(1)}\left(k_{1}\right), \omega _{1}^{(2)}\left(k_{1}\right)\right), \quad k_{1}=-\mathrm{i}\alpha t+a_{1}, \quad \alpha\in \mathbb{R}.
\end{equation}
According to Eqs. \eqref{m} and \eqref{r2s2}, $\bm{M}_{1}$ can be expressed as
\begin{equation}\label{1m1}
		\bm{M}_{1}=m_{1,1}
		=\frac{\delta_{1}^{(1)}\left(k_{1}\right) \omega _{2}^{(1)}\left(g_{1}\right)+\delta _{1}^{(2)}\left(k_{1}\right) \omega _{2}^{(2)}\left(g_{1}\right)}{k_{1}-g_{1}}
		= \frac{\sum_{n=1}^{2} \delta_{1}^{(n)}\left(k_{1}\right)\left(\omega_{1}^{(n)}\left(k_{1}\right)\right)^{*}}{k_{1}+k_{1}^{*}},
\end{equation}
where $\delta_{1}^{(n)}\left(k_{1}\right)$ and $\omega_{1}^{(n)}\left(k_{1}\right)$ are defined in \eqref{1xjdy}. Substituting Eqs. \eqref{rs1} and \eqref{1m1} into \eqref{zj11}, we have
\begin{equation}\label{Qij}
	\begin{aligned}	
		Q_{i j}&=\left(\omega _{1}^{(i)}\left(k_{1}\right)\right)^{*}\cdot \left(1+\left|m_{1,1}\right|^{2}\right)^{-1} \cdot\left(\delta _{1}^{(j)}\left(k_{1}\right)\right)^{*} \\
		&=\frac{4\left| \operatorname{Re}\left[a_{1}\right]\right|^{2}\left(\omega _{1}^{(i)}\left(k_{1}\right)\right)^{*} \left(\delta _{1}^{(j)}\left(k_{1}\right)\right)^{*}}{4\left| \operatorname{Re}\left[a_{1}\right]\right|^{2}+\left|\sum_{n=1}^{2} \delta _{1}^{(n)}\left(k_{1}\right)\left(\omega _{1}^{(n)}\left(k_{1}\right)\right)^{*}\right|^{2}} .
	\end{aligned}
\end{equation}
We set
\begin{equation}\label{fh1}
	a_{1}=c_{1}+\mathrm{i}d_{1}, \quad \exp \left(\varphi _{1}^{(m)}\left(k_{1}\right)\right)=\lambda _{m}, \quad \exp \left(\phi _{1}^{(n)}\left(k_{1}\right)\right)=\mu _{n},
\end{equation}
where  $c_{1}, d_{1}\in \mathbb{R}$, $c_{1}\ne 0$ and $\lambda _{m}, \mu _{n}\in \mathbb{C}$. Then, by using Eq. \eqref{1xjdy}, we find
\begin{subequations}
\begin{align*}
		\left( \omega _{1}^{(n)}\left(k_{1}\right)\right)^{*}\left( \delta_{1}^{(m)}\left(k_{1}\right)\right)^{*} &=\exp \left(2k_{1}(t)^{*}x+\frac{{2k_{1}^4(t)}^{*}}{\mathrm{i}\alpha }  \right) \cdot \exp \left(\varphi _{1}^{(m)}\left(k_{1}\right)^{*}+\phi _{1}^{(n)}\left(k_{1}\right)^{*}\right)	\\	&=\exp \left(2\left(\mathrm{i}\alpha t+a_{1}^{*} \right) x+\frac{2\left({a_{1}^{*}}^{4}+\alpha^{4}t^{4}-6{a_{1}^{*}}^{2}\alpha^{2}t^{2}+4\mathrm{i}\alpha ta_{1}^{*}-4\mathrm{i}\alpha^{3} a_{1}^{*}t^{3}\right)}{\mathrm{i}\alpha } \right)\cdot \lambda _{m}^{*} \mu _{n}^{*},
	\\
		\delta_{1}^{(m)}\left(k_{1}\right) \left( \omega _{1}^{(n)}\left(k_{1}\right)\right)^{*}&=\exp \left(\left(k_{1}(t)+k_{1}(t)^{*}\right)x+\frac{{k_{1}^4(t)}^{*}-k_{1}^4(t)}{\mathrm{i}\alpha }  \right) \cdot \exp \left(\varphi _{1}^{(m)}\left(k_{1}\right)+\phi _{1}^{(n)}\left(k_{1}\right)^{*}\right)	\\	&=\exp \left(2c_{1} x+8c_{1}\left(c_{1}^{2}-3d_{1}^{2}\right)t+24\alpha c_{1}d_{1}t^{2}-8c_{1}\alpha^{2}t^{3} +\frac{8}{\alpha}c_{1}d_{1}\left(d_{1}^{2}-c_{1}^{2}\right)\right)\cdot \lambda _{m} \mu _{n}^{*}.
\end{align*}
\end{subequations}
According to Eq. \eqref{lmnb2}, we have $\lambda_{1} \mu_{2}=\lambda_{2} \mu_{1}$,  so we obtain
\begin{equation*}
\begin{aligned}
&\left|\sum_{n=1}^{2}\delta_{1}^{(m)}\left(k_{1}\right) \left( \omega _{1}^{(n)}\left(k_{1}\right)\right)^{*}\right|^{2}\\&= \exp \left(4c_{1} x+16c_{1}\left(c_{1}^{2}-3d_{1}^{2}\right)t+48\alpha c_{1}d_{1}t^{2}-16c_{1}\alpha^{2}t^{3} +\frac{16}{\alpha}c_{1}d_{1}\left(d_{1}^{2}-c_{1}^{2}\right)\right)\cdot \left(\left|\lambda _{1} \mu _{1}\right|^{2}+\left|\lambda _{2} \mu _{2}\right|^{2}+2\left|\lambda _{1}\mu _{2}\right|^{2}\right).
\end{aligned}
\end{equation*}
Thus, the one-soliton solutions can be written as 
\begin{equation}\label{Qij1}
	\begin{aligned}	
		Q_{i j}=\frac{\exp \left(2\left(\mathrm{i}\alpha t+a_{1}^{*} \right) x+\frac{2\left({a_{1}^{*}}^{4}+\alpha^{4}t^{4}-6{a_{1}^{*}}^{2}\alpha^{2}t^{2}+4\mathrm{i}\alpha ta_{1}^{*}-4\mathrm{i}\alpha^{3} a_{1}^{*}t^{3}\right)}{\mathrm{i}\alpha } \right)\cdot \lambda _{j}^{*} \mu _{i}^{*}}{1+ \exp \left( 4c_{1} x+16c_{1}\left(c_{1}^{2}-3d_{1}^{2}\right)t+48\alpha c_{1}d_{1}t^{2}-16c_{1}\alpha^{2}t^{3} +\frac{16}{\alpha}c_{1}d_{1}\left(d_{1}^{2}-c_{1}^{2}\right)\right)\cdot \xi},
	\end{aligned}
\end{equation}
where
\begin{equation*}
	\xi \doteq \frac{\left|\lambda _{1} \mu _{1}\right|^{2}+\left|\lambda _{2} \mu _{2}\right|^{2}+2\left|\lambda _{1}\mu _{2}\right|^{2}}{4 c_{1}^{2}}> 0.
\end{equation*}
Then, taking the module for $Q_{i j}$, one obtains
\begin{equation}\label{1oneso}
		\left|Q_{i j}\right|
	=\frac{|c_{1}|\left|\lambda _{j} \mu _{i}\right|}{\sqrt{\left|\lambda _{1} \mu _{1}\right|^{2}+\left|\lambda _{2} \mu _{2}\right|^{2}+2\left|\lambda _{1} \mu _{2}\right|^{2}}} \operatorname{sech}\left(2c_{1} x+8c_{1}\left(c_{1}^{2}-3d_{1}^{2}\right)t+24\alpha c_{1}d_{1}t^{2}-8c_{1}\alpha^{2}t^{3} +\frac{8}{\alpha}c_{1}d_{1}\left(d_{1}^{2}-c_{1}^{2}\right)+\frac{1}{2} \ln (\xi)\right).
\end{equation}

\begin{figure}[!ht]
	\centering
	\subfigure[]{
		\includegraphics[width=6cm,height=4cm]{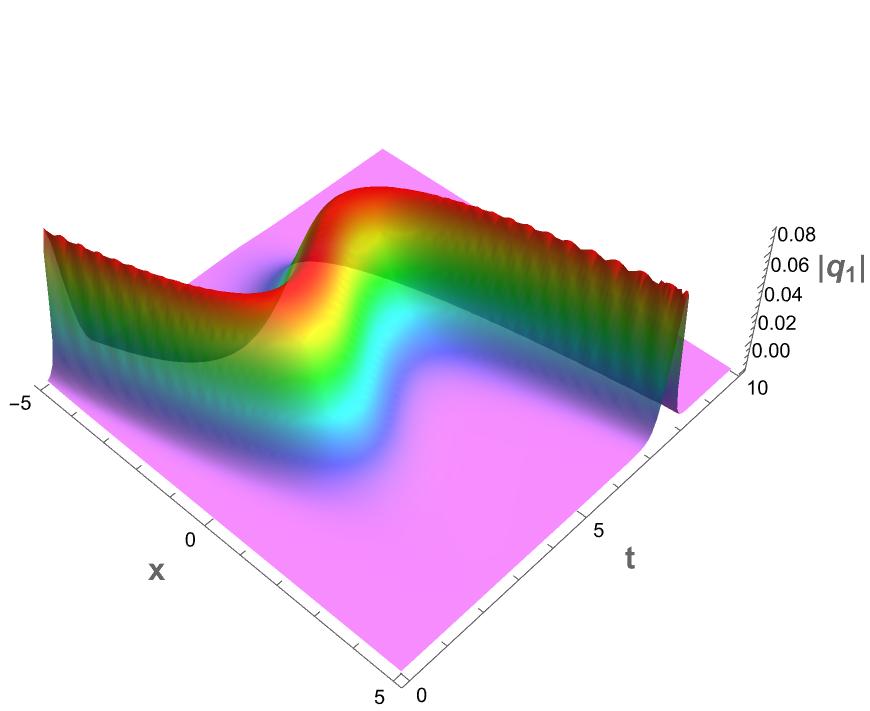} \label{Fig.1(a)}}
	\hspace{1.5mm}
	\subfigure[]{
		\includegraphics[width=6cm,height=4cm]{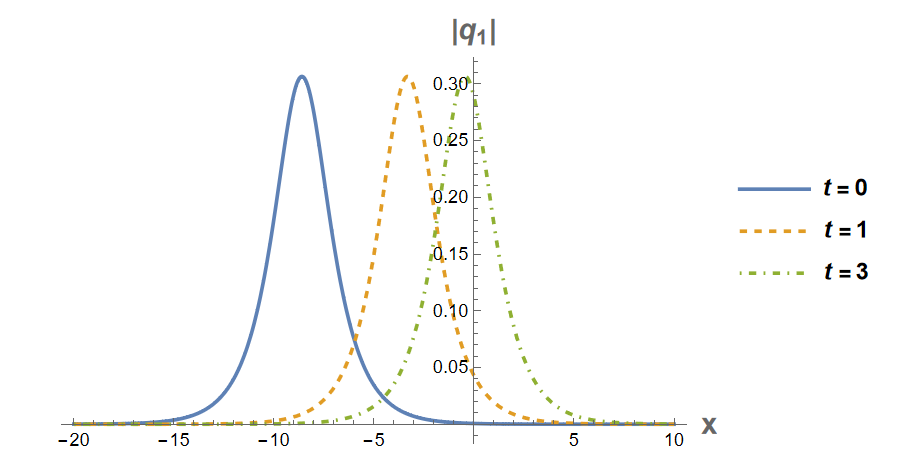} \label{Fig.1(b)}}	
	\caption{The one-soliton solution of the NTCmKdV-$\mathrm{\uppercase\expandafter{\romannumeral1}}$ equation \eqref{NTCmKdV1}. (a) Shape and motion of $|q_{1}|$ with $\alpha=0.2$, $a_{1}=0.4+0.8\mathrm{i}$,  $\lambda_{1}=-3+\mathrm{i}$, $\mu_{1}=2.5-0.4\mathrm{i}$ and $\lambda_{2}=0.7+1.6\mathrm{i}$. (b) Shapes of $|q_{1}|$ with the parameters as in (a) for $t=0$, $t=1$, and $t=3$.}
	\label{tp1}
	\end{figure}

From \eqref{1oneso} we can find that the one-soliton is characterized as follows.
\begin{center}
	\begin{varwidth}{\linewidth}
		\raggedright 
	fixed amplitude: $\frac{|c_{1}|\left|\lambda _{j} \mu _{i}\right|}{\sqrt{\left|\lambda _{1} \mu _{1}\right|^{2}+\left|\lambda _{2} \mu _{2}\right|^{2}+2\left|\lambda _{1} \mu _{2}\right|^{2}}},$ 
	
	initial phase: $ -\frac{4d_{1}}{\alpha }\left(d_{1}^{2}-c_{1}^{2}\right) -\frac{\ln \left(\xi\right)}{4c_{1}},$ 
	
	top trace:  $x(t)=-4\left(c_{1}^{2}-3d_{1}^{2}\right)t-12\alpha d_{1}t^{2}+4\alpha^{2}t^{3}-\frac{4d_{1}}{\alpha }\left(d_{1}^{2}-c_{1}^{2}\right)-\frac{\ln (\xi)}{4c_{1}} $, $\alpha\ne 0, c_{1}\ne 0,$ 
	
	velocity: $x^{\prime}(t)=-4\left(c_{1}^{2}-3d_{1}^{2}\right)-24\alpha d_{1}t+12\alpha^{2}t^{2}$. 
	\end{varwidth}
\end{center}
It can also be seen that $|q_{1}|$, $|q_{2}|$ and $|q_{0}|$ maintain a linear relation, so in the following we  only take $|q_{1}|$ as an example to illustrate dynamic behaviors of the solutions. From Fig. \ref{tp1}, it can be seen that the trace of the one-soliton solution is mainly influenced by the $t^{3}$ term. 

\subsubsection{Two-soliton solution}\label{sec512}
For $N=2$, we have
 	\begin{equation}\label{cj2}
 	\bm{K}_{1}=\left(\begin{array}{cc}
 		k_{1} & \\
 		& k_{2}
 	\end{array}\right), \quad \bm{r}_{1}=\left(\begin{array}{cc}
 		\delta_{1}^{(1)}\left(k_{1}\right) & \delta _{1}^{(2)}\left(k_{1}\right) \\
 		\delta_{1}^{(1)}\left(k_{2}\right) & \delta_{1}^{(2)}\left(k_{2}\right)
 	\end{array}\right), \quad \bm{s}_{1}=\left(\begin{array}{ll}
 		\omega_{1}^{(1)}\left(k_{1}\right) & \omega _{1}^{(2)}\left(k_{1}\right) \\
 		\omega_{1}^{(1)}\left(k_{2}\right) & \omega_{1}^{(2)}\left(k_{2}\right)
 	\end{array}\right), 
 \end{equation}
and
\begin{equation}
 k_{1}=-\mathrm{i}\alpha t+a_{1}, \quad k_{2}=-\mathrm{i}\alpha t+a_{2},\quad
\left(\bm{M}_{1}\right)_{i, j}
=\frac{\delta_{1}^{(1)}\left(k_{i}\right) \omega _{2}^{(1)}\left(g_{j}\right)+\delta _{1}^{(2)}\left(k_{i}\right) \omega _{2}^{(2)}\left(g_{i}\right)}{k_{i}-g_{j}},\quad i,j=1,2.
\end{equation}
For convenience, we set
\begin{subequations}
	\begin{equation}\label{note21}
		\bm{M}_{1}=\left(\begin{array}{ll}
			m_{11} & m_{12} \\
			m_{21} & m_{22}
		\end{array}\right),\quad
		a_{1}=c_{1}+\mathrm{i}d_{1}, \quad a_{2}=c_{2}+\mathrm{i}d_{2},
	\end{equation}
	\begin{equation}\label{note22}
		\exp \left(\varphi _{1}^{(m)}\left(k_{1}\right)\right)=\exp \left(\varphi _{1}^{(m)}\left(k_{2}\right)\right)=\lambda _{m}, \quad \exp \left(\phi  _{1}^{(n)}\left(k_{1}\right)\right)=\exp \left(\phi _{1}^{(n)}\left(k_{2}\right)\right)=\mu _{n}.
	\end{equation}
\end{subequations}
According to the relation \eqref{r2s2}, we have
\begin{equation}\label{mijtwo1}
	m_{i, j}=\frac{\sum_{n=1}^{2} \delta_{1}^{(n)}\left(k_{i}\right)\left(\omega_{1}^{(n)}\left(k_{j}\right)\right)^{*}}{k_{i}+k_{j}^{*}},\quad i,j=1,2.
\end{equation}
Substituting Eq. \eqref{1xjdy} into Eq. \eqref{mijtwo1}, the elements $m_{i, j}$  of the matrix $\bm{M}$ are displayed as 
\begin{equation}
	\begin{array}{l}
		m_{11}= \frac{1}{2 c_{1}}\cdot \exp \left(2c_{1} x+8c_{1}\left(c_{1}^{2}-3d_{1}^{2}\right)t+24\alpha c_{1}d_{1}t^{2}-8c_{1}\alpha^{2}t^{3} +\frac{8}{\alpha}c_{1}d_{1}\left(d_{1}^{2}-c_{1}^{2}\right)\right) \cdot\left(\lambda _{1} \mu_{1}^{*}+\lambda_{2} \mu_{2}^{*}\right), \\
		m_{12}= \frac{1}{k_{1}\left(t\right)+k_{2}^{*}\left(t\right)} \cdot\exp \left(\left(k_{1}\left(t\right)+k_{2}^{*}\left(t\right)\right)x+\frac{{k_{2}^*(t)}^{4}-k_{1}(t)^4}{\mathrm{i}\alpha}\right) \cdot\left(\lambda _{1} \mu_{1}^{*}+\lambda_{2} \mu_{2}^{*}\right), \\
		m_{21}= \frac{1}{k_{2}\left(t\right)+k_{1}^{*}\left(t\right)} \cdot\exp \left(\left(k_{2}\left(t\right)+k_{1}^{*}\left(t\right)\right)x+\frac{{k_{1}^*(t)}^{4}-k_{2}(t)^4}{\mathrm{i}\alpha}\right) \cdot\left(\lambda _{1} \mu_{1}^{*}+\lambda_{2} \mu_{2}^{*}\right),\\
		m_{22}= \frac{1}{2 c_{2}}\cdot \exp \left(2c_{2} x+8c_{2}\left(c_{2}^{2}-3d_{2}^{2}\right)t+24\alpha c_{2}d_{2}t^{2}-8c_{2}\alpha^{2}t^{3} +\frac{8}{\alpha}c_{2}d_{2}\left(d_{2}^{2}-c_{2}^{2}\right)\right) \cdot\left(\lambda _{1} \mu_{1}^{*}+\lambda_{2} \mu_{2}^{*}\right).
	\end{array}
\end{equation}
Here, we adopt the notation $\operatorname{adj}(\bm{A})$ to represent the adjoint matrix of $\bm{A}$. In the explicit expression \eqref{1zj1} of $\bm{Q}$, we use $\left(\bm{I}+ \bm{M}_{1}^{*} \bm{M}_{1}\right)^{-1}=\frac{\operatorname{adj}(\bm{I}+ \bm{M}_{1}^{*} \bm{M}_{1})}{|	\bm{I}+ \bm{M}_{1}^{*} \bm{M}_{1}|}$ to compute the inverse of the matrix. With some direct calculations, we have
\begin{subequations}\label{njz}
	\begin{align}
	&\begin{aligned}
		\operatorname{adj}(\bm{I}+ \bm{M}_{1}^{*} \bm{M}_{1})=\left(\begin{array}{ll}
				1+m_{21}^{*}m_{12}+|m_{22}|^{2} & 
				-m_{11}^{*}m_{12}-m_{12}^{*}m_{22}\\
				-m_{21}^{*}m_{11}-m_{22}^{*}m_{21}&
				1+m_{12}^{*}m_{21}+|m_{11}|^{2}
			\end{array}\right),
		\end{aligned}\\
	&|\bm{I}+ \bm{M}_{1}^{*} \bm{M}_{1}|=1+|m_{11}|^{2}+|m_{22}|^{2}+|m_{11}m_{22}-m_{12}m_{21}|^{2}+2\operatorname{Re}[m_{12}m_{21}^{*}].
\end{align}
\end{subequations}
Finally, the two-soliton can be written as
\begin{equation}\label{twoj}
	\begin{aligned}
		Q_{i j}= \frac{1}{|\bm{I}+ \bm{M}_{1}^{*} \bm{M}_{1}|} \cdot&  \left( \left(\omega  _ { 1 } ^ { ( i ) } ( k _ { 1 } )\right) ^ { * } \left(1+\left|m_{22}\right|^{2}+m_{21}^{*}m_{12} \right)\left(\delta_{1}^{(j)}\left(k_{1}\right)\right)^{*}\right. \\
		& -\left(\omega_{1}^{(i)}\left(k_{1}\right)\right)^{*}
		\left(m_{21}^{*}m_{11} +m_{22}^{*}m_{21} \right)\left(\delta _{1}^{(j)}\left(k_{1}\right)\right)^{*} \\
		&-\left(\omega_{1}^{(i)}\left(k_{1}\right)\right)^{*}
		\left( m_{11}^{*}m_{12}+ m_{12}^{*}m_{22}\right)
		\left(\delta_{1}^{(j)}\left(k_{1}\right)\right)^{*}\\
		&+\left.\left(\omega_{1}^{(i)}\left(k_{1}\right)\right)^{*}\left(1+\left|m_{11}\right|^{2}+m_{12}^{*}m_{21} \right)\left(\delta_{1}^{(j)}\left(k_{1}\right)\right)^{*}\right).
	\end{aligned}
\end{equation}

\begin{figure}[!ht]
	\centering
	\subfigure[]{
		\includegraphics[width=6cm,height=4cm]{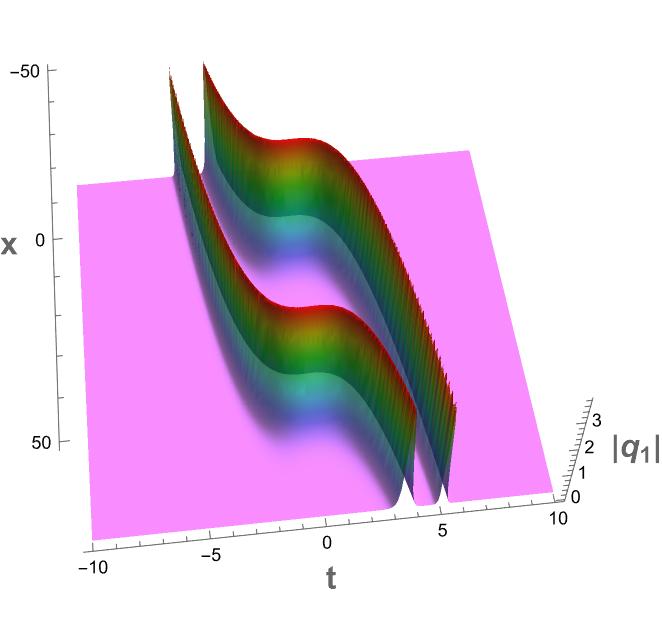} \label{Fig.2(a)}}
\hspace{1.5mm}
	\subfigure[]{
		\includegraphics[width=6cm,height=4cm]{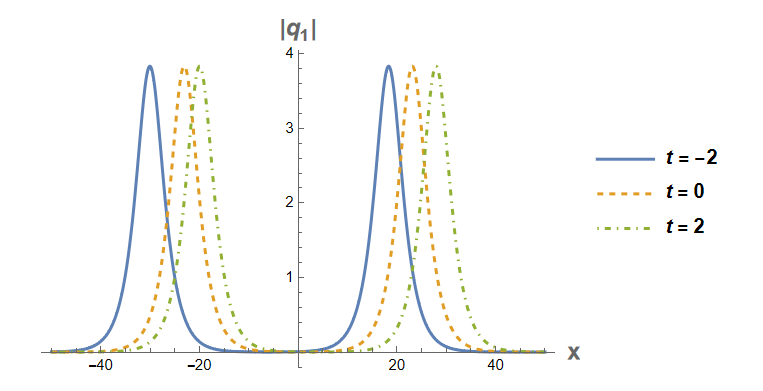} \label{Fig.2(b)}}	

	\caption{The two-soliton solution of the NTCmKdV-$\mathrm{\uppercase\expandafter{\romannumeral1}}$ equation \eqref{NTCmKdV1}. (a) Shape and motion of $|q_{1}|$ with $\alpha=0.4$, $a_{1}=0.2+0.1\mathrm{i}$, $a_{2}=-0.2$, $\lambda_{1}=-3+\mathrm{i}$, $\mu_{1}=-2.5-0.4\mathrm{i}$ and $\lambda_{2}=0.7+1.6\mathrm{i}$. (b) Shapes of $|q_{1}|$ with the parameters as in (a) for $t=-2$, $t=0$, and $t=2$.}
	\label{tp2}
	\end{figure}

The shape and motion of $|q_{1}|$ are illustrated in Fig. \ref{tp2}. From Figure \ref{Fig.2(a)}, it is observed that the two-soliton solution still exhibits a curved trace due to the influence of the $t^{3}$ term.

\subsubsection{Double-pole solution}\label{sec513}
For the double pole-solution, we take
\begin{equation}\label{cj3}
\bm{K}_{1}=\left(\begin{array}{cc}k_{1} & 0 \\ \partial_{a_{1}} k_{1} & k_{1}\end{array}\right),\quad \bm{r}_{1}=\left(\begin{array}{cc}\delta _{1}^{(1)}\left(k_{1}\right) & \delta _{1}^{(2)}\left(k_{1}\right) \\ \partial_{a_{1}} \delta _{1}^{(1)}\left(k_{1}\right) & \partial_{a_{1}} \delta _{1}^{(2)}\left(k_{1}\right)\end{array}\right), \quad \bm{s}_{1}=\left(\begin{array}{cc}\omega_{1}^{(1)}\left(k_{1}\right) & \omega_{1}^{(2)}\left(k_{1}\right) \\ \partial_{a_{1}} \omega_{1}^{(1)}\left(k_{1}\right) & \partial_{a_{1}} \omega _{1}^{(2)}\left(k_{1}\right)\end{array}\right).
\end{equation}
Then matrix $\bm{M}_{1}^{(i)}$ can be constructed according to Eqs. \eqref{m12jn} and \eqref{r2s2} as 
\begin{equation}
\bm{M}_{1}^{(i)}=\left(\begin{array}{cc}\delta _{1}^{(i)}\left(k_{1}\right) & 0 \\ \partial_{a_{1}} \delta _{1}^{(i)}\left(k_{1}\right) & \delta _{1}^{(i)}\left(k_{1}\right)\end{array}\right)\left(\begin{array}{cc}\frac{1}{k_{1}+k_{1}^{*}} & -\frac{\left(\partial_{a_{1}} k_{1}\right)^{*}}{\left(k_{1}+k_{1}^{*}\right)^{2}} \\ -\frac{\partial_{a_{1}} k_{1}}{\left(k_{1}+k_{1}^{*}\right)^{2}} & \frac{2\left(\partial_{a_{1}} k_{1}\right)\left(\partial_{a_{1}} k_{1}\right)^{*}}{\left(k_{1}+k_{1}^{*}\right)^{3}}\end{array}\right)\left(\begin{array}{cc}\left(\omega_{1}^{(i)}\left(k_{1}\right)\right)^{*} & \left(\partial_{a_{1}} \omega_{1}^{(i)}\left(k_{1}\right)\right)^{*} \\ \left(\partial_{a_{1}} \omega_{1}^{(i)}\left(k_{1}\right)\right)^{*} & 0\end{array}\right).
\end{equation}
Let
\begin{equation*}
	\bm{M}_{1}^{(i)}=\left(\begin{array}{ll}
		\left(\bm{M}_{1}^{(i)}\right)_{1, 1} &\left(\bm{M}_{1}^{(i)}\right)_{1, 2} \\
		\left(\bm{M}_{1}^{(i)}\right)_{2, 1} & \left(\bm{M}_{1}^{(i)}\right)_{2, 2}
	\end{array}\right),
\end{equation*}
and using $\bm{M}_{1}=\bm{M}_{1}^{(1)}+ \bm{M}_{1}^{(2)}=\left(\begin{array}{ll}
	m_{11} & m_{12} \\
	m_{21} & m_{22}
\end{array}\right)$ in Eq. \eqref{m12j}, we have
\begin{equation}\label{mijys}
	\begin{aligned}
		m_{11}= & \sum_{i=1}^{2}\left(\frac{\delta_{1}^{(i)}\left(k_{1}\right)\left(\omega_{1}^{(i)}\left(k_{1}\right)\right)^{*}}{k_{1}+k_{1}^{*}}-\frac{\delta_{1}^{(i)}\left(k_{1}\right) \left(\partial_{a_{1}} k_{1}\right)^{*} \left(\partial_{a_{1}} \omega _{1}^{(i)}\left(k_{1}\right)\right)^{*}}{\left(k_{1}+k_{1}^{*}\right)^{2}}\right),\\
		m_{12}= & \sum_{i=1}^{2}\frac{\delta _{1}^{(i)}\left(k_{1}\right)\left(\partial_{a_{1}} \omega _{1}^{(i)}\left(k_{1}\right)\right)^{*}}{k_{1}+k_{1}^{*}}, \\
			m_{21}= & \sum_{i=1}^{2}\left(\frac{\partial_{a_{1}} \delta _{1}^{(i)}\left(k_{1}\right)\left(\omega _{1}^{(i)}\left(k_{1}\right)\right)^{*}}{k_{1}+k_{1}^{*}}-\frac{\delta_{1}^{(i)}\left(k_{1}\right)\partial_{a_{1}} k_{1}\left(\omega _{1}^{(i)}\left(k_{1}\right)\right)^{*}}{\left(k_{1}+k_{1}^{*}\right)^{2}}\right. \\
			&\left.-\frac{\partial_{a_{1}} \delta _{1}^{(i)}\left(k_{1}\right)\left(\partial_{a_{1}} k_{1}\right)^{*}\left(\partial_{a_{1}} \omega _{1}^{(i)}\left(k_{1}\right)\right)^{*}}{\left(k_{1}+k_{1}^{*}\right)^{2}}+\frac{2 \delta _{1}^{(i)}\left(k_{1}\right)\partial_{a_{1}} k_{1}\left(\partial_{a_{1}} k_{1}\right)^{*}\left(\partial_{a_{1}} \omega _{1}^{(i)}\left(k_{1}\right)\right)^{*}}{\left(k_{1}+k_{1}^{*}\right)^{3}} \right), \\
		m_{22}= & \sum_{i=1}^{2}\left(\frac{\partial_{a_{1}} \delta _{1}^{(i)}\left(k_{1}\right)\left(\partial_{a_{1}} \omega _{1}^{(i)}\left(k_{1}\right)\right)^{*}}{k_{1}+k_{1}^{*}}-\frac{\delta_{1}^{(i)}\left(k_{1}\right)\partial_{a_{1}} k_{1}\left(\partial_{a_{1}} \omega _{1}^{(i)}\left(k_{1}\right)\right)^{*}}{\left(k_{1}+k_{1}^{*}\right)^{2}} \right) .
	\end{aligned}
\end{equation}
Since $\bm{Q}$ can be expressed as $\bm{s}_{1}^{\dagger}\left(\bm{I}_{2}+\bm{M}_{1}^{*} \bm{M}_{1}\right)^{-1} \bm{r}_{1}^{*}$, combined with Eq. \eqref{njz}, we have
\begin{equation}\label{dej}
	\begin{aligned}
		Q_{i j}=  \frac{1}{|\bm{I}+\bm{M}_{1}^{*} \bm{M}_{1}|} \cdot& \left( \left(\omega  _ { 1 } ^ { ( i ) } ( k _ { 1 } )\right) ^ { * } \left(1+\left|m_{22}\right|^{2}+m_{21}^{*}m_{12} \right)\left(\delta_{1}^{(j)}\left(k_{1}\right)\right)^{*}\right. \\
		&- \left(\partial_{a_{1}} \omega _{1}^{(i)}\left(k_{1}\right)\right)^{*}\left(m_{21}^{*}m_{11} +m_{22}^{*}m_{21}\right)\left(\delta_{1}^{(j)}\left(k_{1}\right)\right)^{*}\\
		&-\left(\omega_{1}^{(i)}\left(k_{1}\right)\right)^{*}			\left(m_{11}^{*}m_{12}+m_{12}^{*}m_{22}\right)\left(\partial_{a_{1}} \delta _{1}^{(j)}\left(k_{1}\right)\right)^{*} \\
		& +\left.\left(\partial_{a_{1}} \omega _{1}^{(i)}\left(k_{1}\right)\right)^{*} \left(1+\left|m_{11}\right|^{2}+m_{12}^{*}m_{21} \right)\left(\partial_{a_{1}} \delta _{1}^{(j)}\left(k_{1}\right)\right)^{*} \right),
	\end{aligned}
\end{equation}
where 
\begin{equation*}
	\begin{aligned}
		m_{11}= & \sum_{i=1}^{2}\left(\frac{\lambda _{i} \mu_{i}^{*}\cdot \exp{\left(2c_{1}x+D\right)}}{2c_{1}}-\frac{\lambda _{i} \mu_{i}^{*}\cdot \exp{\left(2c_{1}x+D\right)}}{4c_{1}^{2}}\left(x+\frac{{4k_{1}^{*}}^{3}}{\mathrm{i}\alpha } \right)\right),\\
		m_{12}= & \sum_{i=1}^{2}\frac{\lambda _{i} \mu_{i}^{*}\cdot \exp{\left(2c_{1}x+D\right)}}{2c_{1}}\left(x+\frac{{4k_{1}^{*}}^{3}}{\mathrm{i}\alpha } \right), \\
		m_{21}= & \sum_{i=1}^{2}\left(\frac{\lambda _{i} \mu_{i}^{*}\cdot \exp{\left(2c_{1}x-D\right)}}{2c_{1}}-\frac{\lambda _{i} \mu_{i}^{*}\cdot \exp{\left(2c_{1}x+D\right)}}{4c_{1}^{2}}\right. \\
		&\left.-\frac{\lambda _{i} \mu_{i}^{*}\cdot \exp{\left(2c_{1}x+D\right)}}{4c_{1}^{2}}\left(x-\frac{{4k_{1}}^{3}}{\mathrm{i}\alpha } \right)\left(x+\frac{{4k_{1}^{*}}^{3}}{\mathrm{i}\alpha } \right)+\frac{\lambda _{i} \mu_{i}^{*}\cdot \exp{\left(2c_{1}x+D\right)}}{4c_{1}^{3}}\left(x+\frac{{4k_{1}^{*}}^{3}}{\mathrm{i}\alpha } \right)\right), \\
		m_{22}= & \sum_{i=1}^{2}\left(\frac{\lambda _{i} \mu_{i}^{*}\cdot \exp{\left(2c_{1}x+D\right)}}{2c_{1}}\left(x+\frac{{4k_{1}^{*}}^{3}}{\mathrm{i}\alpha } \right)\left(x-\frac{{4k_{1}}^{3}}{\mathrm{i}\alpha } \right)-\frac{\lambda _{i} \mu_{i}^{*}\cdot \exp{\left(2c_{1}x+D\right)}}{4c_{1}^{2}}\left(x+\frac{{4k_{1}^{*}}^{3}}{\mathrm{i}\alpha } \right)\right) .
	\end{aligned}
\end{equation*}
with
\begin{equation*}
D=8c_{1}\left(c_{1}^{2}-3d_{1}^{2}\right)t+24\alpha c_{1}d_{1}t^{2}-8c_{1}\alpha^{2}t^{3} +\frac{8}{\alpha}c_{1}d_{1}\left(d_{1}^{2}-c_{1}^{2}\right).
\end{equation*}

\begin{figure}[!ht]
	\centering
	\subfigure[]{
		\includegraphics[width=6cm,height=4cm]{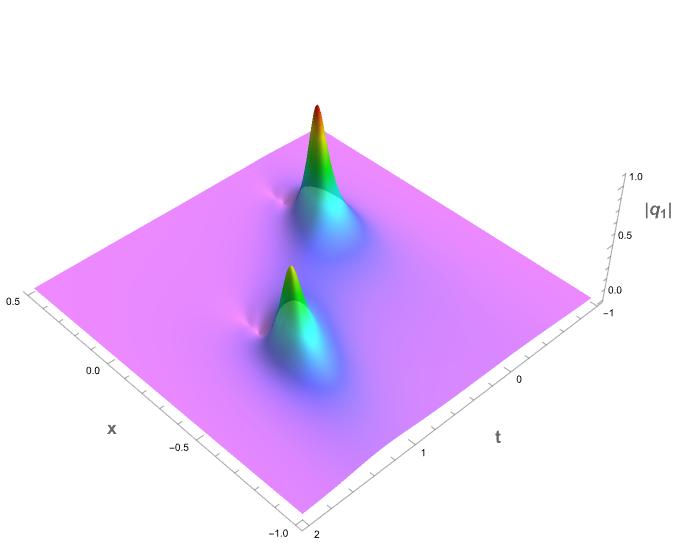} \label{Fig.3(a)}}
\hspace{1.5mm}
	\subfigure[]{
		\includegraphics[width=6cm,height=4cm]{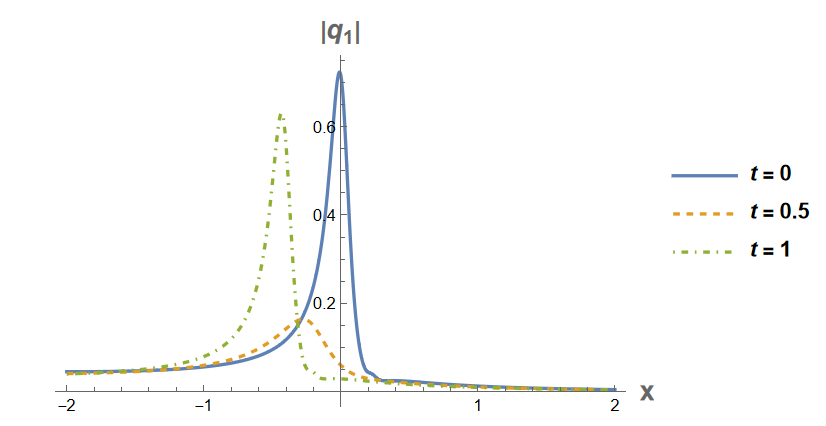} \label{Fig.3(b)}}	
	
	\caption{The double-pole solution of the NTCmKdV-$\mathrm{\uppercase\expandafter{\romannumeral1}}$ equation \eqref{NTCmKdV1}. (a) Shape and motion of $|q_{1}|$ with $\alpha=0.2$, $a_{1}=0.2+0.1\mathrm{i}$, $\lambda_{1}=-3+\mathrm{i}$, $\mu_{1}=2.5-0.4\mathrm{i}$ and $\lambda_{2}=0.7+1.6\mathrm{i}$. (b) Shapes of $|q_{1}|$ with the parameters as in (a) for $t=0$, $t=0.5$, and $t=1$. 	}
	\label{tp3}
\end{figure}

Therefore, the double-pole solution of the NTCmKdV-$\mathrm{\uppercase\expandafter{\romannumeral1}}$ equation \eqref{NTCmKdV1} is given from Eq. \eqref{dej}. For the shape and motion of $|q_{1}|$ can be  observed in Fig.\ref{tp3}. As can be seen in Fig. \ref{Fig.3(b)}, the amplitude of $|q_{1}|$ varies with time. 

\subsection{Solutions to the NTCmKdV-$\mathrm{\uppercase\expandafter{\romannumeral2}}$}\label{sec52}
\subsubsection{One-soliton solution}\label{sec521}
For $N=1$, we have
\begin{equation}\label{cj4}
	\bm{K}_{1}=k_{1}, \quad \bm{r}_{1}=\left(\delta _{1}^{(1)}\left(k_{1}\right), \delta_{1}^{(2)}\left(k_{1}\right)\right), \quad \bm{s}_{1}=\left(\omega _{1}^{(1)}\left(k_{1}\right), \omega _{1}^{(2)}\left(k_{1}\right)\right), \quad k_{1}=a_{1}\mathrm{e}^{-\beta t}, \quad \beta\in \mathbb{R}.
\end{equation}
The $\bm{M}_{1}$ is defined in \eqref{1m1}. According to \eqref{zj11}, $	Q_{i j}$ can be expressed as
\begin{equation}\label{Qij2}
		Q_{i j}=\frac{4 \mathrm{e}^{-2\beta t}\left| \operatorname{Re}\left[a_{1}\right]\right|^{2}\left(\omega _{1}^{(i)}\left(k_{1}\right)\right)^{*} \left(\delta _{1}^{(j)}\left(k_{1}\right)\right)^{*}}{4\mathrm{e}^{-2\beta t}\left| \operatorname{Re}\left[a_{1}\right]\right|^{2}+\left|\sum_{n=1}^{2} \delta _{1}^{(n)}\left(k_{1}\right)\left(\omega _{1}^{(n)}\left(k_{1}\right)\right)^{*}\right|^{2}} .
\end{equation}
We continue to use the notations $c_{1}$, $d_{1}$, $\lambda _{m}$ and $\mu _{n}$ in Section \ref{sec511}, and from Eq. \eqref{2xjdy} we obtain
\begin{subequations}
	\begin{align*}
		\left( \omega _{1}^{(n)}\left(k_{1}\right)\right)^{*}\left( \delta_{1}^{(m)}\left(k_{1}\right)\right)^{*} &=\exp \left(2a_{1}^{*} \mathrm{e}^{-\beta t}x-\frac{8{a_{1}^{*}}^{3}\mathrm{e}^{-3\beta t}}{3\beta}-\beta t\right)\cdot \lambda _{m}^{*} \mu _{n}^{*},
		\\
		\delta_{1}^{(m)}\left(k_{1}\right) \left( \omega _{1}^{(n)}\left(k_{1}\right)\right)^{*}&=\exp \left(2c_{1} \mathrm{e}^{-\beta t}x-\frac{8c_{1}\left(c_{1}^{2}-3d_{1}^{2}\right)\mathrm{e}^{-3\beta t}}{3\beta}-\beta t\right)\cdot \lambda _{m} \mu _{n}^{*}.
		\end{align*}
\end{subequations}
Hence, we have 
\begin{equation*}
		\left|\sum_{n=1}^{2}\delta_{1}^{(m)}\left(k_{1}\right) \left( \omega _{1}^{(n)}\left(k_{1}\right)\right)^{*}\right|^{2}= \exp \left(4c_{1} \mathrm{e}^{-\beta t}x-\frac{16c_{1}\left(c_{1}^{2}-3d_{1}^{2}\right)\mathrm{e}^{-3\beta t}}{3\beta}-2\beta t\right)\cdot \left(\left|\lambda _{1} \mu _{1}\right|^{2}+\left|\lambda _{2} \mu _{2}\right|^{2}+2\left|\lambda _{1}\mu _{2}\right|^{2}\right).
\end{equation*}
Thus, the one-soliton solutions can be written as 
\begin{equation}\label{Qij21}
	Q_{i j}=\frac{\exp \left(2a_{1}^{*} \mathrm{e}^{-\beta t}x-\frac{8{a_{1}^{*}}^{3}\mathrm{e}^{-3\beta t}}{3\beta}-3\beta t\right)\cdot \lambda _{m}^{*} \mu _{n}^{*}}{\exp\left(-2\beta t\right)+\exp \left(4c_{1} \mathrm{e}^{-\beta t}x-\frac{16c_{1}\left(c_{1}^{2}-3d_{1}^{2}\right)\mathrm{e}^{-3\beta t}}{3\beta}-2\beta t\right)\cdot\xi},
\end{equation}
where $\xi \doteq \frac{\left|\lambda _{1} \mu _{1}\right|^{2}+\left|\lambda _{2} \mu _{2}\right|^{2}+2\left|\lambda _{1}\mu _{2}\right|^{2}}{4 c_{1}^{2}}$.

Then, the $\left|Q_{i j}\right|$ can be written as
\begin{equation}\label{2oneso}
\left|Q_{i j}\right|=\frac{|c_{1}|\left|\lambda _{j} \mu _{i}\right|\mathrm{e}^{-\beta t}}{\sqrt{\left|\lambda _{1} \mu _{1}\right|^{2}+\left|\lambda _{2} \mu _{2}\right|^{2}+2\left|\lambda _{1} \mu _{2}\right|^{2}}} \operatorname{sech}\left(2c_{1} \mathrm{e}^{-\beta t}x-\frac{8c_{1}\left(c_{1}^{2}-3d_{1}^{2}\right)}{3\beta}\mathrm{e}^{-3\beta t}+\frac{1}{2} \ln (\xi)\right).
\end{equation}

\begin{figure}[!ht]
	\centering
	\subfigure[]{
		\includegraphics[width=6cm,height=4cm]{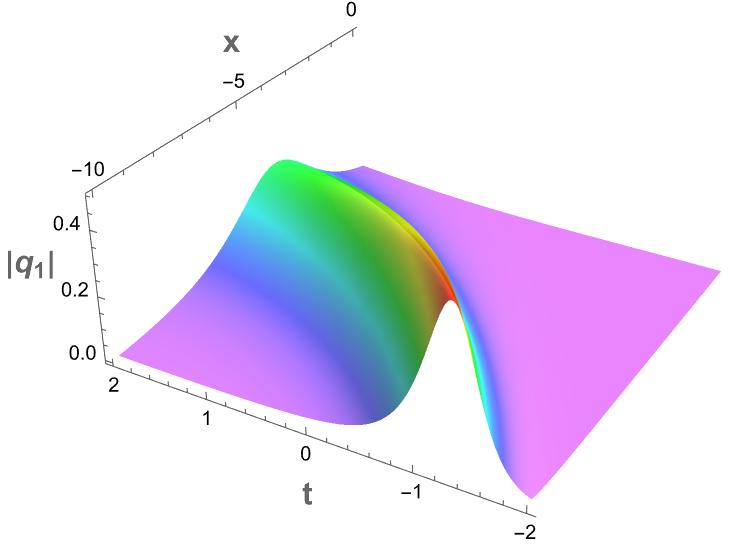} \label{Fig.4(a)}}
	\hspace{1.5mm}
	\subfigure[]{
		\includegraphics[width=6cm,height=4cm]{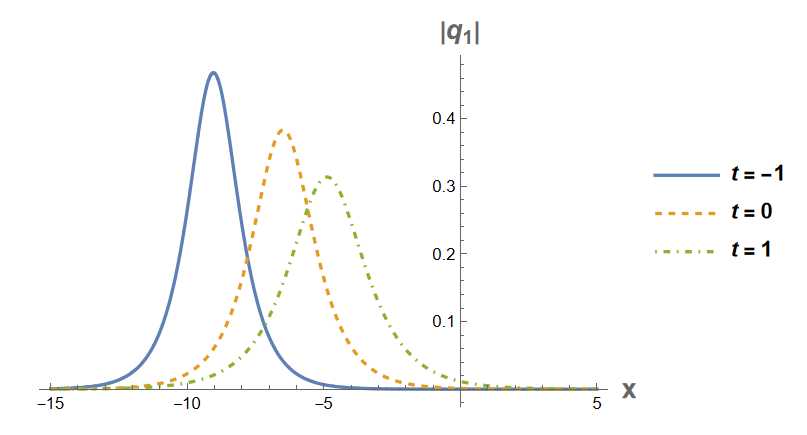} \label{Fig.4(b)}}	
	
	\caption{The one-soliton solution of the NTCmKdV-$\mathrm{\uppercase\expandafter{\romannumeral2}}$ equation \eqref{NTCmKdV2}. (a) Shape and motion of $|q_{1}|$ with $\beta=0.2$, $a_{1}=0.5+0.6\mathrm{i}$,  $\lambda_{1}=-3+\mathrm{i}$, $\mu_{1}=2.5-0.4\mathrm{i}$ and $\lambda_{2}=0.7+1.6\mathrm{i}$. (b) Shapes of $|q_{1}|$ with the parameters as in (a) for $t=-1$, $t=0$, and $t=1$.}
	\label{tp4}
\end{figure}

From \eqref{2oneso} we can find that 
the one-soliton is characterized as follows.
\begin{center}
	\begin{varwidth}{\linewidth}
		\raggedright 
		amplitude: $\frac{|c_{1}|\left|\lambda _{j} \mu _{i}\right|}{\sqrt{\left|\lambda _{1} \mu _{1}\right|^{2}+\left|\lambda _{2} \mu _{2}\right|^{2}+2\left|\lambda _{1} \mu _{2}\right|^{2}}}\mathrm{e}^{-\beta t},$ 
		
		top trace:  $x(t)=\frac{4\left(c_{1}^{2}-3d_{1}^{2}\right)}{3\beta}\mathrm{e}^{-2\beta t}-\frac{\ln (\xi)}{4c_{1}}\mathrm{e}^{\beta t}$, $\beta\ne 0, c_{1}\ne 0,$ 
		
		velocity: $x^{\prime}(t)=-\frac{8\left(c_{1}^{2}-3d_{1}^{2}\right)}{3}\mathrm{e}^{-2\beta t}-\frac{\beta\ln (\xi)}{4c_{1}}\mathrm{e}^{\beta t}$. 
	\end{varwidth}
\end{center}
In the solution given by Eq. \eqref{2oneso}, the  top trace is influenced by the exponential function, resulting in an exponentially shaped trace of the one-soliton solution, as can be observed in Fig. \ref{Fig.4(a)}. It can also be seen that the amplitude function varies over time. Fig. \ref{Fig.4(b)} further reveals time-dependent amplitude variations.

\subsubsection{Two-soliton solution}\label{sec522}
For $N=2$, we have
\begin{equation}\label{cj5}
	\bm{K}_{1}=\left(\begin{array}{cc}
		k_{1} & \\
		& k_{2}
	\end{array}\right), \quad \bm{r}_{1}=\left(\begin{array}{cc}
		\delta_{1}^{(1)}\left(k_{1}\right) & \delta _{1}^{(2)}\left(k_{1}\right) \\
		\delta_{1}^{(1)}\left(k_{2}\right) & \delta_{1}^{(2)}\left(k_{2}\right)
	\end{array}\right), \quad \bm{s}_{1}=\left(\begin{array}{ll}
		\omega_{1}^{(1)}\left(k_{1}\right) & \omega _{1}^{(2)}\left(k_{1}\right) \\
		\omega_{1}^{(1)}\left(k_{2}\right) & \omega_{1}^{(2)}\left(k_{2}\right)
	\end{array}\right), 
\end{equation}
and
\begin{equation}
	k_{1}=a_{1}\mathrm{e}^{-\beta t},\quad	k_{2}=a_{2}\mathrm{e}^{-\beta t}.
\end{equation}
The explicit formula for the two-soliton solution in Eq. \eqref{twoj} can be obtained, where $m_{i, j}$  are represented as
\begin{equation}
	\begin{array}{l}
		m_{11}= \frac{1}{2 c_{1}\mathrm{e}^{-\beta t}}\cdot \exp \left(2c_{1} \mathrm{e}^{-\beta t}x-\frac{8c_{1}\left(c_{1}^{2}-3d_{1}^{2}\right)}{3\beta}\mathrm{e}^{-3\beta t}-\beta t\right) \cdot\left(\lambda _{1} \mu_{1}^{*}+\lambda_{2} \mu_{2}^{*}\right), \\
		m_{12}= \frac{1}{k_{1}\left(t\right)+k_{2}^{*}\left(t\right)} \cdot\exp \left(\left(k_{1}\left(t\right)+k_{2}^{*}\left(t\right)\right)x-\frac{4\left(k_{1}(t)^3+{k_{2}^*(t)}^{3}\right)}{3\beta}-\beta t\right) \cdot\left(\lambda _{1} \mu_{1}^{*}+\lambda_{2} \mu_{2}^{*}\right), \\
		m_{21}= \frac{1}{k_{2}\left(t\right)+k_{1}^{*}\left(t\right)} \cdot\exp \left(\left(k_{2}\left(t\right)+k_{1}^{*}\left(t\right)\right)x-\frac{4\left(k_{2}(t)^3+{k_{1}^*(t)}^{3}\right)}{3\beta}-\beta t\right) \cdot\left(\lambda _{1} \mu_{1}^{*}+\lambda_{2} \mu_{2}^{*}\right),\\
		m_{22}= \frac{1}{2 c_{2}\mathrm{e}^{-\beta t}}\cdot \exp \left(2c_{2} \mathrm{e}^{-\beta t}x-\frac{8c_{2}\left(c_{2}^{2}-3d_{2}^{2}\right)}{3\beta}\mathrm{e}^{-3\beta t}-\beta t\right) \cdot\left(\lambda _{1} \mu_{1}^{*}+\lambda_{2} \mu_{2}^{*}\right).
	\end{array}
\end{equation}

\begin{figure}[!ht]
	\centering
	\subfigure[]{
		\includegraphics[width=6cm,height=4cm]{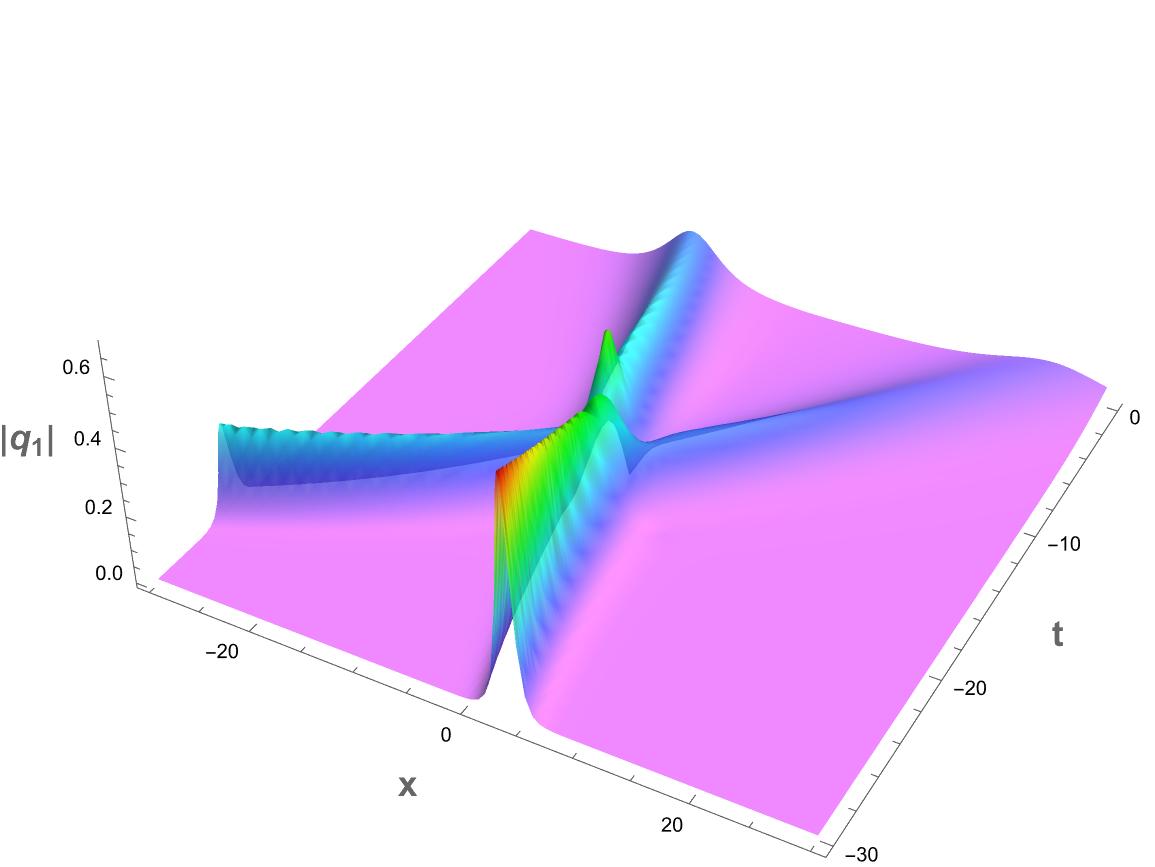} \label{Fig.5(a)}}
	\hspace{1.5mm}
	\subfigure[]{
		\includegraphics[width=6cm,height=4cm]{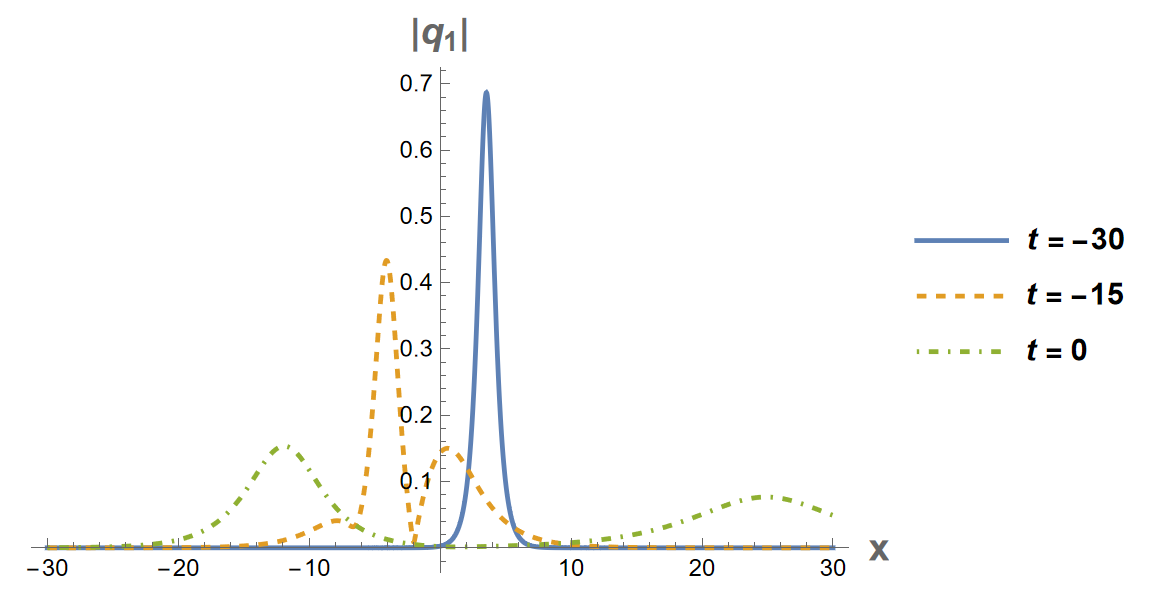} \label{Fig.5(b)}}	
	
		\subfigure[]{
		\includegraphics[width=6cm,height=4cm]{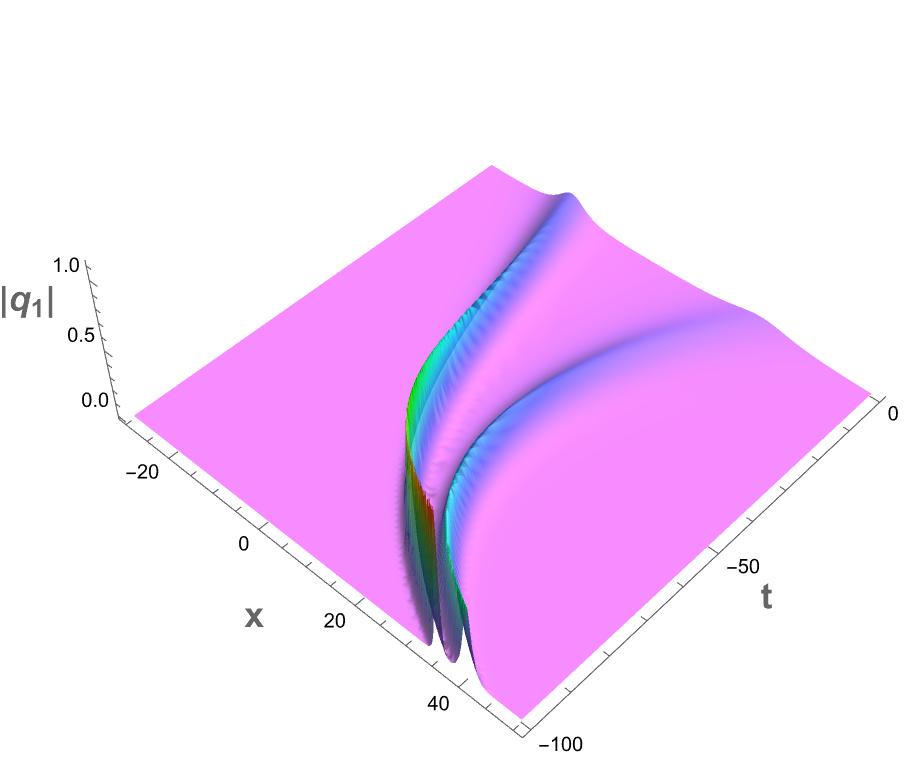} \label{Fig.5(c)}}
	\hspace{1.5mm}
	\subfigure[]{
		\includegraphics[width=6cm,height=4cm]{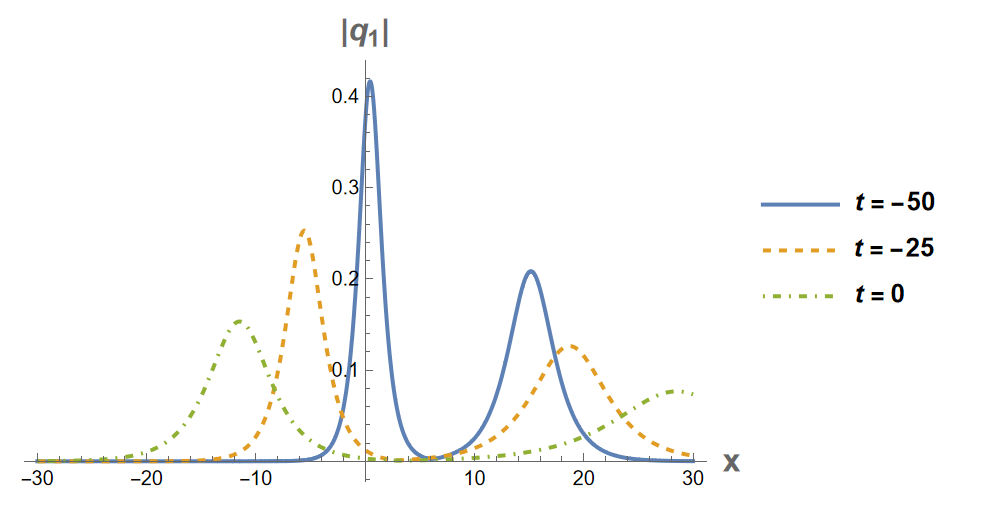} \label{Fig.5(d)}}	
	
	\caption{The two-soliton solution of the NTCmKdV-$\mathrm{\uppercase\expandafter{\romannumeral2}}$ equation \eqref{NTCmKdV2}. (a) Shape and motion of $|q_{1}|$ with $\beta=0.05$, $a_{1}=0.2+0.1\mathrm{i}$, $a_{2}=-0.1+0.2\mathrm{i}$, $\lambda_{1}=-3+\mathrm{i}$, $\mu_{1}=2.5-0.4\mathrm{i}$ and $\lambda_{2}=0.7+1.6\mathrm{i}$. (b) Shapes of $|q_{1}|$ with the parameters as in (a) for $t=-30$, $t=-15$, and $t=0$. (c) Shape and motion of $|q_{1}|$ with $\beta=0.05$, $a_{1}=0.2+0.1\mathrm{i}$, $a_{2}=-0.1$, $\lambda_{1}=-3+\mathrm{i}$, $\mu_{1}=2.5-0.4\mathrm{i}$ and $\lambda_{2}=0.7+1.6\mathrm{i}$. (d) Shapes of $|q_{1}|$ with the parameters as in (a) for $t=-50$, $t=-25$, and $t=0$.}
	\label{tp5}
\end{figure}

Fig. \ref{Fig.5(a)} and \ref{Fig.5(c)} show the interaction between two solitons with different parameters. It is also observable that  $\operatorname{Im}[a_{2}]=d_{2}\ne 0$ leads to the emergence of strong  interaction as shown in Fig. \ref{Fig.5(a)}. It is interesting that their amplitudes display time-dependent characteristic as shown in Fig. \ref{Fig.5(b)} and \ref{Fig.5(d)} .

\subsubsection{Double-pole solution}\label{sec523}
The double-pole solution of the NTCmKdV-$\mathrm{\uppercase\expandafter{\romannumeral2}}$ equation \eqref{NTCmKdV2} can be  obtained by formula \eqref{dej}, where the definition of $\bm{K}_{1}$, $\bm{r}_{1}$ and $\bm{s}_{1}$ are shown in \eqref{cj3} and \eqref{kjgj2}. $m_{ij}$ can be directly calculated through \eqref{mijys} as follows
\begin{equation*}
	\begin{aligned}
		m_{11}= & \sum_{i=1}^{2}\left(\frac{\lambda _{i} \mu_{i}^{*}\cdot \exp{\left(E\right)}}{2c_{1}}-\frac{\lambda _{i} \mu_{i}^{*}\cdot \exp{\left(E-\beta t\right)}}{4c_{1}^{2}}\left(x-\frac{{4k_{1}^{*}}^{2}}{\beta} \right)\right),\\
		m_{12}= & \sum_{i=1}^{2}\frac{\lambda _{i} \mu_{i}^{*}\cdot \exp{\left(E-\beta t\right)}}{2c_{1}}\left(x-\frac{{4k_{1}^{*}}^{2}}{\beta} \right), \\
		m_{21}= & \sum_{i=1}^{2}\left(\frac{\lambda _{i} \mu_{i}^{*}\cdot \exp{\left(E-\beta t\right)}}{2c_{1}}\left(x-\frac{4k_{1}^{2}}{\beta} \right)-\frac{\lambda _{i} \mu_{i}^{*}\cdot \exp{\left(E\right)}}{4c_{1}^{2}}\right. \\
		&\left.-\frac{\lambda _{i} \mu_{i}^{*}\cdot \exp{\left(E-2\beta t\right)}}{4c_{1}^{2}}\left(x-\frac{4k_{1}^{2}}{\beta} \right)\left(x-\frac{{4k_{1}^{*}}^{2}}{\beta} \right)+\frac{\lambda _{i} \mu_{i}^{*}\cdot \exp{\left(E-\beta t\right)}}{4c_{1}^{3}}\left(x-\frac{{4k_{1}^{*}}^{2}}{\beta}\right)\right), \\
		m_{22}= & \sum_{i=1}^{2}\left(\frac{\lambda _{i} \mu_{i}^{*}\cdot \exp{\left(E-2\beta t\right)}}{2c_{1}}\left(x-\frac{{4k_{1}^{*}}^{2}}{\beta} \right)\left(x-\frac{{4k_{1}}^{2}}{\beta } \right)-\frac{\lambda _{i} \mu_{i}^{*}\cdot \exp{\left(E-\beta t\right)}}{4c_{1}^{2}}\left(x-\frac{{4k_{1}^{*}}^{2}}{\beta} \right)\right) .
	\end{aligned}
\end{equation*}
with
\begin{equation*}
	E=2c_{1} \mathrm{e}^{-\beta t}x-\frac{8c_{1}\left(c_{1}^{2}-3d_{1}^{2}\right)}{3\beta}\mathrm{e}^{-3\beta t}.
\end{equation*}

\begin{figure}[!ht]
	\centering
	\subfigure[]{
		\includegraphics[width=6cm,height=4cm]{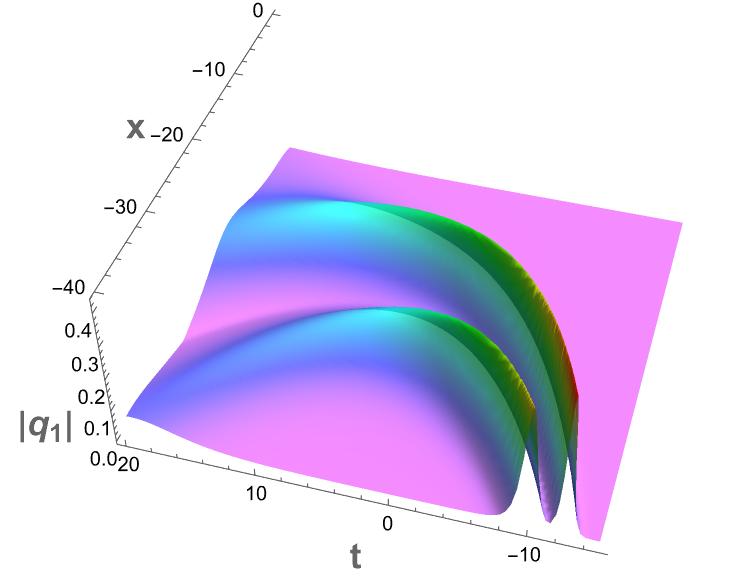} \label{Fig.6(a)}}
	\hspace{1.5mm}
	\subfigure[]{
		\includegraphics[width=6cm,height=4cm]{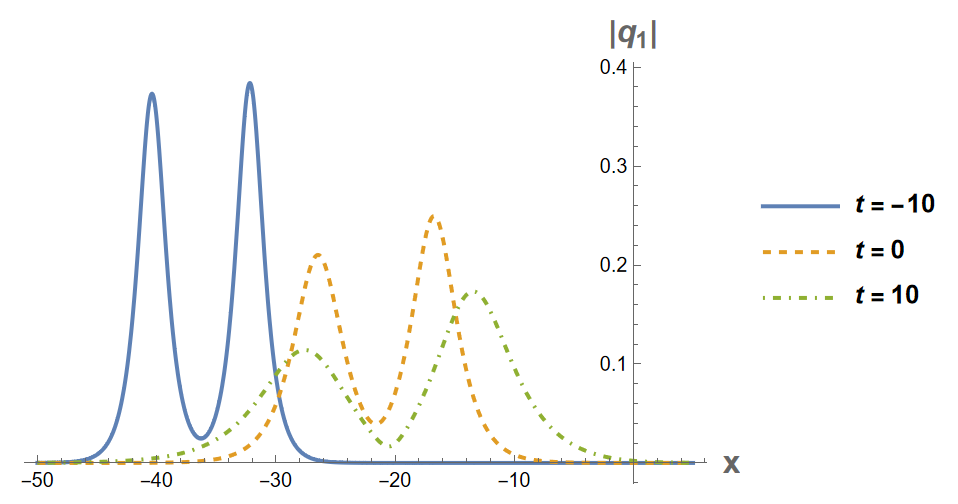} \label{Fig.6(b)}}	
	
	\caption{The double-pole solution of the NTCmKdV-$\mathrm{\uppercase\expandafter{\romannumeral2}}$ equation \eqref{NTCmKdV2}. (a) Shape and motion of $|q_{1}|$ with $\beta=0.05$, $a_{1}=0.3-0.4\mathrm{i}$, $\lambda_{1}=-3+\mathrm{i}$, $\mu_{1}=2.5-0.4\mathrm{i}$ and $\lambda_{2}=0.7+1.6\mathrm{i}$. (b) Shapes of $|q_{1}|$ with the parameters as in (a) for $t=-10$, $t=0$, and $t=10$.
	}
	\label{tp6}
\end{figure}

The shape and motion of $|q_{1}|$ can be  observed in Fig. \ref{tp6}. From Fig. \ref{Fig.6(a)}, it is found that the two solitons move along a curved trace, and in combination with Fig. \ref{Fig.6(b)}, it can be seen that the amplitude is varied with time and the distance between the solitons increases.

\subsection{Solutions to the NTCmKdV-$\mathrm{\uppercase\expandafter{\romannumeral3}}$}\label{sec53}
\subsubsection{One-soliton solution}\label{sec531}
For $N=1$, we have
\begin{equation}\label{cj7}
	\bm{K}_{1}=k_{1}, \quad \bm{r}_{1}=\left(\delta _{1}^{(1)}\left(k_{1}\right), \delta_{1}^{(2)}\left(k_{1}\right)\right), \quad \bm{s}_{1}=\left(\omega _{1}^{(1)}\left(k_{1}\right), \omega _{1}^{(2)}\left(k_{1}\right)\right), \quad k_{1}=\frac{1}{\mathrm{i}\gamma\left(t+a_{1}\right)}, \quad\gamma\in \mathbb{R}.
\end{equation}
From \eqref{zj11}, \eqref{1m1}, we have
\begin{equation}\label{Qij4}
	Q_{i j}=\frac{\left(\omega _{1}^{(i)}\left(k_{1}\right)\right)^{*} \left(\delta _{1}^{(j)}\left(k_{1}\right)\right)^{*}}{\frac{4 \operatorname{Im}\left[a_{1}\right]^{2}}{\gamma^{2}\left(t^{2}+2 \operatorname{Re}\left[a_{1}\right] t+\operatorname{Re}\left[a_{1}\right]^{2}+\operatorname{Im}[a_{1}]^{2}\right)^{2}}+\left|\sum_{n=1}^{2} \delta _{1}^{(n)}\left(k_{1}\right)\left(\omega _{1}^{(n)}\left(k_{1}\right)\right)^{*}\right|^{2}}\cdot \frac{4 \operatorname{Im}\left[a_{1}\right]^{2}}{\gamma^{2}\left(t^{2}+2 \operatorname{Re}\left[a_{1}\right] t+\operatorname{Re}\left[a_{1}\right]^{2}+\operatorname{Im}[a_{1}]^{2}\right)^{2}} .
\end{equation}
From the explicit expressions $\delta$ and $\omega$ representing Eq. \eqref{3xjdy}, we have
\begin{subequations}
	\begin{align*}		\left( \omega _{1}^{(n)}\left(k_{1}\right)\right)^{*}\left( \delta_{1}^{(m)}\left(k_{1}\right)\right)^{*} &=\exp\left(\frac{2x}{-\mathrm{i}\gamma \left(t+a_{1}^{*} \right)}-\frac{4}{\mathrm{i}\gamma ^{3}\left(t+a_{1}^{*} \right)^{2}} \right)\cdot\frac{1}{-\gamma ^{2}\left(t+a_{1}^{*} \right)^{2}} \cdot \lambda _{m}^{*}  \mu _{n}^{*},
		\\
	\delta_{1}^{(m)}\left(k_{1}\right) \left( \omega _{1}^{(n)}\left(k_{1}\right)\right)^{*}&=\exp\left(\frac{-2d_{1}}{\gamma \left(t^{2}+2c_{1}t+c_{1}^{2}+d_{1}^{2}\right)}x -\frac{8d_{1}\left(t+c_{1}\right)}{\gamma ^{3}\left(t^{2}+2c_{1}t+c_{1}^{2}+d_{1}^{2}\right)}\right)\cdot \frac{1}{\gamma^{2} \left(t^{2}+2c_{1}t+c_{1}^{2}+d_{1}^{2}\right)}\cdot \lambda _{m} \mu _{n}^{*}.
	\end{align*}
\end{subequations}
Thus, the one-soliton solutions can be written as 
\begin{equation}\label{Qij41}
	Q_{i j}=\frac{\exp\left(\frac{2x}{-\mathrm{i}\gamma \left(t+a_{1}^{*} \right)}-\frac{4}{\mathrm{i}\gamma ^{3}\left(t+a_{1}^{*} \right)^{2}} \right)\cdot\frac{1}{-\gamma ^{2}\left(t+a_{1}^{*} \right)^{2}} \cdot \lambda _{m}^{*}  \mu _{n}^{*} }{1+\gamma^{2} \left(t^{2}+2c_{1}t+c_{1}^{2}+d_{1}^{2}\right)^{2}\eta \cdot  \exp\left(\frac{-4d_{1}}{\gamma \left(t^{2}+2c_{1}t+c_{1}^{2}+d_{1}^{2}\right)}x -\frac{16d_{1}\left(t+c_{1}\right)}{\gamma ^{3}\left(t^{2}+2c_{1}t+c_{1}^{2}+d_{1}^{2}\right)}\right)\cdot \frac{1}{\gamma^{4} \left(t^{2}+2c_{1}t+c_{1}^{2}+d_{1}^{2}\right)^{2}}},
\end{equation}
where $\eta  \doteq \frac{\left|\lambda _{1} \mu _{1}\right|^{2}+\left|\lambda _{2} \mu _{2}\right|^{2}+2\left|\lambda _{1}\mu _{2}\right|^{2}}{4 d_{1}^{2}}$.

\begin{figure}[!ht]
	\centering
	\subfigure[]{
		\includegraphics[width=6cm,height=4cm]{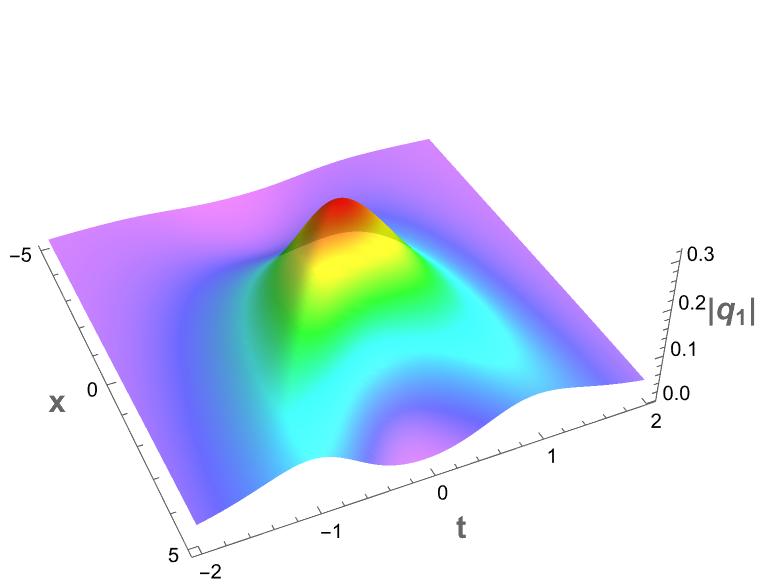} \label{Fig.7(a)}}
	\hspace{1.5mm}
	\subfigure[]{
		\includegraphics[width=6cm,height=4cm]{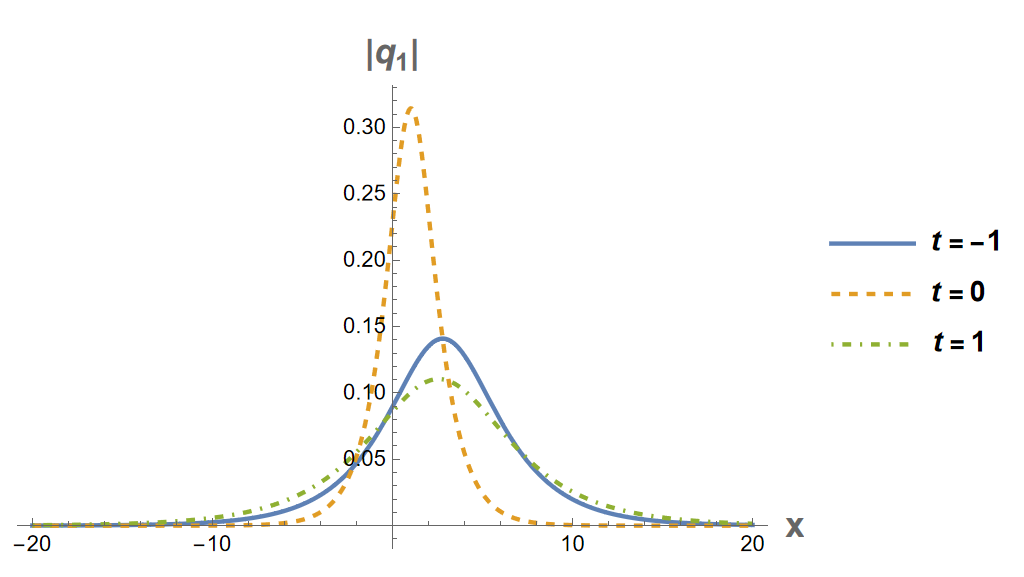} \label{Fig.7(b)}}	
	
	\caption{The one-soliton solution of the NTCmKdV-$\mathrm{\uppercase\expandafter{\romannumeral3}}$ equation \eqref{NTCmKdV3}. (a) Shape and motion of $|q_{1}|$ with $\gamma=3$, $a_{1}=0.1+0.8\mathrm{i}$,  $\lambda_{1}=-3+\mathrm{i}$, $\mu_{1}=2.5-0.4\mathrm{i}$ and $\lambda_{2}=0.7+1.6\mathrm{i}$. (b) Shapes of $|q_{1}|$ with the parameters as in (a) for $t=-1$, $t=0$, and $t=1$.}
	\label{tp7}
\end{figure}

Hence, the $\left|Q_{i j}\right|$ can be written as
\begin{equation}\label{3oneso}
	\begin{aligned}
	\left|Q_{i j}\right|=&\frac{|d_{1}|\left|\lambda _{j} \mu _{i}\right|}{\sqrt{\left|\lambda _{1} \mu _{1}\right|^{2}+\left|\lambda _{2} \mu _{2}\right|^{2}+2\left|\lambda _{1} \mu _{2}\right|^{2}}\cdot \left(\left(t+c_{1}\right)^{2}+d_{1}^{2}\right)\cdot \left|\gamma \right|} \\ &\cdot \operatorname{sech}\left(\frac{-2d_{1}}{\gamma \left(\left(t+c_{1}\right)^{2}+d_{1}^{2}\right)}x -\frac{8d_{1}\left(t+c_{1}\right)}{\gamma ^{3}\left(\left(t+c_{1}\right)^{2}+d_{1}^{2}\right)}+\frac{1}{2} \ln (\eta)-\ln (\gamma)\right).
	\end{aligned}
\end{equation}
From \eqref{3oneso} we can find that the one-soliton is characterized as follows.
\begin{center}
	\begin{varwidth}{\linewidth}
		\raggedright 
		amplitude: $\frac{|d_{1}|\left|\lambda _{j} \mu _{i}\right|}{\sqrt{\left|\lambda _{1} \mu _{1}\right|^{2}+\left|\lambda _{2} \mu _{2}\right|^{2}+2\left|\lambda _{1} \mu _{2}\right|^{2}}\cdot \left(\left(t+c_{1}\right)^{2}+d_{1}^{2}\right)\cdot \left|\gamma \right|},$ 
		
		top trace:  $x(t)=\frac{4\left(t+c_{1}\right)}{\gamma^{2}}+\frac{\gamma \left(\left(t+c_{1}\right)^{2}+d_{1}^{2}\right)\ln (\eta)}{4d_{1}}-\frac{\gamma \left(\left(t+c_{1}\right)^{2}+d_{1}^{2}\right)\ln (\gamma)}{2d_{1}}$, $\gamma\ne 0, d_{1}\ne 0,$ 
		
		velocity: $x^{\prime}(t)=\frac{4}{\gamma^{2}}+\frac{\gamma\left(t+c_{1}\right)\ln (\eta)}{2d_{1}}-\frac{\gamma\left(t+c_{1}\right)\ln (\gamma)}{d_{1}}$. 
	\end{varwidth}
\end{center}
The shape and motion of $|q_{1}|$ can be  observed in Fig. \ref{tp7}. It can be seen that the trace of the one-soliton solution is affected by $t^{-2}$ and moves along a curve.

\subsubsection{Two-soliton solution}\label{sec532}
For $N=2$, we have
\begin{equation}\label{cj6}
	\bm{K}_{1}=\left(\begin{array}{cc}
		k_{1} & \\
		& k_{2}
	\end{array}\right), \quad \bm{r}_{1}=\left(\begin{array}{cc}
		\delta_{1}^{(1)}\left(k_{1}\right) & \delta _{1}^{(2)}\left(k_{1}\right) \\
		\delta_{1}^{(1)}\left(k_{2}\right) & \delta_{1}^{(2)}\left(k_{2}\right)
	\end{array}\right), \quad \bm{s}_{1}=\left(\begin{array}{ll}
		\omega_{1}^{(1)}\left(k_{1}\right) & \omega _{1}^{(2)}\left(k_{1}\right) \\
		\omega_{1}^{(1)}\left(k_{2}\right) & \omega_{1}^{(2)}\left(k_{2}\right)
	\end{array}\right), 
\end{equation}
and
\begin{equation}
	k_{1}=\frac{1}{\mathrm{i}\gamma\left(t+a_{1}\right)},\quad	k_{2}=\frac{1}{\mathrm{i}\gamma\left(t+a_{2}\right)}.
\end{equation}
Thus, the explicit formula for the two-soliton solution in Eq. \eqref{twoj} can be obtained, where $m_{i, j}$  are represented as
\begin{equation}
	\begin{array}{l}
		m_{11}= \exp \left(\frac{-2d_{1}}{\gamma \left(\left(t+c_{1}\right)^{2}+d_{1}^{2}\right)}x -\frac{8d_{1}\left(t+c_{1}\right)}{\gamma ^{3}\left(\left(t+c_{1}\right)^{2}+d_{1}^{2}\right)}\right) \cdot\left(\lambda _{1} \mu_{1}^{*}+\lambda_{2} \mu_{2}^{*}\right)\cdot \frac{-2d_{1}}{\gamma ^{3}\left(\left(t+c_{1}\right)^{2}+d_{1}^{2}\right)}, \\
		m_{12}= \frac{1}{k_{1}\left(t\right)+k_{2}^{*}\left(t\right)} \cdot\exp \left(\left(k_{1}\left(t\right)+k_{2}^{*}\left(t\right)\right)x+\ln(k_{1}k_{2}^{*})+\frac{2\left({k_{2}^*(t)}^{2}-k_{1}(t)^2\right)}{\mathrm{i}\gamma}\right) \cdot\left(\lambda _{1} \mu_{1}^{*}+\lambda_{2} \mu_{2}^{*}\right), \\
		m_{21}= \frac{1}{k_{2}\left(t\right)+k_{1}^{*}\left(t\right)} \cdot\exp \left(\left(k_{2}\left(t\right)+k_{1}^{*}\left(t\right)\right)x+\ln(k_{2}k_{1}^{*})+\frac{2\left({k_{1}^*(t)}^{2}-k_{2}(t)^2\right)}{\mathrm{i}\gamma}\right) \cdot\left(\lambda _{1} \mu_{1}^{*}+\lambda_{2} \mu_{2}^{*}\right),\\
		m_{22}= \exp \left(\frac{-2d_{2}}{\gamma \left(\left(t+c_{2}\right)^{2}+d_{2}^{2}\right)}x -\frac{8d_{2}\left(t+c_{2}\right)}{\gamma ^{3}\left(\left(t+c_{2}\right)^{2}+d_{2}^{2}\right)}\right) \cdot\left(\lambda _{1} \mu_{1}^{*}+\lambda_{2} \mu_{2}^{*}\right)\cdot \frac{-2d_{2}}{\gamma ^{3}\left(\left(t+c_{2}\right)^{2}+d_{2}^{2}\right)}.
	\end{array}
\end{equation}

\begin{figure}[!ht]
	\centering
	\subfigure[]{
		\includegraphics[width=6cm,height=4cm]{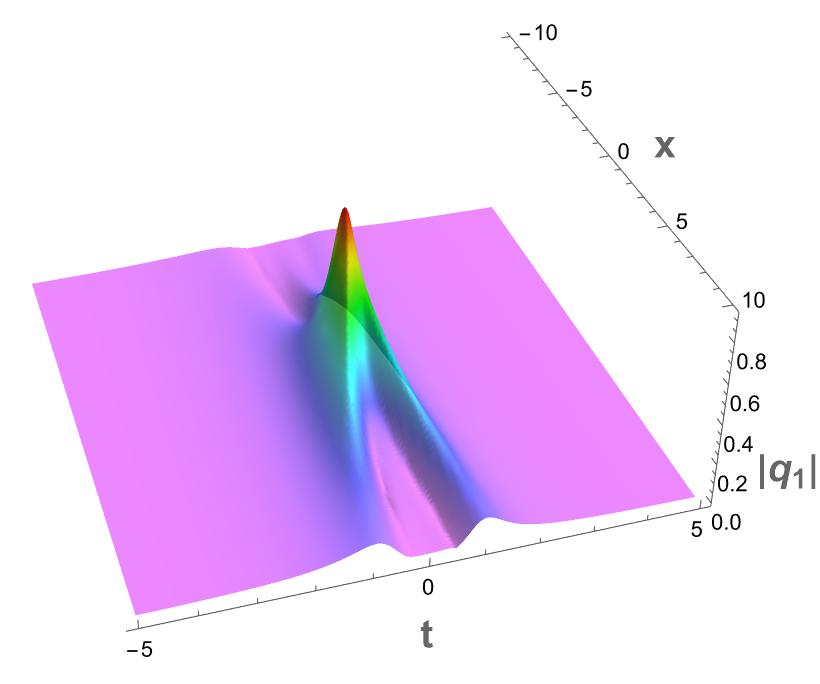} \label{Fig.8(a)}}
	\hspace{1.5mm}
	\subfigure[]{
		\includegraphics[width=6cm,height=4cm]{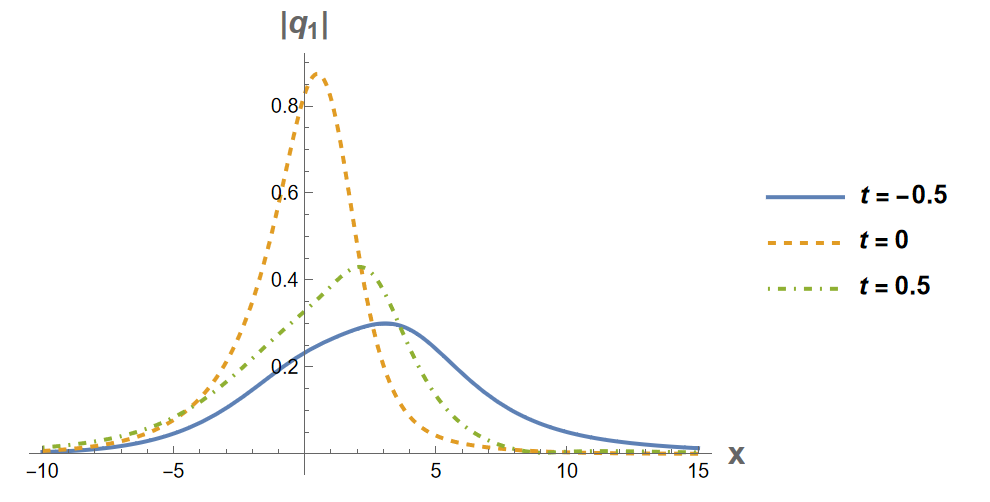} \label{Fig.8(b)}}	
	
	\caption{The two-soliton solution of the NTCmKdV-$\mathrm{\uppercase\expandafter{\romannumeral3}}$ equation \eqref{NTCmKdV3}. (a) Shape and motion of $|q_{1}|$ with $\gamma=5$, $a_{1}=0.1+0.8\mathrm{i}$, $a_{2}=-0.1+0.3\mathrm{i}$, $\lambda_{1}=-3+\mathrm{i}$, $\mu_{1}=-2.5-0.4\mathrm{i}$ and $\lambda_{2}=0.7+1.6\mathrm{i}$. (b) Shapes of $|q_{1}|$ with the parameters as in (a) for $t=-0.5$, $t=0$, and $t=0.5$.}
	\label{tp8}
\end{figure}

The shape and motion of $|q_{1}|$ can be  observed in Fig. \ref{tp8}. It can be seen in Fig. \ref{Fig.8(a)} that the two solitons move along the curve and then collide.

\subsubsection{Double-pole solution}\label{sec533}
The double-pole solution of the NTCmKdV-$\mathrm{\uppercase\expandafter{\romannumeral3}}$ equation \eqref{NTCmKdV2} can be  obtained by formula \eqref{dej}, where the definition of $\bm{K}_{1}$, $\bm{r}_{1}$ and $\bm{s}_{1}$ are shown in \eqref{cj3} and \eqref{kjgj3}. $m_{i,j}$ can be directly calculated through \eqref{mijys} as follows
\begin{subequations}
	\begin{align*}
		&\begin{aligned}
		m_{11}= \sum_{i=1}^{2}\left(\frac{\lambda _{i} \mu_{i}^{*}\cdot \exp{\left(L\right)}}{-2d_{1}\gamma}-\frac{l\lambda _{i} \mu_{i}^{*}\cdot \exp{\left(L\right)}}{4d_{1}^{2}\mathrm{i}\gamma\left(t+a_{1}^{*}\right)^{2}}V_{1}\right),
	   \end{aligned}\\
	   &\begin{aligned}
		m_{12}= \sum_{i=1}^{2}\frac{\lambda _{i} \mu_{i}^{*}\cdot \exp{\left(L\right)}}{-2d_{1}}V_{1}, 
	    \end{aligned}\\
	    &\begin{aligned}
		m_{21}=\sum_{i=1}^{2}\left(\frac{\lambda _{i} \mu_{i}^{*}\cdot \exp{\left(L\right)}}{-2d_{1}}V_{2}-\frac{l\lambda _{i} \mu_{i}^{*}\cdot \exp{\left(L\right)}}{4d_{1}^{2}\mathrm{i}\gamma \left(t+a_{1}\right)^{2}}-\frac{\gamma l \lambda _{i} \mu_{i}^{*}\cdot \exp{\left(L\right)}}{4d_{1}^{2}\mathrm{i}  \left(t+a_{1}\right)^{2}}V_{1}V_{2}-\frac{l^{2}\lambda _{i} \mu_{i}^{*}\cdot \exp{\left(L\right)}}{4d_{1}^{3}\left(t+a_{1}\right)^{2}\left(t+a_{1}^{*}\right)^{2}}V_{1}\right), 
	\end{aligned}\\
	&\begin{aligned}
		m_{22}= \sum_{i=1}^{2}\left(\frac{\gamma l\lambda _{i} \mu_{i}^{*}\cdot \exp{\left(L\right)}}{-2d_{1}}  V_{1}V_{2}+\frac{l\lambda _{i} \mu_{i}^{*}\cdot \exp{\left(L\right)}}{4d_{1}^{2}\mathrm{i} \gamma \left(t+a_{1}\right)^{2}}V_{1}\right).
	\end{aligned}
\end{align*}
\end{subequations}
with
\begin{equation*}
	\begin{aligned}
	&l=\left(t+c_{1}\right)^{2}+d_{1}^{2},\quad L=\frac{-2d_{1}}{\gamma l}x -\frac{8d_{1}\left(t+c_{1}\right)}{\gamma ^{3}l},\\
&V_{1}=\frac{x}{\mathrm{i}\gamma \left(t+a_{1}^{*}\right)^{2}} -\frac{1}{t+a_{1}^{*}} +\frac{4}{\mathrm{i}\gamma^{3} \left(t+a_{1}^{*}\right)^{3}}, \quad V{2}=\frac{-x}{\mathrm{i}\gamma \left(t+a_{1}\right)^{2}} -\frac{1}{t+a_{1}} -\frac{4}{\mathrm{i}\gamma^{3} \left(t+a_{1}\right)^{3}}.
\end{aligned}
\end{equation*}

\begin{figure}[!ht]
	\centering
	\subfigure[]{
		\includegraphics[width=6cm,height=4cm]{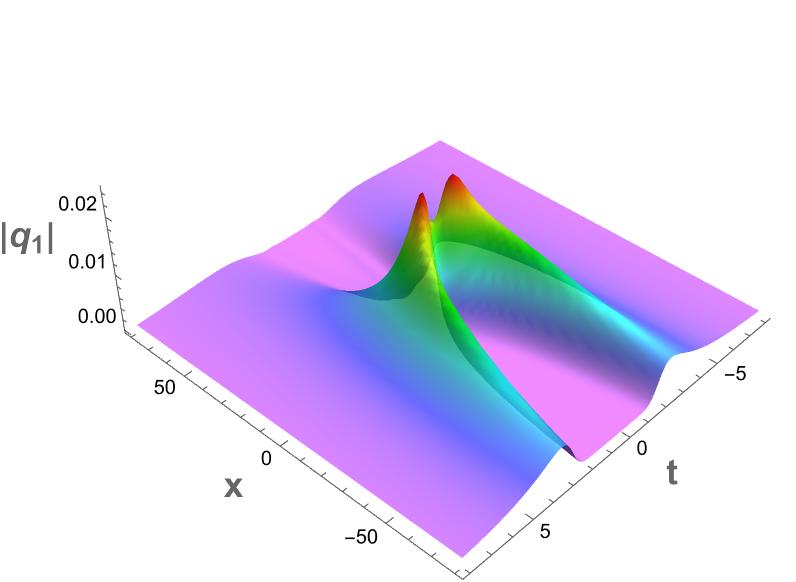} \label{Fig.9(a)}}
	\hspace{1.5mm}
	\subfigure[]{
		\includegraphics[width=6cm,height=4cm]{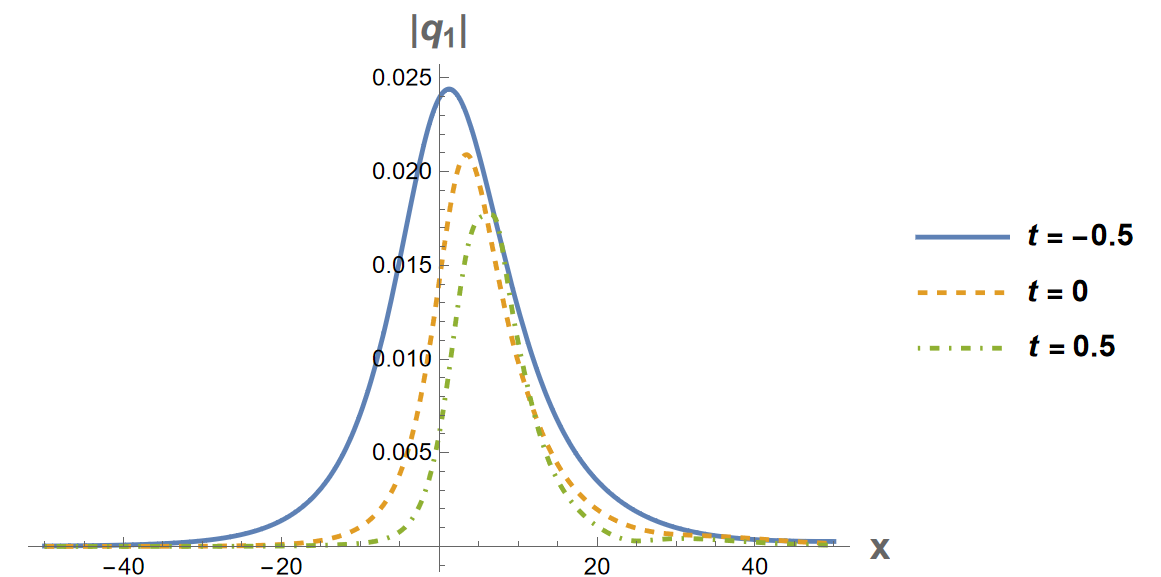} \label{Fig.9(b)}}	
	
	\caption{The double-pole solution of the NTCmKdV-$\mathrm{\uppercase\expandafter{\romannumeral3}}$ equation \eqref{NTCmKdV3}. (a) Shape and motion of $|q_{1}|$ with $\gamma=10$, $a_{1}=-0.4+1\mathrm{i}$, $\lambda_{1}=-3+\mathrm{i}$, $\mu_{1}=2.5-0.4\mathrm{i}$ and $\lambda_{2}=0.7+1.6\mathrm{i}$. (b) Shapes of $|q_{1}|$ with the parameters as in (a) for $t=-0.5$, $t=0$, and $t=0.5$.
	}
	\label{tp9}
\end{figure}

The shape and motion of $|q_{1}|$ can be  observed in Fig. \ref{tp9}. Fig. \ref{Fig.9(a)} reveals a solitons interaction characterized by two pronounced wave peaks, while Fig. \ref{Fig.9(b)} demonstrates a clear evolution of these peaks.

\begin{remark*}
The one-soliton, two-soliton, and double-pole solutions of the three NTCmKdV equations exhibit distinct explicit forms and dynamical behaviors due to differences in their respective time evolution characteristics. (i) One-soliton differences: NTCmKdV-$\mathrm{\uppercase\expandafter{\romannumeral1}}$, NTCmKdV-$\mathrm{\uppercase\expandafter{\romannumeral2}}$, NTCmKdV-$\mathrm{\uppercase\expandafter{\romannumeral3}}$ equations involve the $t^{3}$, $\mathrm{e}^{-t}$, and $t^{-2}$ terms, respectively. These differences result in distinctly different curvature of the soliton traces. (ii) Two-soliton differences: The interaction between solitons is stronger in NTCmKdV-$\mathrm{\uppercase\expandafter{\romannumeral2}}$ and NTCmKdV-$\mathrm{\uppercase\expandafter{\romannumeral3}}$ equations, while in NTCmKdV-$\mathrm{\uppercase\expandafter{\romannumeral1}}$ equation, the interaction is relatively weak. (iii)  Double-pole differences: NTCmKdV-$\mathrm{\uppercase\expandafter{\romannumeral1}}$ equation produces two localized and separate peaks; NTCmKdV-$\mathrm{\uppercase\expandafter{\romannumeral2}}$ equation generates two non-intersecting waves; and NTCmKdV-$\mathrm{\uppercase\expandafter{\romannumeral3}}$ equation exhibits two peaks arising from nonlinear interaction.
\end{remark*}

\section{Conclusion}\label{sec6}
In this paper, we have applied the Cauchy matrix approach to the NTCmKdV equations. We systematically derived four types of time-evolution relations $\bm{K}_{t}=-\mathrm{i}\alpha\bm{I}$, $\bm{K}_{t}=-\beta \bm{K}(t)$, $\bm{K}_{t}=-\mathrm{i}\gamma\bm{K}^{2}(t)$, and $\bm{K}_{t}=4\bm{K}^{3}(t)$ for the matrix function $\bm{K}$ in the Sylvester equation. Based on these relations and the structure of the master function $\bm{S}^{(i, j)}$, we constructed three types of unreduced nonisospectral systems \eqref{wyh1}, \eqref{wyh2} and \eqref{wyh3}. In contrast to the isospectral case, where $\bm{K}$ remains a constant matrix and its canonical forms enable the classification of solutions, the nonisospectral setting no longer permits such classification due to the time-dependence of $\bm{K}$. Therefore, we consider the case where $\bm{K}$ is diagonal to obtain  $N$-soliton solutions and the case where $\bm{K}$ is the lower triangular Toeplitz to obtain multi-pole solutions. This study extends the matrix mKdV framework to the nonisospectral system and enriches the structure of its solutions. Subsequently, we imposed complex reduction on the unreduced matrix nonisospectral systems to obtain the NTCmKdV equations  \eqref{NTCmKdV1}, \eqref{NTCmKdV2}, \eqref{NTCmKdV3}. The main difference between isospectral and nonisospectral equations lies in the fact that the coefficients of nonisospectral equations involve the variable $x$. Specifically, NTCmKdV-$\mathrm{\uppercase\expandafter{\romannumeral1}}$ equation \eqref{NTCmKdV1} can be regarded as an isospectral equation coupled with the terms $-2\mathrm{i}\alpha xq_{i} \left(i=0,1,2\right)$; NTCmKdV-$\mathrm{\uppercase\expandafter{\romannumeral2}}$ \eqref{NTCmKdV2} can be regarded as an isospectral equation coupled with the terms $\beta\left(x q_{i}\right)_{x} \left(i=0,1,2\right)$; and  NTCmKdV-$\mathrm{\uppercase\expandafter{\romannumeral3}}$ equation \eqref{NTCmKdV3} can be regarded as an isospectral equation coupled with the second-order derivative terms $-\frac{1}{2} \mathrm{i} \gamma xq_{i,xx} \left(i=0,1,2\right)$, nonlinear terms $-\mathrm{i} \gamma x|q_{i}|^{2} \left(i=0,1,2\right)$, and the integral terms $\partial^{-1}_{x}$. The solution properties of these three types of equations are different. Finally, we give the explicit formulas and analyze the dynamical behavior of the one-soliton, two-soliton and the double-pole solution of NTCmKdV equations, respectively.

At present, we have considered only one complex reduction. When  $\bm{K}_{2}=\bm{K}_{1}^{*}$, a nonisospectral nonlocal three-component mKdV equation can be derived, and when $\bm{K}_{2}=-\bm{K}_{1}$, a nonisospectral real three-component mKdV equation can be derived. These two cases will be further investigated in our future research. 

\section*{Acknowledgements}
The authors are very grateful for Prof. Dajun Zhang’s guidance. This work was supported by the National Natural Science Foundation of China (Grant. No.12171475) and Beijing Natural Science Foundation (No.1252012).

\end{document}